\documentclass[a4paper,12pt]{article}
\usepackage{harvard}
\usepackage{booktabs}
\usepackage[pdftex]{graphicx}
\usepackage{threeparttable}
\usepackage{hyperref}
\usepackage{authblk}
\usepackage{tabularx}
\usepackage{multirow}
\usepackage{ltxtable}
\usepackage{longtable,pdflscape}
\usepackage[capposition=top]{floatrow}
\usepackage{rotating}
\usepackage{booktabs}
\usepackage{float}
\usepackage[fleqn]{amsmath}
\usepackage{endnotes}
\let\footnote=\endnote
\usepackage[greek,english]{babel}
\usepackage[utf8x]{inputenx}
\usepackage[LGR, T1]{fontenc}
\newcommand{\gr}{\selectlanguage{greek}}

\usepackage[margin=1in]{geometry} 
\usepackage{array}
\usepackage{hyperref}
\usepackage{authblk}
\usepackage[section]{placeins}

\begin{document}

\date{\today}

\author[1]{Bastiaan Bruinsma}
\author[2]{Kostas Gemenis}
\affil[1]{Scuola Normale Superiore}
\affil[2]{University of Twente}
\renewcommand\Authands{ and }

\title{Validating Wordscores}

\maketitle

\begin{abstract}
\noindent \emph{Wordscores} is a popular quantitative text scaling method to estimate parties' positions on a priori specified dimensions, without requiring the researchers to read or even understand the language in the documents they are analysing. This study tries to establish whereas \emph{Wordscores} is able to deliver this promise by conducting a rigorous validation of its output using the Euromanifestos of 164 parties across 23 countries. We assess content validity by looking at the scored words in their context, criterion validity by comparing the \emph{Wordscores} output to expert surveys and other judgemental estimates of party positions, and construct validity by using the \emph{Wordscores} estimates to predict party membership in the European Parliament groups. We conclude that, despite the promises, \emph{Wordscores} fails to deliver valid party positions, and outline three conditions under which its performance can be improved.
\end{abstract}

\noindent \thanks{Paper presented at the `Text in Politics' Workshop PoliticologenEtmaal, Leiden University 1-2 June 2017. Earlier versions of this paper were presented at the Text Analysis Lecture Series, University of Amsterdam, 12 May 2016, at the POLSENT Workshop on Politics and the Media, Dublin City University, 15 July 2015, and at the 5th Annual General Conference of the European Political Science Association, Vienna, 25--27 June 2015. Bastiaan Bruinsma acknowledges financial support from a Twente Graduate School Bridging Grant.}

\newpage

\section*{Introduction}

The empirical evaluation of many theories in comparative politics, ranging from government coalitions to voting behaviour, requires data on the policy positions of political parties. Yet, despite the promise and availability of several cross-national data sources, the methods used to estimate parties' positions continue to be a highly contested area of political science. In the debate regarding the appropriateness of competing methods, the computer-assisted analysis in political text has offered particularly promising insights \cite{Grimmer2013}. One prominent method in this area is the \emph{Wordscores} scaling method as proposed by \citeasnoun{Laver2003}. \emph{Wordscores} can be seen as an application of correspondence analysis to words as data \cite[366--368]{Lowe2008}. In a nutshell, the vocabulary of a set of `reference' texts for which the position on the dimension of interest is known is used as a training set for estimating the unknown positions of another set of `virgin' texts.

To position documents and hence political actors, \emph{Wordscores} makes a series of assumptions regarding the distribution of reference documents across the dimension of interest, the distribution of words across reference documents, and of the use of words as data more generally \cite{Lowe2008}. As \citeasnoun{Grimmer2013} note, however, most of these assumptions might not hold in practice, so it is absolutely important to evaluate the performance of computer-assisted methods for analysing political text. Nevertheless, despite the `validate, validate, validate' recommendation by \citeasnoun{Grimmer2013}, our review of the published studies using \emph{Wordscores} revealed that there are very few studies that assessed the validity of \emph{Wordscores} output. Moreover, most of the few attempts that tried to assess the validity of \emph{Wordscores} in the context of estimating parties' positions were rather limited in terms of their scope. 

In this paper, we present the most rigorous approach to date in validating \emph{Wordscores}.\footnote{Full replication material, including .do files and all associated source documents, will be made available through a public data-verse on publication.} After a short explanation of the \emph{Wordscores} assumptions, we review the previous attempts to validate the \emph{Wordscores} output and outline the design of our study. Our analysis consists of an extensive application of \emph{Wordscores} to estimate the positions of 164 parties across 23 countries over four widely-used policy dimensions. We furthermore check the robustness of our estimation employing multiple reference scores for the reference texts and methods of transforming the raw \emph{Wordscores} output. Following estimation, we attempt a rigorous assessment of validity in the framework laid out by \citeasnoun{Carmines1979}. We conclude that, despite the promise in the original expos\'e \cite{Laver2003}, \emph{Wordscores} cannot produce valid estimates of parties' positions in a cross-national context. Our findings have important implications for those who use \emph{Wordscores} in their empirical analyses.

\section*{\emph{Wordscores} as a popular method of automated text analysis}

The \emph{Wordscores} method was originally proposed by \citeasnoun{Laver2003}. According to the method, it is possible to estimate the positions of documents (called `virgin' texts) on an a priori defined dimension of interest, by comparing them to a set of documents (called `reference' texts) in which their position on the dimension of interest is known. \emph{Wordscores} can therefore be described as a supervised scaling model \cite{Grimmer2013}, in the sense that documents are placed on a priori defined policy scales, that it uses `reference texts' and scores assigned to them akin to a training set in a machine learning framework. As such, \emph{Wordscores} makes the `bag-of-words' assumption by treating individual words as `data' irrespective of their syntactic context, and assumes that the relative frequencies of specific words provide manifestations of underlying political positions \cite[748]{Klemmensen2007}.

Over the years, \emph{Wordscores} has proven to be highly popular due to its ease of use and implementation in two popular statistical programmes (Stata and R). As of October 2016, Google Scholar gives 1021 citations to \citeasnoun{Laver2003} who introduced \emph{Wordscores} (hereafter Laver et al.). Some of the most prominent applications applications of the method involve the analysis of election manifestos to estimate the policy preferences of political parties and use these measurements in order to empirically test a wide range of questions. For instance, \emph{Wordscores} has been used to explain government coalitions at the national and sub-national level \cite{Back2013,Debus2009a,Linhart2010,Proksch2006}, to study party competition by mapping parties in multi-dimensional ideological space \cite{Laver2006}, to study similarity in the context of intra-party politics \cite{Coffe2011,Debus2009b}, to investigate whether parties keep their policy promises \cite{Debus2008}, to explain the success of bills in legislatures \cite{Brunner2008}, the choice of putting the EU's constitutional treaty on a referendum \cite{Hug2007a}, and to establish the policy preferences of sub-national parties and governments \cite{Klingelhofer2014,Muller2009}, or simply to map the positions of political parties across time \cite{Kritzinger2004}. 

Moreover, \emph{Wordscores} has been used extensively to estimate the positions of documents other than party manifestos. These include speeches delivered by MPs in Ireland, Italy, Germany, and Spain \cite{Bernauer2009,Giannetti2009,Laver2002,Leonisio2012}, speeches by US state governors \cite{Weinberg2010}, leaders of Russian regional parliaments \cite{Baturo2013}, delegates at the Convention on the Future of Europe \cite**{Benoit2005} and the head of state in the UK \cite{Hakhverdian2009}. Furthermore, novel applications of \emph{Wordscores} outside comparative politics include analyses of reports from US state lotteries \cite{Charbonneau2009}, Chinese newspaper articles \cite{Chen2011}, public statements by US Senators justifying their votes \cite{Bertelli2006}, advocacy briefs in the US Supreme Court \cite{Evans2007a}, press releases of the European Commission \cite{Kluver2009}, and even open-ended questions in surveys \cite{Baek2011}.

\begin{figure}[!htb]
\caption{Analysis of citations to Laver et al. article}
\includegraphics[width=.45\textwidth]{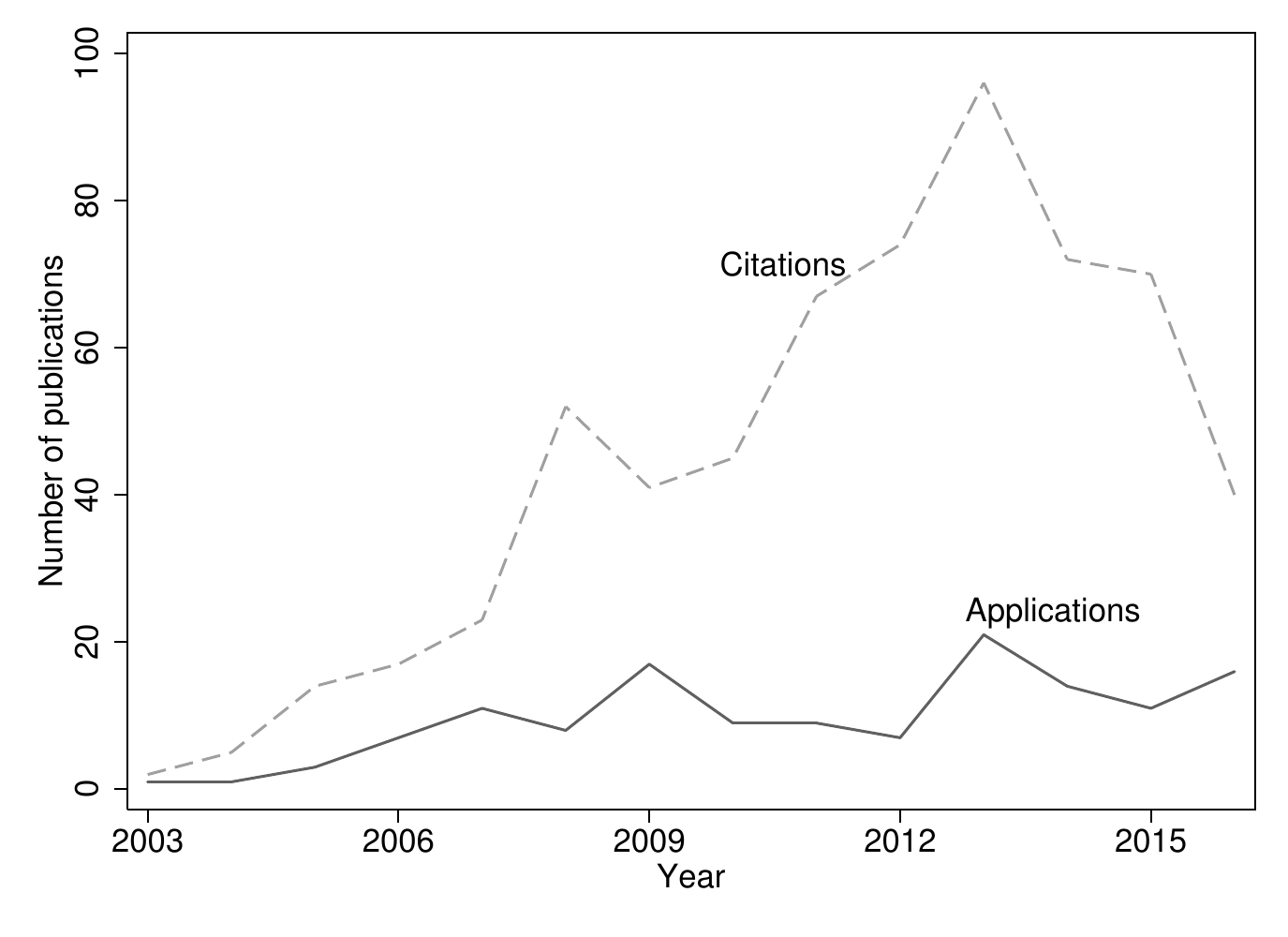}
\includegraphics[width=.45\textwidth]{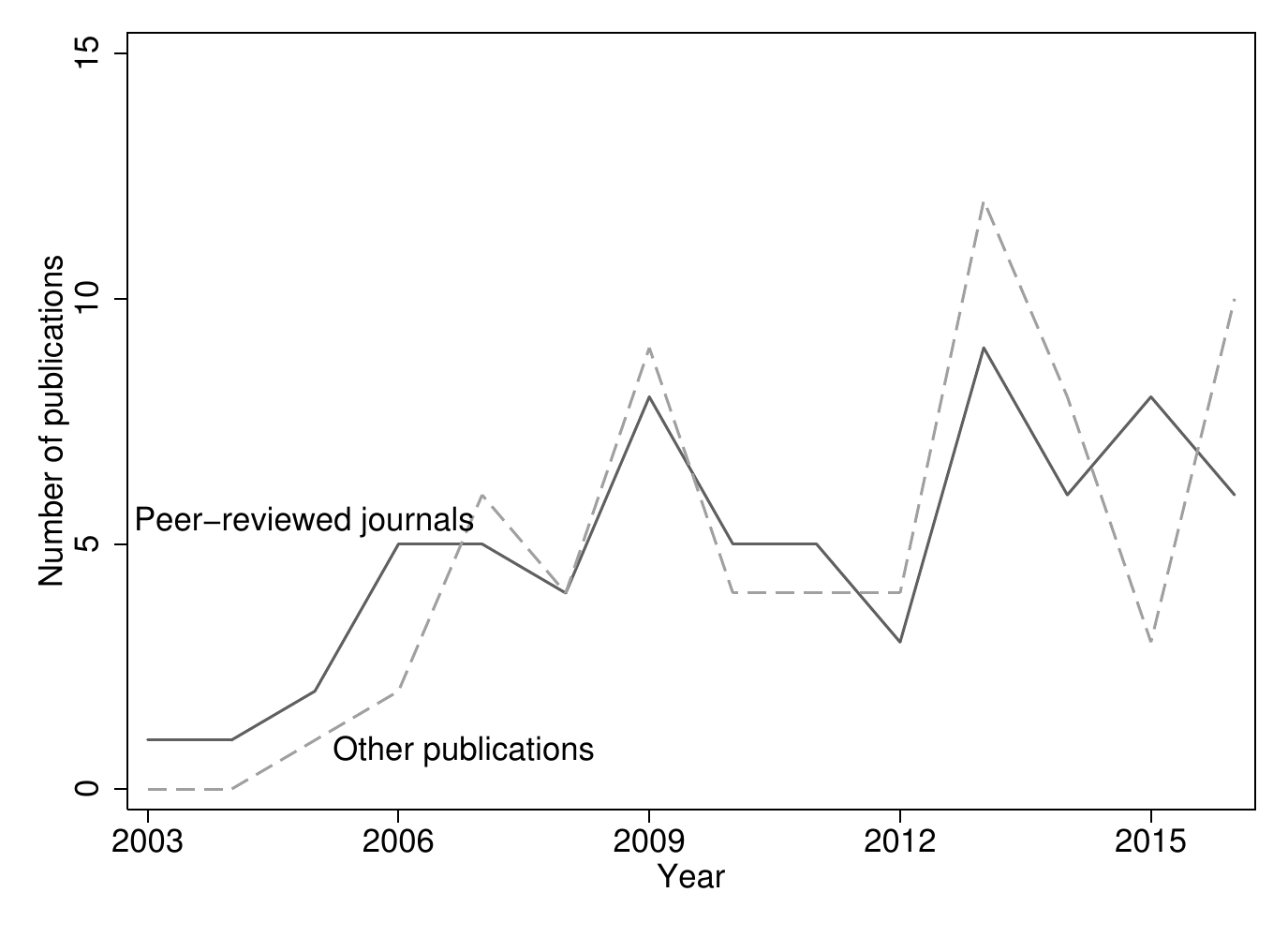}
\floatfoot{Note: The plot on the left shows mere citations compared to empirical applications, while the plot on the right shows the empirical applications published in peer-reviewed journals compared to other outlets.}
\end{figure}

Despite this breath and wealth of applications, one could argue that \emph{Wordscores} is becoming increasingly outdated as a method, especially due to the advent of more sophisticated methods of automated text analysis in political science \citeaffixed{Grimmer2013}{see}. To investigate this possibility, we performed a rigorous review of all the citations to Laver et al. article that were captured by Google Scholar.\footnote{A spreadsheet with the details of the review can be found in the replication materials.} Our review revealed that there are total of 146 uses of \emph{Wordscores} in empirical analyses, 78 of which have been published in peer-review journals, with the remaining appearing in monographs, chapters in edited volumes, working papers, and conference papers. Interestingly, as Figure 1 shows, the publication of empirical analyses using \emph{Wordscores} constitute a relatively stable fraction of the total citations to the Laver et al. article, whereas the trend of the publications of empirical analyses in peer-review journals closely mirrors the trend of publications in other outlets. Finally, as shown in Figure 2, our review shows no evidence that the empirical analyses using \emph{Wordscores} are now published in lesser quality journals (at least judging from their impact factor) compared to previous years. We therefore conclude that, despite the advent of more sophisticated methods of automated text analysis, \emph{Wordscores} deserves a rigorous evaluation in its own right as it remains a popular automated text analysis method in the literature.

\begin{figure}[!htb]
\caption{Journal impact factors of articles using \emph{Wordscores} in empirical analyses.}
\includegraphics[width=.5\textwidth]{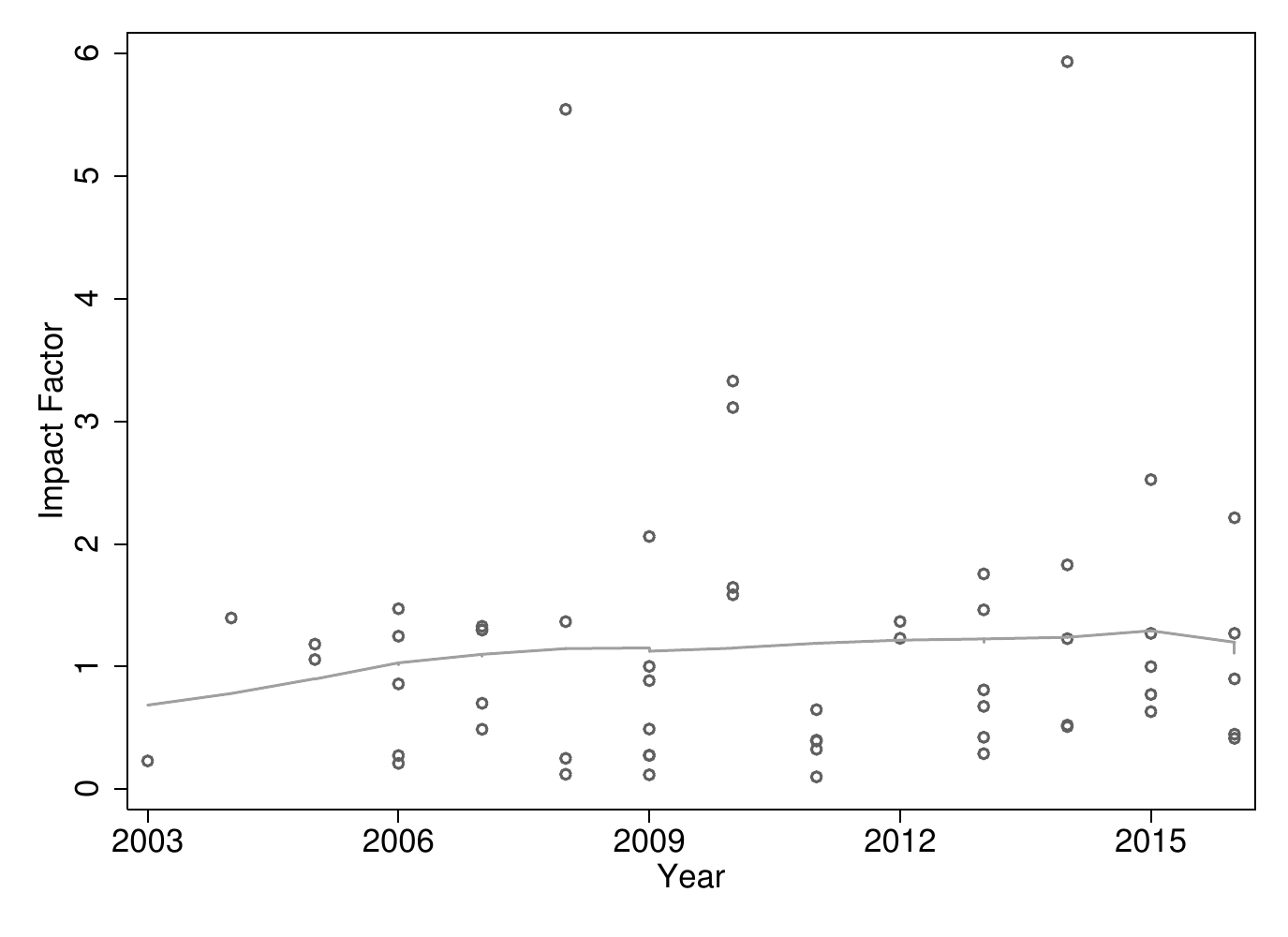}
\floatfoot{Note: Trend line is a locally adjusted regression curve (loess, bandwidth=.7).}
\end{figure}

\section*{Estimation and assumptions}

The estimation process begins with the researcher defining a set of reference texts that have positions on a dimension that we can assume with some confidence (for example, when they are obtained by an expert survey). Reference texts therefore need to be informative with regards to their content (words), and need to have a known position on the dimension of interest. \emph{Wordscores}, implemented as a user-written package in Stata and R, begins by counting the frequency of words in each reference text and assigns a score to each of these words. To do so, \emph{Wordscores} calculates the probability $P$ that a word $w$ appears in reference text $r$ as follows:

\begin{equation}
P_{wr}=\frac{F_{wr}}{\sum_{r}F_{wr}}
\end{equation}

where $F_{wr}$ is the frequency of word $w$ in reference text $r$. Using these probabilities, \emph{Wordscores} calculates a score for each word $w$ on each dimension of interest $d$ as follows:

\begin{equation}
S_{wd}=\sum_{r}P_{wr}A_{rd} 
\end{equation}

where $A_{rd}$ is the known position of reference text $r$ on dimension $d$. To score each virgin text $v$ on dimension $d$, \emph{Wordscores} use the word scores $S_{wd}$ obtained from reference texts as follows:

\begin{equation}
S_{vd}=\sum_{w}F_{wv}S_{wd} 
\end{equation}

According to \citeasnoun[316]{Laver2003}, $F_{wv}$ in equation 3 denotes `the relative frequency of each virgin text word [$w$], as a proportion of the \emph{total number of words in the virgin text} [$v$]' (emphasis added). However, all the statistical packages that have been written to implement \emph{Wordscores},\footnote{These are the \texttt{wordscores} package in Stata (written by Kenneth Benoit), and the \texttt{austin} (written by Will Lowe) and \texttt{quanteda} (written by Kenneth Benoit and Paul Nulty) packages in R.} use a different definition of $F_{wv}$. Here the relative frequency of each virgin text word $w$ is taken as a proportion of the total number of words \emph{co-occurring between the reference and the virgin texts}. This inconsistency between the Laver et al. article and the software implementations is of no particular concern to how \emph{Wordscores} work, but it does challenge the proof-of-concept validation presented in the Laver et al. article as we will see in the following section.

Nevertheless, irrespective of how one defines $F_{wv}$, the $S_{vd}$ scores only indicate the relative position of virgin texts to each other on dimension $d$. To be able to compare the scores of virgin texts to the scores of reference texts, we need one more step. \emph{Wordscores} will transform the raw scores back to the original metric used in the scores used in the reference texts, as this allows us to compare the raw scores of the virgin texts with the assigned scores of the reference texts. In their original paper, Laver et al. suggest the following transformation:

\begin{equation}
S^{*}_{vd}=(S_{vd}-S_{\bar{vd}})\left(\frac{SD_{rd}}{SD_{vd}}\right)+S_{\bar{vd}}
\end{equation}

Here, $S^{*}_{vd}$ is the transformed score, $S_{vd}$ the raw score, $S_{\bar{vd}}$ the average raw score of the virgin texts, and $SD_{rd}$ and $SD_{vd}$ the standard deviations of the reference and virgin text scores respectively. This metric preserves the mean of the virgin text scores, but equals their variance to that of the reference text scores, thus allowing for comparison.

\citeasnoun{Lowe2008} points out that the LBG transformation assumes that the raw virgin text scores have the correct mean, but the incorrect variance. However, due to the large amount of overlapping words, the virgin score mean is invariably close to the reference text mean---an effect called shrinkage. These overlapping words are often words as `the' or `and', and as they occur frequently in all documents, they get centrist scores. As such, the distances between the virgin texts are shrunken, and all texts bounce towards the middle of the scale. Laver et al. fix this by recouping the original variance, but falsely assume that the newly derived mean is correct. This is no problem when the variance and mean are expected to be the same for both reference and virgin texts. However, as \citeasnoun[359--360]{Lowe2008} notes, increasing polarisation between parties, or joint movement to the sides of a set of parties, makes it hard, if impossible, to discern whether the mean of the virgin texts is centrist due to the reference scores or a shrinkage artifact.

\citeasnoun[95--97]{Martin2008} agree with the above criticism and note several more shortcomings of the Laver et al. transformation method. First, as the transformation uses the standard deviation of the virgin text raw scores it depends on the set of virgin texts themselves. This makes the scores non-robust with regard to the virgin texts, and any difference in the set of reference texts automatically leads to a difference in the scores. This way, a researcher could obtain different positions in the virgin texts solely because of a different selection in the reference texts. Second, despite what Laver et al. claim, their method fails to recover the accurate relative distance ratios and therefore put the transformed scores and the virgin scores on the same metric. This is due to shrinkage, as we pointed out above. To combat these problems, \citeasnoun{Martin2008} provide a new transformation based on the idea of relative distance ratios $S_{i}$:

\begin{equation}
S_{i}=\frac{S_{i}-S_{R1}}{S_{R1}-S_{R2}}
\end{equation}

where two `anchoring texts' $S_{R1}$ and $S_{R2}$ are chosen, and the placement of all other texts are expressed in relation to this `standard unit' \cite[97]{Martin2008}. They then use these ratios to construct a new transformation:

\begin{equation}
S^{*}_{vd}=\left((S_{vd}-S_{R1})\frac{A_{R2}-A_{R1}}{S_{R2}-S_{R1}}\right)+A_{R1} 
\end{equation}

Here, $S^{*}_{vd}$ is the transformed score, $S_{vd}$ the raw score, $A_{R1}$ and $A_{R1}$ are the assigned scores to reference texts $R1$ and $R2$ (where $R1$ is located to the left of $R2$), and $S_{R1}$ and $S_{R2}$ are the reference texts' raw scores. In their article, Martin \& Vanberg use two reference texts, or `anchor texts' located to the left and right of virgin texts. As seen in equation (6) above, both the assigned scores for the reference texts are recovered, and the virgin texts are thus placed on the original metric. However, as soon as more than two reference texts are used---as \citeasnoun{Laver2003} strongly advise---not all the original exogenous scores of the reference texts can be recovered exactly, as only two texts can be used to define the metric. MV thus suggest a change to the transformation:

\begin{equation}
S^{*}_{vd}=\left((S_{vd}-S_{Rmin})\frac{A_{Rmax}-A_{Rmin}}{S_{R2}-S_{R1}}\right)+A_{Rmin} 
\end{equation}

Here $A_{Rmin}$ and $A_{Rmax}$ denote the lowest and highest placed reference text on the original metric. The positions of these texts will be recovered exactly, while the scores of the other texts will be distorted as the relative distance ratios of the raw scores do not correspond to the relative distance ratios of the reference scores. Comparison between reference and virgin texts thus becomes difficult and researchers face a trade-off between increased accuracy of the dictionary and internal consistency, and the ability to make valid comparisons \cite{Martin2008} (see Appendix F).

To conclude, while the transformation by \citeasnoun{Laver2003} depends on the virgin texts and is indifferent to the composition of the reference texts, the transformation by \citeasnoun{Martin2008} depends on the reference texts and is indifferent to the composition of the set of virgin texts \cite[360]{Lowe2008}. Moreover, Laver et al. assume that the variances of both the set of reference texts and virgin texts are the same, while the Martin \& Vanberg transformation does not do so \cite[110]{Benoit2008}. In this paper, we use both transformation methods as we have no use for the raw scores and neither of the scores has until now proven to be the most appropriate in all circumstances.

More generally, \citeasnoun{Lowe2008} criticised \emph{Wordscores} for its heavy dependence on reference texts. \citeasnoun[366--368]{Lowe2008} views \emph{Wordscores} as an approximation to correspondence analysis and goes on to treat the method as a statistical ideal point model for words. In doing so, he identified six conditions that \emph{Wordscores} needs to fulfil in order to ensure consistent and unbiased estimation of the parameters of the ideal point model:

\begin{enumerate}
\item The word scores of the virgin texts need to be equally spaced and extend over the whole range of word scores for the reference texts
\item The word scores of the virgin texts need to be spaced relative to the informativeness term (all texts are thus informative)
\item The reference scores of the reference texts need to be equally spaced and extend past each word score of the virgin texts in both directions
\item The word scores of the reference texts need to be spaced relative to the informativeness term (all texts are thus informative)
\item All the words need to be equally informative
\item The probability of seeing a word needs to be the same for all words
\end{enumerate}

According to \citeasnoun[369]{Lowe2008}, conditions 5 and 6 will never hold for word count data because any text exhibits a highly skewed word frequency distribution, regardless of the genre, and contain many uninformative words. Nevertheless, we can significantly reduce these problems by filtering out uninformative words such as stop words, function words that do not convey meaning but primarily serve grammatical functions, very uncommon words, and words which appear in less than 1\% and more than 99\% of documents in the corpus \cite{Grimmer2013}. Doing this makes the probability of seeing a word more equal, and removes non-informative words. 

Conditions 1 and 2 will be less likely to hold when there is not enough overlap between word distributions between the reference documents. However, by using many documents as reference texts (as Laver et al. advised), the conditions might be well approximated. Condition 2, however, suffers from the fact that some documents are small, and thus contain very little to no information. This does not only increase the confidence intervals around the estimates, but also creates a large bias in the estimates, negatively influence the validity of the virgin documents scores.

Conditions 3 and 4 are similar to 1 and 2, but as words are more plentiful then texts, the changes of insufficient overlap are considerably lower, and the conditions are thus less important. Lowe even states `we might hope that they [words] may relatively evenly spread out across a policy dimension' \cite[369]{Lowe2008}, which makes the conditions even more plausible. Last, \citeasnoun[369]{Lowe2008} considers that conditions 1 and 3 can never hold simultaneously, as this would require an infinite data set---and thus concludes that bias in \emph{Wordscores} is inevitable.

\section*{Previous validation attempts and their shortcomings}

Considering the comprehensive critique of \citeasnoun{Lowe2008} one could conclude that \emph{Wordscores} could find little use in political science. However, as \citeasnoun[270]{Grimmer2013} note, the question is not whether computer-assisted methods satisfy assumptions with regards to how language works and texts are generated, but to evaluate methods on the basis of `their ability to perform some useful social scientific task'. In this respect, we should not focus on the assumptions, but on validation. As \citeasnoun[271]{Grimmer2013} note, validation in supervised methods such as \emph{Wordscores} should involve demonstrating that the computer-assisted method can reproduce the results in a set of documents for which the true scores of the quantity of interest are known. When true scores are not known, the output of computer-assisted methods can be validated against human judgement \citeaffixed{Lowe2013}{see, for instance the validation of another method by}. 

Validation, however, is more difficult in the case of parties' ideological positions because the true scores of the quantity of interest are unknown and it is difficult to estimate them reliably using human judgement \citeaffixed{Mikhaylov2012}{see}. In such instances, researchers often resort to assessing the `face validity' of estimates of party positions, in other words whether positions `appear' to be valid in the eyes of the researcher. As \citeasnoun[363]{Sartori2007} pointed out, however, demonstrating a measure's face validity might be comforting when other types of validity cannot be employed due to the lack of resources, but this strategy is not adequate. Face validity should be seen as a necessary but not sufficient condition for good measurement. In the absence of face validity, one could certainly question the usefulness of the measuring instrument. However, face validity by itself is not enough, and researchers need to assess additional types of validity as outlined in Table 1 \cite{Carmines1979}. These three additional types of validity should not be considered interchangeable \cite[537]{Adcock2001}. If we fail to validate a measure in one type of validity, this cannot be compensated by showing that the measure fares well in terms of another. 

\begin{table}[!htb]\footnotesize{
\caption{Types of validity and their assessment.}
\begin{tabularx}{\textwidth}{l X l}
\toprule
Type& Assesses the degree to which our measure\ldots        & The assessment is\ldots \\
\midrule
Face             & \ldots appears to be valid in light of heuristic knowledge                                           & \ldots qualitative                      \\
Content          & \ldots contains indicators that reflect the construct that is being measured                         & \ldots qualitative                      \\
Criterion        & \ldots correlates with other known measures of the concept that is being measured                    & \ldots quantitative                     \\
Construct        & \ldots is associated with measures of other concepts in a way that conforms to the theoretical expectations & \ldots quantitative              \\
\bottomrule
\multicolumn{3}{l}{Adapted from \citeasnoun{Carmines1979} and \citeasnoun{Sartori2007}}\\
\end{tabularx}
}
\end{table}

More specifically, in the case of estimating parties' ideological positions \citeasnoun[271]{Grimmer2013} argue that validation `requires numerous and substance-based evaluations', and propose that `scholars must combine experimental, substantive, and statistical evidence' to demonstrate that the output of computer-assisted methods such as \emph{Wordscores} can be considered to be valid. Nevertheless, while these recommendations have been stated in classic works in social \cite{Zeller1980} and political science \cite{Adcock2001} measurement, and content analysis \cite{Krippendorff2004}, our review of the literature showed that most of the published studies have used the \emph{Wordscores} routines in Stata or R without validating the output. 

As expected, the first study that attempted to validate the \emph{Wordscores} output was the original article by \citeasnoun{Laver2003}. In their article, Laver et al. use the 1992 manifestos of British and Irish parties as reference texts and assign to them reference scores from  expert surveys conducted in 1992 in order to estimate parties' positions of the 1997 election manifestos in both economic and social policy dimensions. Laver et al. then assess the criterion validity of the estimates by comparing the \emph{Wordscores} output against the estimates of an expert survey conducted in 1997. Laver et al. also used a similar approach to estimate parties' positions for the German election of 1994 but, in lack of comparable expert survey data, only assessed the German estimates in terms of face validity. Our replication of the Laver et al. analysis not only revealed the inconsistencies between the definitions in the article and the way \emph{Wordscores} is implemented in R and Stata, but more importantly, that the results presented in the article are not particularly robust. More specifically, we found that the addition of manifestos of smaller parties in the analysis drastically change the estimates provided by \emph{Wordscores}, making them inconsistent in comparison to expert survey estimates. We report in detail these findings in Appendix A. Furthermore, we argue that if \emph{Wordscores} aims to be a useful tool for estimating parties' positions on policy dimensions, its validity needs to be evaluated beyond such simple `proof of concept' demonstrations, especially when these demonstrations are shown not to be robust. 

In this respect, \citeasnoun{Budge2007} compared the estimates given by \emph{Wordscores} to those of the Manifesto Project on the left-right dimension for British parties across time. Their results were unfavourable as they found that \emph{Wordscores} produces flat scores across time compared to the Manifesto Project estimates. However, in a response, \citeasnoun{Benoit2007} dismissed these findings because \emph{Wordscores} was not properly implemented (Budge \& Pennings merged several manifestos before using them as reference texts) and because the Manifesto Project estimates were used as a benchmark, something which, the authors argue, can easily be contested. 

\citeasnoun{Klemmensen2007} performed a similar evaluation by using \emph{Wordscores} to estimate the positions of Danish parties on the left-right dimension. Although their article has been widely cited as a successful validation of \emph{Wordscores}, a closer investigation of the results shows that this is not actually the case. The correlations reported by Klemmensen et al. show that \emph{Wordscores} performs worse than the Manifestos Project estimates when compared to a common benchmark (expert surveys). If the proponents of \emph{Wordscores} argue that the Manifesto Project estimates are problematic because they do not always correlate with expert surveys \citeaffixed{Benoit2007,Benoit2007a}{e.g.}, then it should follow that \emph{Wordscores} estimates are even worse.

Most recently, \citeasnoun**{Hjorth2015} have repeated this exercise in both Denmark and Germany, by validating the \emph{Wordscores} output against placements by experts and voters using rank order correlations. The results of this validation pointed that the  \emph{Wordscores} output correlated better with independent measures of party positions compared to the output produced by another popular text scaling method (\emph{Wordfish}). However, the rank order correlations examined by the authors produced a far too lenient test on a method which promises to deliver interval level measurements of party positions (point estimates with associated 95\% confidence intervals).

The most comprehensive validation so far has been conducted by \citeasnoun{Brauninger2013} who used \emph{Wordscores} to estimate parties' left-right positions across 13 West European countries between 1980 and 2010 in a study specifically aimed to assess the validity of the technique. Their results were mixed, concluding that \emph{Wordscores} estimates correlated well with the Manifesto Project in some countries, but not in others. We note that the results of this comparative study were far more cautious compared to the earlier investigations based on single countries (including the original proof of concept in Laver et al.). The Br\"auninger et al. study, however, had its own limitations namely that it only assessed estimates on a single dimension (left-right), using a single benchmark (the Manifesto Project data) which is controversial in itself as previously argued.\footnote{\citeasnoun{Ruedin2013} and \citeasnoun{Hug2007} compared \emph{Wordscores} estimates against many other methods aiming to measure parties' positions. Their comparisons, however, did not focus on \emph{Wordscores} as such but rather showed how results might differ across the various methods.} 

In general, all of the previous studies attempted to assess the validity of \emph{Wordscores} in the context of party positions, looked at criterion validity, neglecting other, equally important, types of validity as discussed above. Moreover, the correlation coefficients used to assess criterion validity were either Pearson's product-moment or Spearman's rank-order, which do not take into account systematic measurement error. Finally, none of the studies attempted to investigate the robustness of estimation by using difference sources for the reference scores and different transformation methods. Our study addresses all these limitations and provides the most rigorous validation approach to date. We use \emph{Wordscores} to estimate parties' positions in 23 countries, across four different policy/ideological dimensions, using three different sets of reference scores, and two different transformation methods, and we assess the estimates in terms of content, criterion, and construct validity using appropriate statistical measures.

\section*{Study design}

We applied \emph{Wordscores} to the manifestos of political parties published on the occasion of the 2009 elections to the European Parliament (hereafter we refer to these documents as `Euromanifestos') across 23 countries using the 2004 EP elections Euromanifestos as reference texts.\footnote{The countries in our study include all EU member-states up to 2009 with the exclusion of Luxembourg and Malta where no appropriate reference scores were available for 2004. The names of parties used in the study can be found in Appendix B.} We chose the elections to the EP over national elections to improve the comparability of estimates across countries. National elections contain more idiosyncratic parameters in the campaigning and use of political text compared to elections to the EP that take place at the same time and within a shared political context. Moreover, we avoid stretching the comparison across time (unlike Br{\"a}uninger et al.) in order to ensure that our comparisons are not affected by changes in the political discourse. This way we provide a very favourable context to test the validity of \emph{Wordscores}, much like Laver et al. have done so.

Instead of tracking down all these documents ourselves, we rely on an off-the-shelf collection provided by the Euromanifestos Project.\footnote{The collection can be accessed at http://www.ees-homepage.net/. The names of the documents used can be found in Appendix B. Moreover, following the advice by \citeasnoun[272--273]{Grimmer2013}, we processed these documents to make them suitable for computer-assisted analysis. We present our processing method in Appendix C.} These are the documents collected and coded (according to a hand-coding scheme similar to the Manifesto Project) by country-specific coders of the Euromanifestos Project \cite{Braun2010}. As also shown in the case of the Manifesto Project \cite{Gemenis2012,Hansen2008}, the collection of these documents is fraught with problems. Along with `genuine' Euromanifestos, the collection includes all sorts of documents of dubious usefulness in terms of estimating parties' positions. Amongst them, there are small pamphlets that do not present a broad policy profile, and documents that contain irrelevant or misleading sections (e.g. references to \emph{other} parties' positions). As evident, such documents would be highly problematic to use with computer-assisted methods for content analysis \citeaffixed{Proksch2009}{see}. We nevertheless decided to use this off-the-shelf database in order to test the method in a realistic context as researchers are more likely to rely on off-the-shelf collections for their cross-country comparative analyses than constructing their own using country experts \citeaffixed[328]{Hug2007a,Pennings2006}{e.g.}.

Unlike all the previous studies we do not limit our validation to the left-right dimension, but estimate parties' positions on three additional dimensions: European integration, economic left-right, and the socio-cultural liberal-conservative dimension. These are dimensions that have been used extensively to analyse party competition in the context of (elections to) the EP \cite{Hix1999,Hix2006a,Hooghe1999,Hooghe2002,McElroy2007}. In addition, unlike previous studies, we use a variety of sources for reference scores, and also various sources of party positions to compare the \emph{Wordscores} estimates against. To begin with, we do not use the estimates from the Manifesto Project as we agree with Benoit and Laver \citeyear{Benoit2007,Benoit2007a} that they are fraught with measurement error and, as such, should not be used as a `gold standard' for evaluating the validity of other methods. The reasons for doing so are further explained elsewhere \citeaffixed{Gemenis2013}{see}. Instead, we use expert survey estimates as Laver et al. and most of the empirical applications that we cited earlier on have done. Of course, expert surveys have their own problems, so we cross-validate the \emph{Wordscores} estimates using estimates from an alternative, less used, but highly useful approach: the judgemental estimation of party positions using manifestos and other document sources. For the advantages and shortcomings of the judgemental approach to coding see \citeasnoun[2293--2296]{Gemenis2015b}. We \emph{further} cross-validate the findings by employing two different data sources within each approach. For expert surveys, we use the 2003 \citeasnoun{Benoit2006} and the 2002 and 2010 Chapel Hill Expert Surveys \cite**{Bakker2012,Hooghe2010}, and for judgemental coding, the \emph{overall} position coders assigned to the party on the basis of the whole document in the Euromanifestos Project dataset \cite{Braun2010}, and the estimates from the 2009 EU Profiler dataset \cite{Trechsel2010} as scaled in \citeasnoun{Gemenis2013c}. Table 2 gives a summary of these sources, while the exact wording of questions and scales used in our our study are presented in Appendix D.

\begin{table}[!htb]\footnotesize{
\caption{Party position data sources used in this study.}
\begin{tabularx}{\textwidth}{X X X}
\toprule
Source type& Used for reference scores (2004) & Used for the validation (2009) \\
\midrule
Expert survey&BL 2003&-\\
Expert survey&CHES 2002&CHES 2010\\
Judgemental coding&EMP 2004&EMP 2009\\
Judgemental coding&-&EUP 2009\\
\bottomrule
\end{tabularx}
}
\end{table}

Finally, unlike previous studies we cross-validate the results by employing two different transformations for each set of \emph{Wordscores} estimates: the transformation originally proposed by Laver et al. (hereafter referred to as LBG) and the alternative transformation proposed by \citeasnoun{Martin2008}, hereafter referred to as MV.\footnote{Following, Laver et al. we use all available documents for 2004 as reference texts when using the LBG transformation. This way, the texts more or less extend over the whole range as required by the first assumption made by \emph{Wordscores} (see section on \emph{Wordscores} assumptions). In Appendix E, we show which two documents we selected for each country to serve as anchors for estimation according to the MV transformation.} The use of all of these sources and methods for transforming the raw scores allows us to perform the most extensive validation of \emph{Wordscores} to date.

\section*{Results}

The combination of different sources of reference scores and transformation over the examined methods and countries implies that we ran the \emph{Wordscores} scaling model a whooping 600 times for the validation: 25 countries/territories (including separate analyses for Flanders, Wallonia, and Northern Ireland)*4 dimensions*3 sources of reference scores*2 transformation methods. All the \emph{Wordscores} estimates from these analyses were copied to a meta-dataset with parties as the unit of analysis and merged with estimates from the sources listed in the last column of Table 2. This meta-dataset was used for the subsequent analyses presented below.

\subsection*{Content validity}

According to \citeasnoun{Carmines1979}, content validity refers to whether the method used for measuring a latent construct represents all of its facets. If one uses multiple indicators that are scaled in a single index, then these indicators should represent all facets of the construct. Alternatively, if one uses a single indicator (for instance as done in surveys asking for a left-right placement) then this indicator has to capture all different facets of the construct. Moreover, a measure that includes facets that do not belong to the construct would be problematic in terms of content validity. As noted in the section about the previous validation attempts, the evaluation of content validity is usually of qualitative nature, so it would be difficult to see how it could be assessed in the context of the output presented by \emph{Wordscores}. We propose a workaround this problem by conceptualising the construct in the context of \emph{Wordscores} as being represented by the words used in the reference texts.

When \emph{Wordscores} places virgin texts on a dimension of interest it does so by calculating a wordscore for each of the words occurring in the reference texts. As \emph{Wordscores} is non-discriminating and scores all words on all dimensions, treating all words as equally informative of the dimension of interest is problematic in terms of content validity. This is because we should not expect each and every word in a reference text to be associated with a dimension of interest, no matter what this dimension is. This problem of \emph{Wordscores} is known, of course, but here we are interested in quantifying the degree of content validity in order to investigate how big of a problem it is for estimating parties' positions.

To do so, we decided to treat each of the words scored in the reference texts as an indicator of the latent concept, and evaluated whether these words relate to the latent concept/dimension of interest. To assess this, following \citeasnoun[101--102]{Krippendorff2004} we looked at the context in which these word appear. For example, the word `committee' can be indicative of a party's position in the dimension of EU integration when it refers to an EU committee, but not when it refers to other types of committees. We therefore hand-coded \emph{each and every word} in the reference texts to see how many of the words used to score the virgin texts were actually used in the context of the dimension of interest. As this is a particularly time-consuming process, we restricted this analysis to British documents and the European integration dimension. Our choice of British parties should be fair for \emph{Wordscores} given that British Euromanifestos are some of the best documents in terms of relevance for assessing parties' positions on European integration. For our hand-coding exercise we defined the context as a natural sentence that starts with a capital letter, and end with one of the following delimiters: `.', `?', `!', `;'  \cite**[942]{Daubler2012}. Items in (bullet-pointed) lists were considered as separate sentences. Each word was coded as one (1) when it was used in a context referring to European integration and zero (0) otherwise. In Figure 3 we plot the distribution of the average hand-coding evaluation for among all the words used in each virgin document of each British party. What is clear from the figure is that the vast majority of words used by \emph{Wordscores} to estimate party positions are not particularly informative if one looks at the context in which they appear. It appears that \emph{Wordscores} uses far more noise than signal to estimate party positions.

\begin{figure}[!htb]
\caption{Assessing content validity in the European integration dimension.}
\includegraphics[width=1\textwidth]{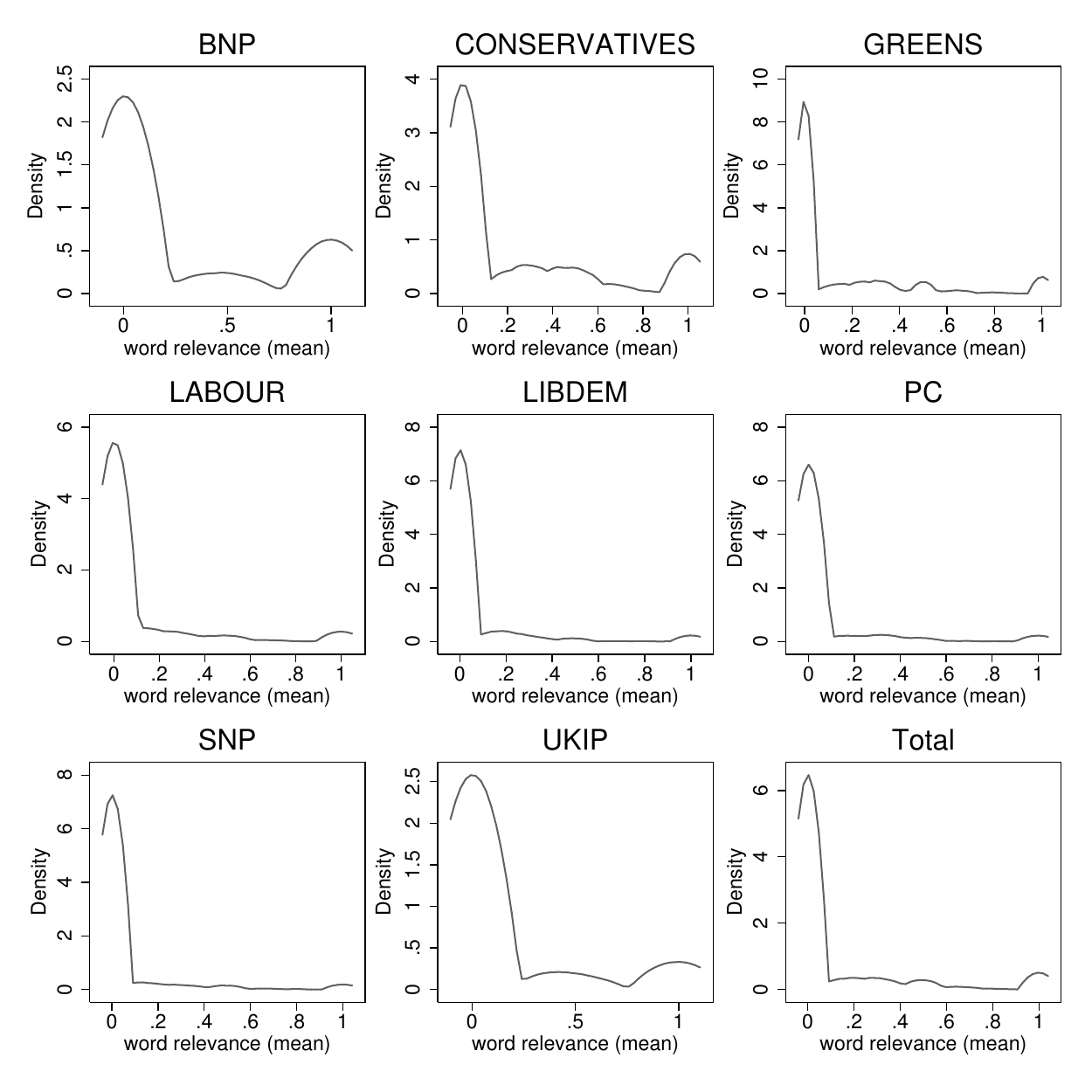}
\floatfoot{Note: The horizontal axis refers to the rate in which words were considered by the hand-coding to be relevant.}
\end{figure}

If one considers that all this noise brought by the non-informative words which are automatically used in \emph{Wordscores} moves party positions towards the middle of the scale, one can understand the logic behind the LBG transformation which stretches the party scores towards the end points of the scale. Although we agree that one needs to make some kind of transformation to account for the presence of noise that leads to the centrist bias in party positions, we do not agree that such a fundamental problem in content validity present in \emph{Wordscores} can be solved by a simple transformation of the raw scores. To give an example, we examine closely the wordscoring of the 2009 UKIP manifesto. UKIP is well-known for its extreme anti-EU stance which should leave no doubts about where the party should be placed. The \emph{Wordscores} raw placement for UKIP is 11.5 [11.2, 11.8] and the LBG transformed one is 9.3 [5.5, 13]. In either case, the party is placed in the middle of the scale. The transformation only improves this placement by specifying that this counter-intuitive middle placement is estimated with a lot of uncertainty. \emph{Wordscores} tells us that UKIP could be placed on either side of the scale even though one should not have much difficulty in establishing the position of the party simply by looking at the UKIP Euromanifesto.

One could argue of course, that this is a problem of the 2009 UKIP Euromanifesto being very short. However, the size of the document should only contribute to making the confidence interval around the point estimate larger. However, the problem here is that the UKIP point estimate is counter-intuitively estimated in the middle of the scale. This is not because the UKIP document is short, but because \emph{Wordscores} is unable to accurately estimate the party position due to all the noise that was introduced by the scoring of non-informative words. This is clearly shown in Figure 4, where we plotted all the words scored in the UKIP 2009 Euromanifesto according to their wordscore. Most of the words scored by \emph{Wordscores} are not informative with regards to placing UKIP on the European integration dimension and since most of the words have wordscores near the middle of the scale, the point estimate for UKIP was counter-intuitively given at 11.5 (transformed by LBG to 9.3).

\begin{figure}[!htb]
\caption{Wordscoring the UKIP 2009 Euromanifesto on the European integration dimension.}
\includegraphics[width=1\textwidth]{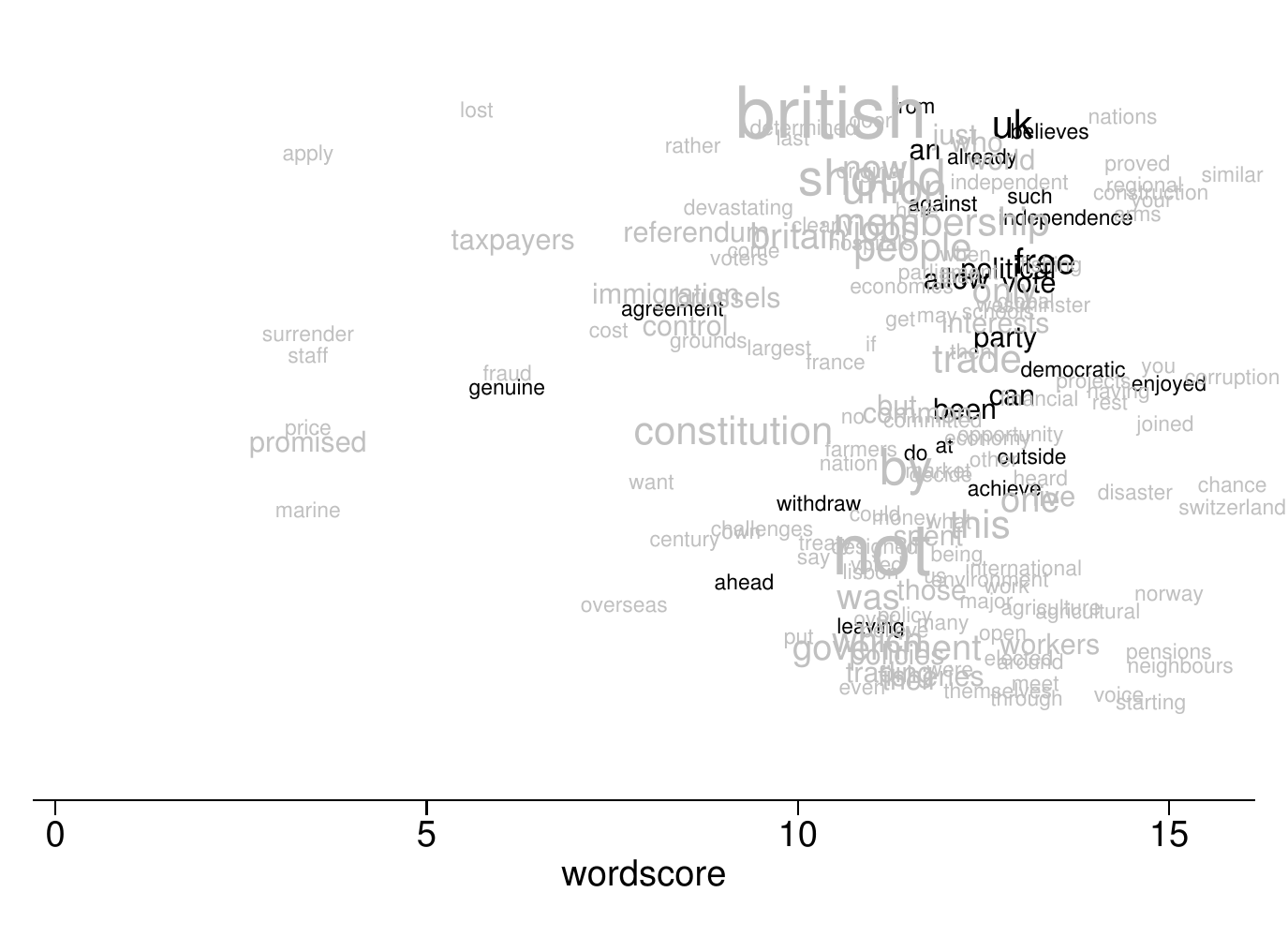}
\floatfoot{Note: Word size corresponds to the frequency of appearance in the UKIP virgin text; words that were hand-coded as being relevant at least 50\% of the instances are plotted in black.}
\end{figure}

The problem is therefore deeper than the uncertainty that comes with the size of the documents, and this can be established simply by looking at the cases of parties with much larger documents than UKIP. The fundamental problem lies in the content validity of \emph{Wordscores}. The lack of content validity brought by scoring each and every word irrespective of its relevance in providing information about the dimension of interest, pushes scores towards the middle of the scale. Transforming the raw scores will pull the estimates towards the endpoints of the scale, but there is no guarantee that the estimates will be pulled to the right direction. This will become evident in the next section where we examine the criterion validity of the \emph{Wordscores} estimates across all countries.

\FloatBarrier

\subsection*{Criterion validity}

Criterion validity refers to the extent to which a measure correlates with another measure which reflects the same concept \cite{Carmines1979}. Here, we assess the criterion validity of \emph{Wordscores} by comparing its estimate to alternative measures of party positions on each dimension as outlined in the study design section. As we have argued, this comparison needs to be made using appropriate correlation coefficients. Neither Pearson's product-moment correlation coefficient nor Spearman's rank-order correlation coefficient are able to capture the presence of systematic measurement error.

As has been pointed out by \cite[144]{Krippendorff1970}, both Pearson's and Spearman's coefficients, are based on the presumption of linearity ($Y=bX$) which is not the same as agreement between two measurements ($Y=X$). It is therefore possible for two measures to correlate perfectly (according to Pearson's or Spearman's coefficients) without them being identical measures. Therefore all the studies that have used such coefficients to assess the criterion validity of measures of party positions (including all previous validation studies involving \emph{Wordscores}) are likely to \emph{overestimate} the degree of validity in case of the presence of systematic measurement error. In order to overcome these problems, we use the concordance correlation coefficient \cite{Lin1989} defined as:

\begin{equation}
\rho_{c}=\frac{2\rho\sigma_{x}\sigma_{y}}{\sigma^{2}_{x}+\sigma^{2}_{y}+(\mu_{x}-\mu_{y})^{2}}
\end{equation}

Where $\mu_{x}$ and $\mu_{y}$ are the means for the two measures and $\sigma_{x}$ and $\sigma_{y}$ are the corresponding variances, and $\rho$ is Pearson's product-moment correlation coefficient between the two measures. Put more simply, CCC is conceptualised as

\begin{equation}
\rho_{c}=\rho C_{b}
\end{equation}

or, in other words, as the product between Pearson's product-moment correlation coefficient $\rho$ that measures dispersion (i.e. the degree of random measurement error) and a bias correction factor $C_{b}$ that measures the deviation from the 45 degrees line of perfect concordance. A $\rho_{c}$ of 0 denotes absence of concordance, a $\rho_{c}$ of 1 denotes perfect concordance, and a $\rho_{c}$ of -1 perfect negative concordance.

To estimate and interpret the CCC, we further need to consider two complicating factors. Firstly, CCC requires for both measures to be on the same scale. Normally, one could rescale all estimates of party positions from 0 to 1 using the well-known $\frac{estimate - min}{max - min}$ formula. Although this is straightforward using the expert survey and judgemental coding data where the scale minimum and maximum are clearly defined, this is not the case with \emph{Wordscores} estimates. Despite the promise made by the LBG transformation that it puts the estimates on the same metric of the reference texts \cite[317]{Laver2003}, this does not always happen in practice. For instance, our \emph{Wordscores} estimates on the left-right range from -2.09 to 22.45 when the BL expert survey that was used for the reference scores ranges from 0 to 20. The question is thus how to treat such counter-intuitive results. Following other studies that used the CCC with the Manifesto Project estimates that suffer from the same problem \cite{Gemenis2012,Gemenis2013}, we use the empirical scale minimum and maximum as given in the \emph{Wordscores} output. In one approach, we do this per dimension (in the aforementioned example, we use -2.09 and 22.45 as min and max in the formula respectively), and in another we implement this process per individual country. This way, we can check whether our inferences are robust to this rescaling.

Secondly, we need to set beforehand an objective criterion of what will be considered the minimum accepted correlation for criterion validity. Unfortunately, all previous studies have interpreted correlation coefficients (as strong, moderate, etc) entirely on subjective criteria. Given that Lin's original strength-of-agreement criterion $\rho_{c}>.9$ is too stringent for social science measurement, we use as the criterion the CCC between various estimates to which we compare the \emph{Wordscores} estimates to.\footnote{We would like to thank Oliver Treib for suggesting this.} This way, we have a clear, precise, and objective criterion for our assessment. If \emph{Wordscores} promises to estimate party positions accurately, then these positions should correlate with other measures of party positions at least as high as these other measures correlate with one another. Finally, we introduce a measure of uncertainty for the CCC, based on 95\% z-transformed confidence intervals. To be as lenient as possible, we consider successful in terms of criterion validity when the upper CI (not the point estimate) of the CCC is higher than three CCCs possible when comparing the three other datasets of party positions to one another.

Despite the objective but lenient terms of our evaluation, Figures 5 and 6 clearly show that the \emph{Wordscores} estimates cannot be considered as valid estimates of party positions in terms of criterion validity (for a detailed overview of the concordance correlations see Appendix G). No matter the dimension (left-right, European integration, economic, or socio-cultural), the source of reference scores (BL, CHES, or EMP), the method of transformation (LBG or MV), rescaling to estimate the CCC (whole dimension or per country), or the dataset to which we compared them to (CHES, EMP, or EUP), the correlation of \emph{Wordscores} with other datasets never attained a CCC as high as the other datasets attained when compared to one another.\footnote{Detailed results and additional figures are available in Appendix G.} To be sure, one could argue that this pessimistic conclusion could be due to the constraints put by rescaling and calculating of the CCC. Nevertheless, the simple Pearson's $r$ correlation coefficients on the estimates before the rescaling needed for CCC (available in Appendix H) were also very low.

\begin{figure}[ht]
\caption{Assessing criterion validity on left-right and European integration dimensions.}
\includegraphics[width=.8\textwidth]{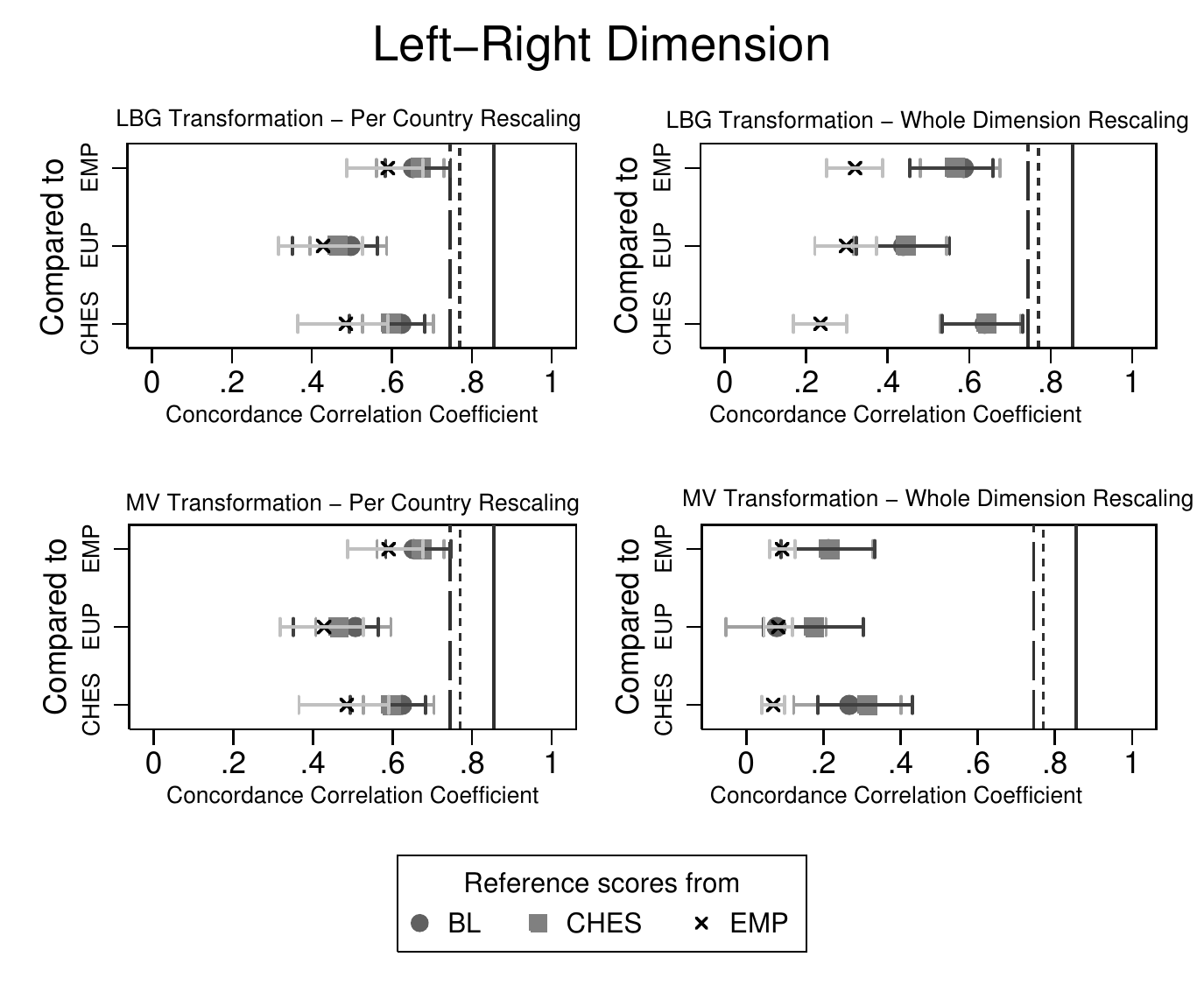}
\includegraphics[width=.8\textwidth]{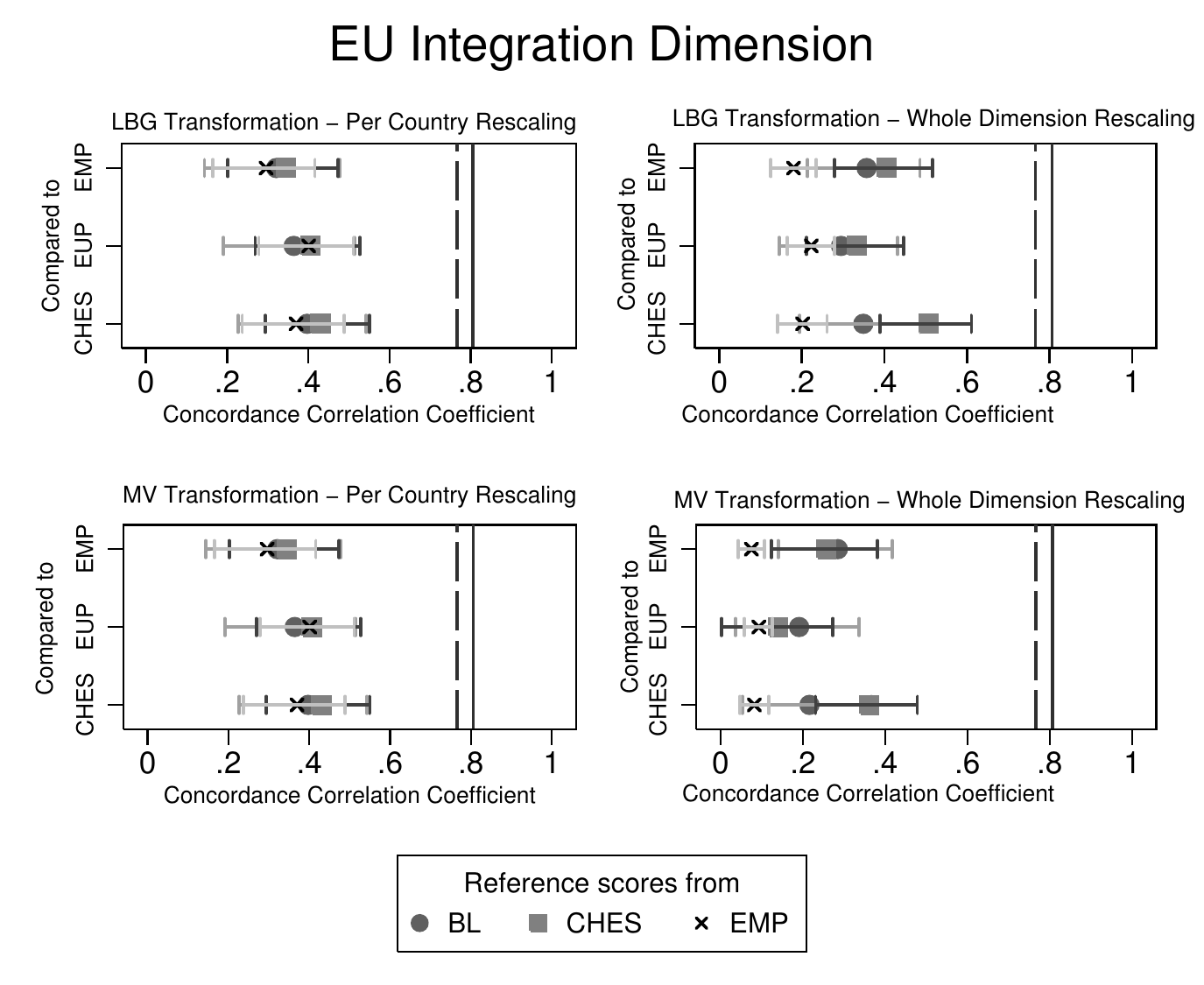}
\floatfoot{Note: Vertical lines represent the CCC between CHES/EMP (solid), CHES/EUP (dotted), EMP/EUP (dash).}
\end{figure}

\begin{figure}[ht]
\caption{Assessing criterion validity on economic and socio-cultural dimensions.}
\includegraphics[width=.8\textwidth]{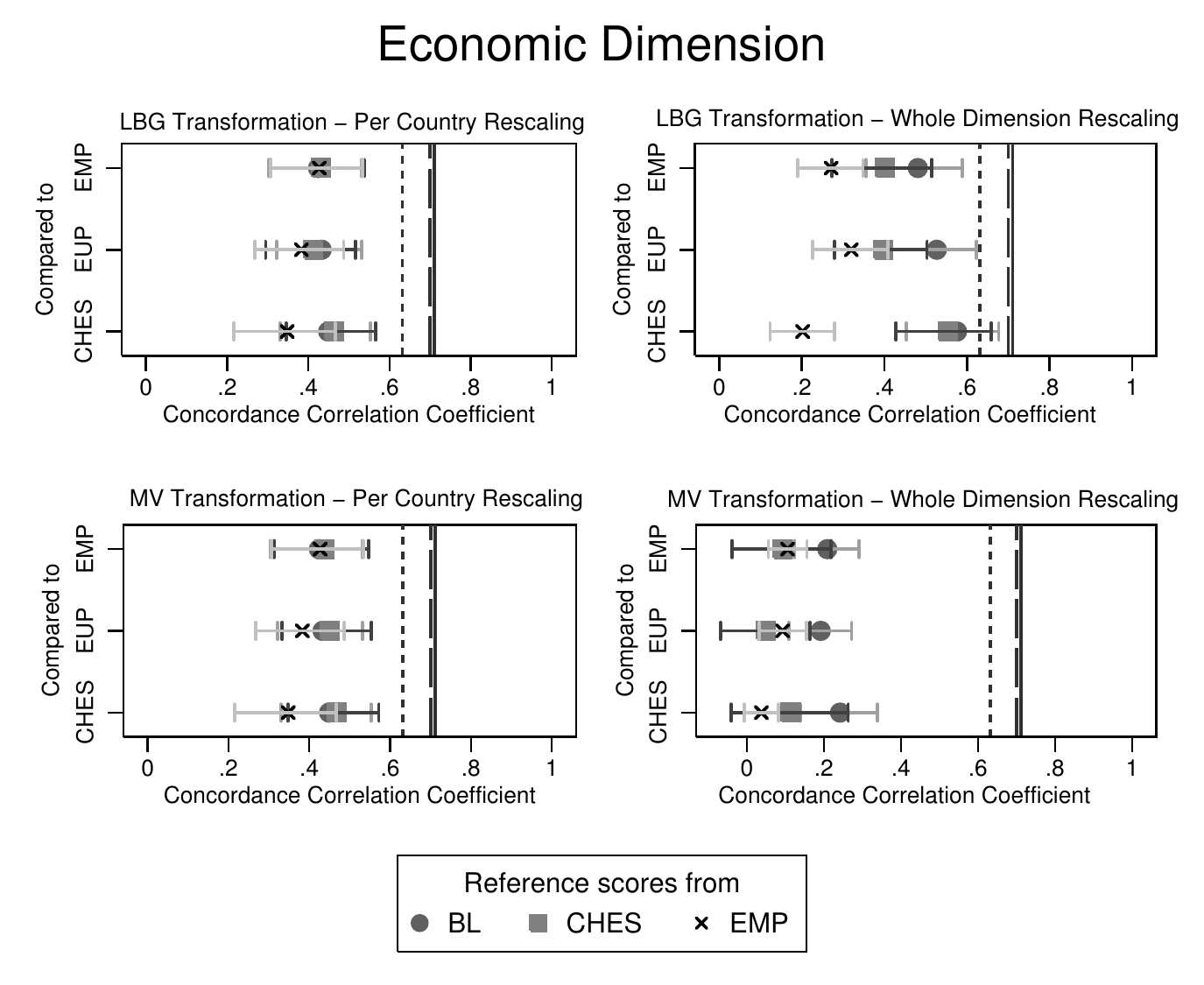}
\includegraphics[width=.8\textwidth]{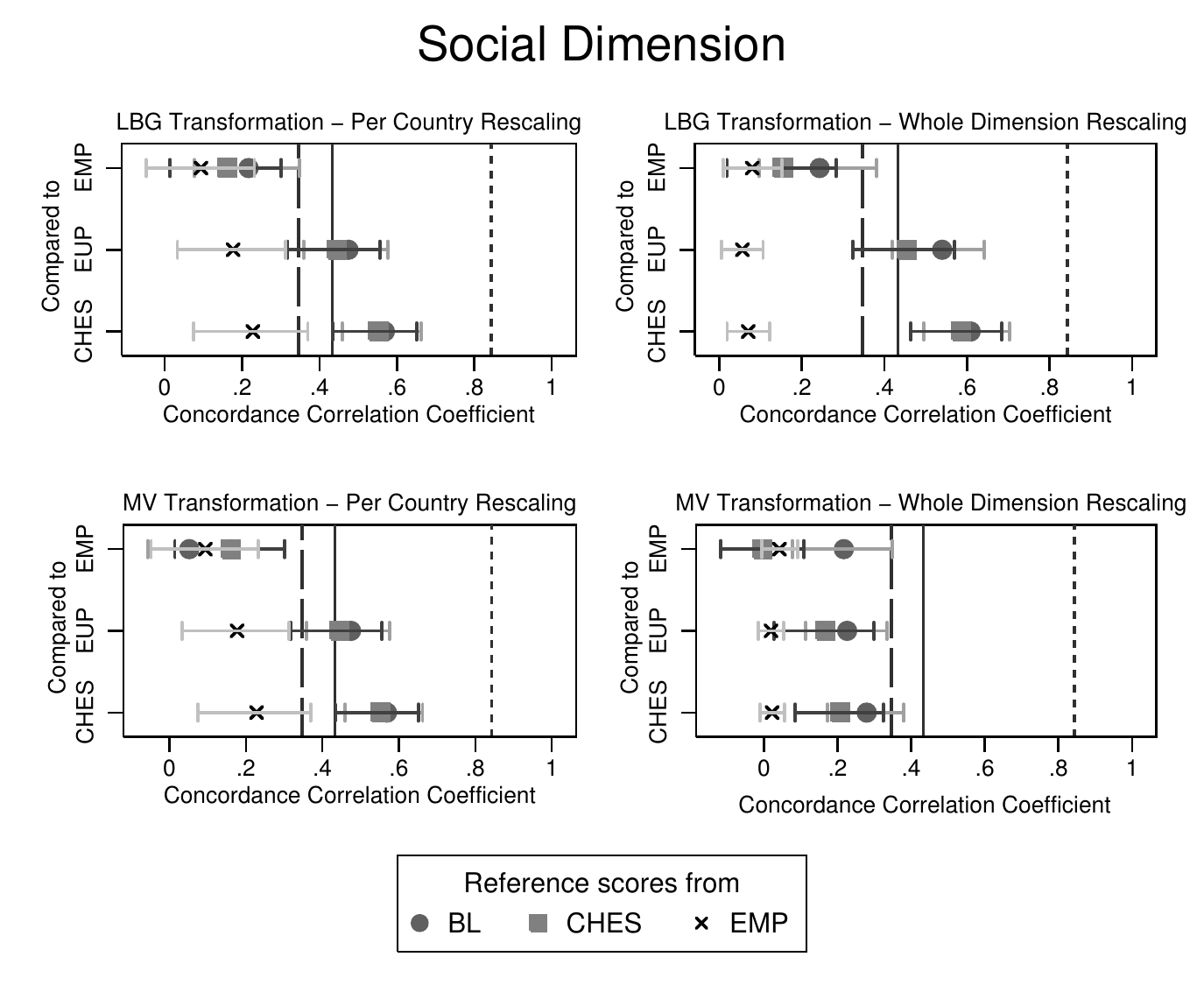}
\floatfoot{Note: Vertical lines represent the CCC between CHES/EMP (solid), CHES/EUP (dotted), EMP/EUP (dash).}
\end{figure}

\FloatBarrier

\subsection*{Construct validity}

Construct validity refers to the extent to which our measure behaves as expected within a given theoretical context. To assess construct validity, we formulate a simple hypothesis, about the relationship between party positions and membership in the political groups of the EP. This relationship has been used before to illustrate the use of the Manifesto Project \cite[36--39]{Klingemann2007}, and expert survey \cite{McElroy2007} data. In this paper, we take this hypothesis a step further, arguing that we can predict with some confidence party membership in the political groups of the EP on the basis of national parties positions on the socio-economic and European integration dimensions. To do so, we estimate a multinomial regression model, where the dependent variable takes eight values, one for each of the seven party groups in the EP (as of 2009) with non-attached parties forming the eighth group.

To assess the explanatory power of the model we use count $R^{2}$ which is simply the proportion of correct predictions, as well as McFadden's pseudo-$R^{2}$ which compares the explanatory power added by the independent variables compared to a model that includes only the intercept. We compare the explanatory power of the model using the three predictor variables as estimated by \emph{Wordscores} (using all possible configurations of reference scores and transformations) to the explanatory power of models using exactly the same predictors as measured by three alternative datasets as shown in Table 2: the 2010 Chapel Hill Expert Survey, and the judgemental coding of the Euromanifestos Project and EU Profiler. 

As can be seen from Figure 7, in none of the cases do the \emph{Wordscores} estimates perform better than estimates from other datasets in predicting membership in the EP party groups. To avoid misleading evaluations as to how much better one model is compared to the other, we use the Bayesian Information Criterion (BIC) as a measure of overall fit. In every case, the difference in BIC between models using the \emph{Wordscores} estimates and models using estimates from the other datasets is larger than 10. This indicates `very strong' evidence \citeaffixed[87]{Long2001}{see} against the model using the \emph{Wordscores} estimates. What does this imply for \emph{Wordscores}? According to \citeasnoun[82]{Zeller1980}, construct validation requires `a pattern of consistent findings' across different hypotheses and studies in order for a measure to establish a high degree of construct validity. Our study did not provide such extensive evidence, but it is rather instructive that \emph{Wordscores} failed the very simple construct validation test that has been used elsewhere in the literature.

\begin{figure}[ht]
\caption{Assessing construct validity by predicting membership in EP party groups.}
\includegraphics[width=.6\textwidth]{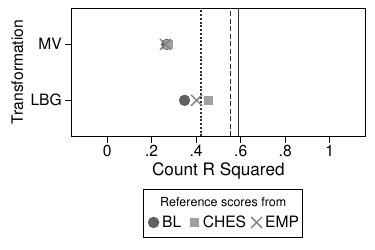}
\includegraphics[width=.6\textwidth]{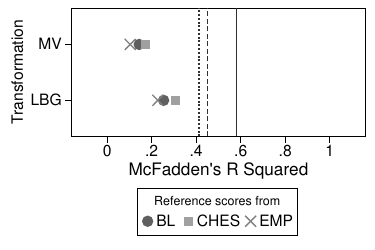}
\floatfoot{Note: Vertical lines represent the R squared of models using estimates from CHES (solid), EUP (dashed), and EMP (dotted).}
\end{figure}

\FloatBarrier

\section*{Conclusions}

In their proof-of-concept \citeasnoun[329]{Laver2003} promised that \emph{Wordscores} can deliver `effective' estimates of political actors' policy positions in a matter of seconds. Our replication of Laver et al. revealed inconsistencies in the software implementations of \emph{Wordscores} and showed that the results presented in their proof-of-concept are not particularly robust. Following Grimmer \& Stewart's \citeyear{Grimmer2013} advice to `validate, validate, validate', we subjected \emph{Wordscores} to a rigorous validation on conditions that should be favourable to the method. Hence, we focused on a cross-sectional rather than longitudinal \citeaffixed{Brauninger2013}{cf.} comparison where we should not expect significant changes in the discourse that could compromise the effectiveness of the method. Moreover, we used an `off-the-shelf' collection of documents and data from expert surveys and the judgemental coding of party manifestos, which are consistent with how the method is used in practice.

In contrast to what was promised by Laver et al. our findings showed that the \emph{Wordscores} estimates of party positions cannot be considered valid. The examination of content validity showed that the \emph{Wordscores} estimates are compounded by the scoring of irrelevant words and this cannot be corrected by the LBG rescaling method. The examination of criterion validity showed that the \emph{Wordscores} estimates correlate far lower with other estimates of party positions than the other estimates correlate with one another. Moreover, the examination of construct validity showed that \emph{Wordscores} estimates have significantly lower predictive power when used in statistical models compared to other estimates of parties' positions. Finally, these findings were shown to be robust across different configurations of reference scores and rescaling methods.

In general our overall negative conclusions imply that \emph{Wordscores} should not be used to estimate parties' policy positions using electoral manifestos as reference and virgin texts. However, we need to qualify this conclusion. As the performance of \emph{Wordscores} has shown to vary widely depending on the circumstances of estimation \citeaffixed{Brauninger2013}{see}, we outline three ways in which the \emph{Wordscores} estimates can be improved, namely by careful document selection, pre-processing, and parsing.

With regards to document selection, we note that our results could be driven by the fact that we used Euromanifestos rather than national election manifestos. However, the most comprehensive validation study using national election manifestos, found mixed results \citeaffixed{Brauninger2013}{see}. It seems that the problem is not so much the electoral context in which the documents are produced, but rather the quality of the documents as sources of party positions. In our validation we used the off-the-shelf collection of the Euromanifestos Project which is less than ideal. One could possibly improve the validity of \emph{Wordscores} estimates by carefully selecting the documents to be analysed, as already pointed out by \citeasnoun{Proksch2009} for the case of Germany.

Second, researchers can further improve the validity of \emph{Wordscores} estimates by using a more rigorous document pre-processing procedure than the one we used in this paper. Instead of removing the most frequently occurring words as we did, researchers could consider removing stop words even more rigorously using a pre-defined list. Removing stop words would reduce the amount of noise, which tends to push \emph{Wordscores} estimates towards the middle of the scale irrespective of the informative content of the documents. It is also worth mentioning this this problem has already been accounted for by another popular scaling method, \emph{Wordfish}, which applies weights `capturing the importance of [words] in discriminating between party positions' \cite[709]{Slapin2008}.

Third, researchers should consider using only those parts of the documents they are interested in. So, when the object of investigation is foreign policy, only the paragraphs directly dealing with foreign policy should be used, and not the document as a whole. Parsing documents to different policy areas depending on the estimated policy dimension is required in text scaling methods like \emph{Wordfish} that assume that the text is unidimensional \cite{Slapin2008}. The same logic can be taken to \emph{Wordscores} assuming that the content of policy areas one is not interested in would only add noise to the estimates.

Nevertheless, while these three suggestions can improve the validity of the estimates they come at the expense of considerable investment in time and resources. Document selection requires considerable expertise in terms of party politics, and is often difficult to assemble and manage in a cross-national project. Lists of stop words are often context dependent, while compound words can cause considerable problems in identifying stop words by automated software. Moreover, parsing documents into policy-related sections requires knowledge of the language the documents were written, something which goes against the promise of \emph{Wordscores} as a method where it is `not necessary for an analyst [using the technique] to understand or even read the text to which the technique is applied' \cite[329]{Laver2003}.

\emph{Wordscores} could potentially produce valid estimates of party positions, but only after some serious investment in time, language- and country-related expertise. We leave to the reader the question whether this investment negates the original promise of a quick and easy method \cite[226, 312]{Laver2003}. What we showed here is that, when the method is used as a language-blind and quick way to estimate party positions, it does not deliver what it promises. Therefore, any researcher who wishes to use \emph{Wordscores} `as is' should always demonstrate the validity of the output using a carefully designed validation study as shown here.

\newpage

\theendnotes

\newpage

\section*{Appendix A: Reanalysis of Laver, Benoit \& Garry (2003)}

Much of the initial validation for \emph{Wordscores} rested on scoring the 1997 Irish manifestos on a social and economic dimension using the 1992 manifestos as reference texts \cite{Laver2003}. We attempted to replicate the findings in the paper using the  manifestos, code, and reference scores as available on the \emph{Wordscores} website \url{http://www.tcd.ie/Political_Science/wordscores/index.html}. Unfortunately, we were not able to replicate the results published in Laver et al. using the materials from the website. Upon closer examination we realized that replication is not possible for two reasons. 

First, the reference texts provided in the \emph{Wordscores} website are not the same as the ones used in the Laver et al. article. As is clear from the number of words, the documents provided in the website have been cleaned differently compared to the documents used in the Laver et al. article. This cleaning refers to the removal of numbers, special characters, document formatting content (tables of contents, headers, footers), and occasionally stop words which is an important step in computer-assisted text analysis. Moreover, the website, includes in the set of reference texts the manifestos of two additional parties (Greens and Sinn Fein), unlike the Laver et al. article which uses as reference texts the manifestos of only five parties. 

Second, and most importantly, the current (as of \date{\today}) `23-June-2009' version of \texttt{wordscores} for Stata gives different results than the older version `v0.36' that was used to produce the results in the Laver et al. article. The differences in the output given by these two versions can be attributed to changes in the code with regards to how $F_{wv}$ (equation 3 in the main text) is calculated. According to \citeasnoun[316]{Laver2003}, $F_{wv}$ denotes `the relative frequency of each virgin text word [w], as a proportion of the total number of words in the virgin text [v]' (emphasis added). This is what has been implemented in the `23-June-2009' version of the Stata \texttt{wordscores} package. Conversely, `v0.36' and the two packages that can implement Wordscores in R (`austin' and `quanteda'), define Fwv as the relative frequency of each virgin text word w is taken as a proportion of the total number of words co-occurring between the reference and the virgin texts. In an e-mail communication, Kenneth Benoit clarified that the `correct' implementation of \emph{Wordscores} is in the R packages and `v0.36' version of \texttt{wordscores} for Stata. This implies that the definition of $F_{wv}$ given in Laver et al. is incorrect. It also implies that all those who used the `23-June-2009' version in their (published) papers got the `wrong' \emph{Wordscores} results. In our communication, Kenneth Benoit also indicated that the change in how $F_{wv}$ is defined does not make much difference as the results correlate highly.

\begin{figure}[!htbp]
	\caption{Comparing the results of the two implementations of $F_{wv}$ in \texttt{wordscores} for Stata.}
		\includegraphics[width=1\textwidth]{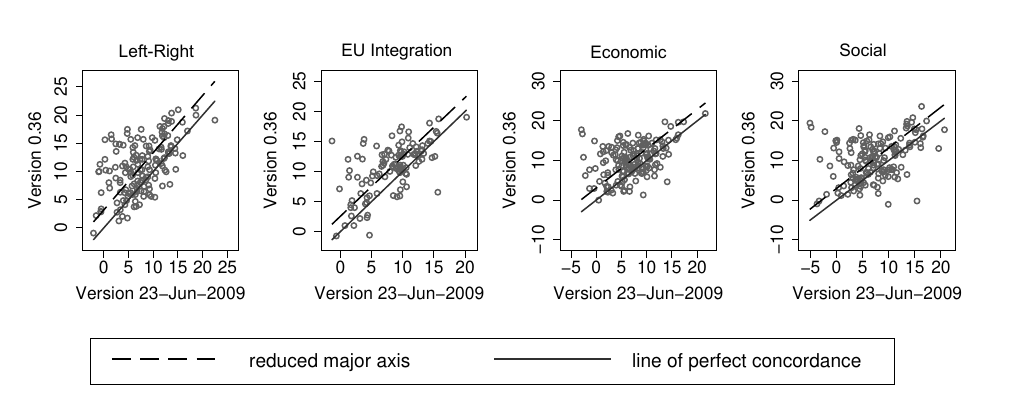}
	\end{figure}

We tested this claim by implementing the two versions of \texttt{wordscores} (v0.23 and `23-June-2009') for Stata across all the parties in our analysis for four different dimensions (left-right, European integration, economic, social) using the \citeasnoun{Benoit2006} expert survey for the reference text scores and the LBG transformation. Figure A1 shows the results which clearly contradict the claim that the results of the two implementations correlate would highly (`about .97'). The concordance between the two scores measured by the concordance correlation coefficient are .44 (left-right), .53 (European integration), .33 (economic), .32 (social). The respective Pearson correlation coefficients are .55, .62, .41, .38. The correlations are similar when different sources for the reference text scores were used. This is clear evidence that changing the definition of $F_{wv}$ changes the \emph{Wordscores} estimates radically.

Nevertheless, the most important point here is that the inconsistency between the Laver et al. article and the software implementations challenges the proof-of-concept validation presented in the Laver et al. article. In the figures presented in Table 1 below, we show how the \emph{Wordscores} estimates for Irish party positions vary when one uses different sets of documents for reference texts (five parties as in the Laver et al. article versus seven parties as in the replication material found in the \emph{Wordscores} website) and different implementations of \texttt{wordscores} for Stata (`v0.36' versus `23-June-2009') lead to substantially different results. 

The results in the top left quartile of Table 1 attempt to replicate the findings of Laver et al. by using the manifestos of five Irish parties (FF, FG, Labour, DL, PD) and the `v0.36' \texttt{wordscores} for Stata (which is identical to the \texttt{wordscores} and \texttt{quanteda} packages in R). They are almost identical save some minor differences due to the way the documents were cleaned for the analysis in Laver et al. As pointed out in that article, the results look reasonable and consistent with how the parties have been placed in expert surveys (e.g. DL and Labour on the economic left, the other parties on the economic right).

However, when we change the definition of $v$ from `the total number of words in the virgin text' as stated in the original article \citeasnoun[316]{Laver2003} to `the proportion of the total number of words co-occurring in the virgin and reference texts' as was done in the `23-Jun-2009' version of \texttt{wordscores} for Stata, we get the much different results presented in the bottom left quartile. It is clear from the figure that changing the definition of $v$ produces estimates that move parties in a way that does not make much sense (for instance, Fianna Fail as the most economically left party) and otherwise makes it impossible to distinguish between the parties given the confidence intervals of the estimates.

The change in the definition of $v$ that was implemented on 23 June 2009 will produce party positions that appear reasonable and intuitive only if one adds the manifestos of Greens and Sinn Fein in the set of reference texts as shown in the bottom right quartile. However, if we add these two manifestos in the set of reference texts, but keep the definition of $v$ as in the Laver et al. article, we will get the results in the top right quartile. Again, these results do not make much sense, since the confidence intervals overlap significantly and many of the point estimates are rather implausible (e.g. the Greens and Sinn Fein are in the middle of both scales. 

We find it strange that the documents for Greens and Sinn Fein were not included in the APSR article, but were included in the replication of the article as implemented in the \emph{Wordscores} website which contained a different Stata \texttt{wordscores} code. Why did the authors not include the SF and Greens documents in their original analysis as presented in the APSR article? We believe that this was not done because the addition of these two parties in 2003 under the alternative definition of $v$ which is used in R and is favoured by Kenneth Benoit (as per our e-mail communication) would have given results that are inconsistent with expert surveys. Similarly, when the \texttt{wordscores} code was changed and the results appeared to be implausible, the two documents were added as reference texts in the replication materials in the \emph{Wordscores} to improve the validity of the results. Since the positions of parties under the Laver et al. transformation (which is used in the APSR article) are sensitive to the inclusion/exclusion of virgin texts as shown by \citeasnoun{Martin2008}, we ask whether the exclusion of SF and the Greens from the analysis in Laver et al. but their inclusion in the `replication' of the analysis in the \emph{Wordscores} website does not constitute an attempt to `cherry pick' among different possible results in a way that supports the argument in favour of \emph{Wordscores}.

\begin{landscape}
 
\begin{table}[htbp]
\caption{Replication of the original scores}
\centering
\begin{tabular}{m{2.5cm}cc}
\toprule
\multicolumn{3}{c}{Number of Parties}\\
Stata Version & 5 parties & 7 parties\\
\midrule
  \multirow{9}{*}{0.36}           &                                             &                                              \\
        & \includegraphics[scale=.9]{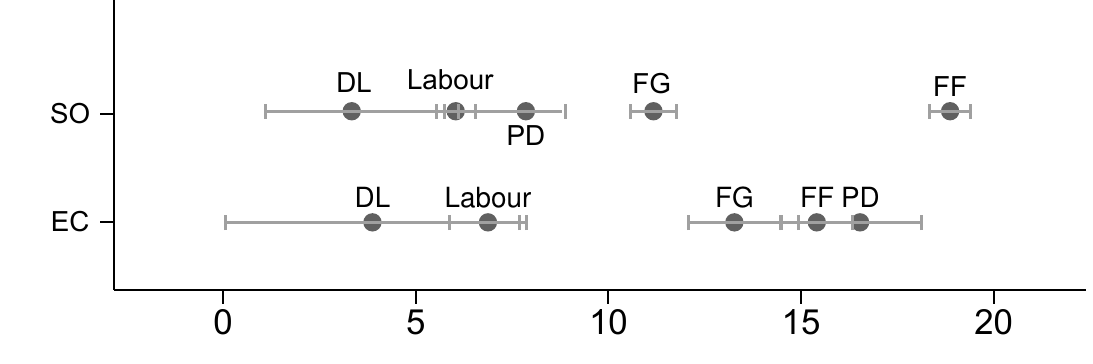}     &  \includegraphics[scale=.9]{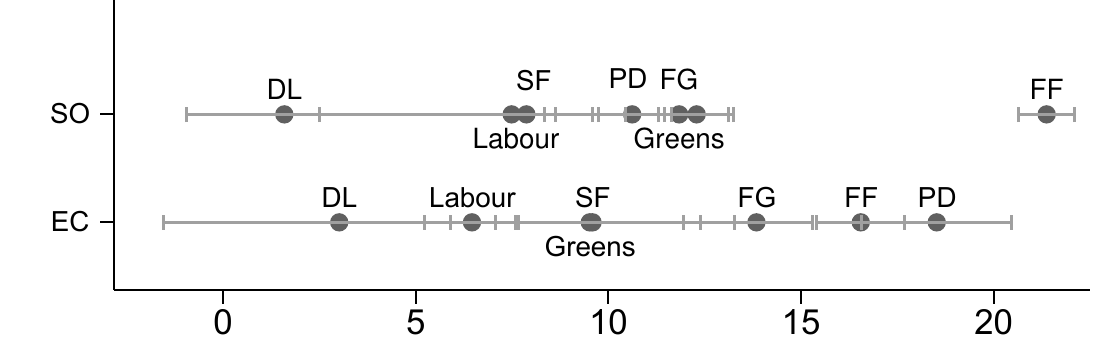}    \\
             & Laver et al. (2003)                   &                                              \\
             &                                              \\
 \multirow{-8}{*}{23-Jun-2009} & \includegraphics[scale=.9]{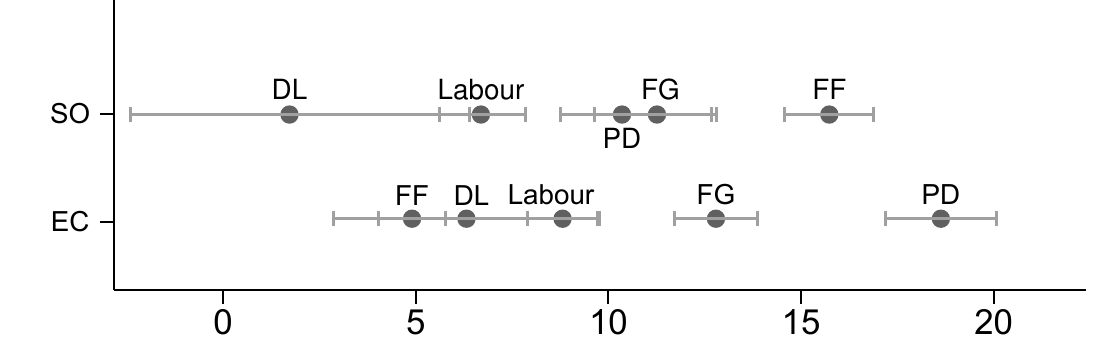}  &  \includegraphics[scale=.9]{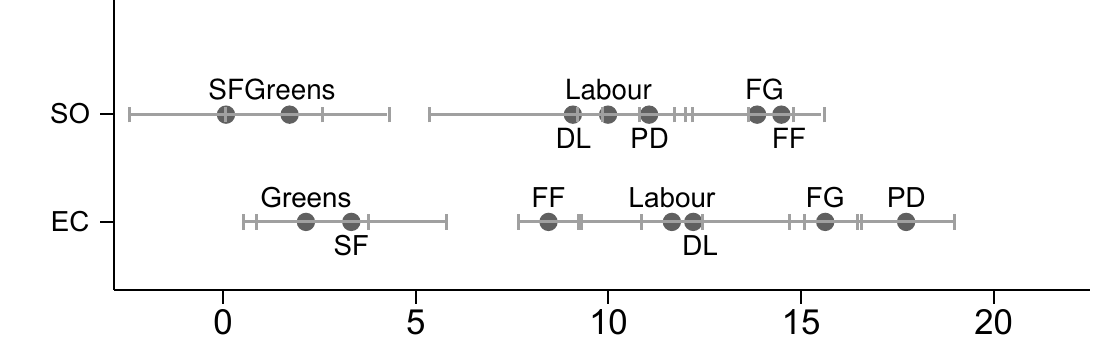} \\
             &                                             &  Laver et al. (2003) Replication Material     \\
    \bottomrule
\end{tabular}
\end{table}

\end{landscape}

\newpage

\begin{landscape}
\section*{Appendix B: Documents used in the analysis}

    \small
    \begin{longtable}{m{1.8cm}m{1cm}m{3cm}m{7cm}m{7.2cm}m{1.5cm}m{1.5cm}}
    \toprule
    Country & Year & Party & Full Name & Title & Total Words* & Unique Words* \\
\midrule
\endhead

    AT & 2004 & FP\"{O}& Freiheitliche Partei \"{O}sterreichs & T\"{u}rkei in der EU?  & 1236 & 792 \\
    AT & 2009 & FP\"{O} & Freiheitliche Partei \"{O}sterreichs & Echte Volksvertreter statt EU-Verr\"{a}ter & 704 & 448 \\
    AT & 2004 & Gr\"{u}nen & Die Gr\"{u}nen – Die Gr\"{u}ne Alternative & Bestimmen Sie! Ihre Zukunt in Europa  & 3699 & 1894 \\
    AT & 2009 & Gr\"{u}nen & Die Gr\"{u}nen – Die Gr\"{u}ne Alternative & Vorw\"{a}rts Gr\"{u}n!& 3585 & 1830 \\
    AT & 2009 & HPM & Liste Hans-Peter Martin & Nur er kontrolliert die M\"{a}chtigen  & 119 & 106 \\
    AT & 2009 & LF & Liberales Forum & Europa als Chance ergreifen  & 5335 & 2308 \\
    AT & 2004 & \"{O}VP & \"{O}sterreichische Volkspartei & Europa-Manifest zur Europawahl 2004  & 2226 & 1145 \\
    AT & 2009 & \"{O}VP & \"{O}sterreichische Volkspartei & Wahlmanifest Zur Europawahl 2009  & 4238 & 1822 \\
    AT & 2004 & SP\"{O} & Sozialdemokratische Partei \"{O}sterreichs & \"{O}sterreich Muss Wieder Geh\"{o}rt Werden!  & 985 & 570 \\
    AT & 2009 & SP\"{O} & Sozialdemokratische Partei \"{O}sterreichs & Wahlmanifest SP\"{O}  & 2268 & 1197 \\
    BE (FR) & 2004 & CDH & Centre D\'{e}mocrate Humaniste & Programme europ\'{e}en 2004 du CDH  & 11184 & 3341 \\
    BE (FR) & 2009 & CDH & Centre D\'{e}mocrate Humaniste & Un autre monde, une autre Europe!  & 15247 & 3995 \\
    BE (FR) & 2004 & ECOLO & Ecolo & Projet pour l'Europe & 4665 & 1969 \\
    BE (FR) & 2009 & ECOLO & Ecolo & Programme Ecole \'{E}lections 2009 & 7760 & 2741 \\
    BE (FR) & 2009 & FN & Front National & Le Manifeste du FN  & 7004 & 2846 \\
    BE (FR) & 2004 & MR & Mouvement R\'{e}formateur & 25 Propositions pour l'Europe  & 3346 & 1486 \\
    BE (FR) & 2009 & MR & Mouvement R\'{e}formateur & Le Programme Complet du Mouvement R\'{e}formateur \'{e}lections 2009  & 9592 & 3041 \\
    BE (FR) & 2004 & PS & Parti Socialiste & Programme du PS pour les \'{e}lections europ\'{e}ennes  & 15640 & 3836 \\
    BE (FR) & 2009 & PS & Parti Socialiste & Programme Union Europ\'{e}enne 2009 & 12213 & 3522 \\
    BE (NL) & 2004 & CD\&V & Christen-Democratisch en Vlaams & Europees verkiezingsprogramma CD\&V 13 Juni 2004  & 5391 & 1976 \\
    BE (NL) & 2009 & CD\&V & Christen-Democratisch en Vlaams & Europa op maat van de globalisering  & 3237 & 1435 \\
    BE (NL) & 2004 & Groen! & Groen! & Europa kan zoveel beter - Jij beslist!  & 6945 & 2612 \\
    BE (NL) & 2009 & Groen! & Groen! & Groene wegen voor een beter Europa & 14811 & 4434 \\
    BE (NL) & 2009 & LDD & Libertair, Direct, Democratisch - Lijst Dedecker & Europees Programma LDD - LDD, de Eurorealisten  & 6353 & 2452 \\
    BE (NL) & 2004 & NVA & Nieuw-Vlaamse Alliantie & Verkiezingsprogramma N-VA Europese verkiezingen 13 juni 2004  & 1774 & 867 \\
    BE (NL) & 2009 & NVA & Nieuw-Vlaamse Alliantie & NVA Europees programma 2009  & 10955 & 3387 \\
    BE (NL) & 2004 & SPA & Socialistische Partij Anders & Europees programme 13 juni 2004  & 7247 & 2433 \\
    BE (NL) & 2009 & SPA & Socialistische Partij Anders & Mensen op 1 - Een eerlijke koers voor Europa  & 5535 & 1860 \\
    BE (NL) & 2004 & VB & Vlaams Belang & Vlaamse Staat, Europese Natie  & 15197 & 4429 \\
    BE (NL) & 2009 & VB & Vlaams Belang & Dit is ons land  & 10178 & 3451 \\
    BE (NL) & 2004 & VLD & Vlaamse Liberalen en Democraten & Programma VLD - Vlaamse en Europese verkiezingen 13 juni 2004  & 748 & 503 \\
    BE (NL) & 2009 & VLD & Vlaamse Liberalen en Democraten & Top 15 van de Europese Liberalen voor de verkiezingen van het Europees parlement  & 4696 & 1956 \\
CY & 2004 & AKEL & \gr{Ανορθωτικό Κόμμα Εργαζόμενου Λαού} & \gr{Προγραμματικη Διακηρυξη}  & 2155 & 1180 \\
    CY & 2009 & AKEL & \gr{Ανορθωτικό Κόμμα Εργαζόμενου Λαού} & \gr{Στην Ευρώπη Διεκδικητές και όχι Χειροκροτητές}  & 989 & 638 \\
    CY & 2004 & DIKO & \gr{Δημοκρατικό Κόμμα} & \gr{Ισχυρή Κύπρο στην Ευρώπη!}  & 1698 & 1002 \\
    CY & 2009 & DIKO & \gr{Δημοκρατικό Κόμμα} & \gr{Στείλε καθαρό µήνυµα στην Ευρώπη }  & 1092 & 643 \\
    CY & 2004 & DISY & \gr{Δημοκρατικός Συναγερμός} & \gr{Η καλύτερη ομάδα}  & 1769 & 985 \\
    CY & 2009 & DISY & \gr{Δημοκρατικός Συναγερμός} & \gr{Πρόταση Πολιτικής }   & 1796 & 1055 \\
    CY & 2004 & EDEK &  \gr{Κίνημα Σοσιαλδημοκρατών ΕΔΕΚ} &  \gr{Έχουμε θέση στην Ευρώπη}  & 465 & 282 \\
    CY & 2009 & EDEK & \gr{Κίνημα Σοσιαλδημοκρατών ΕΔΕΚ} & \gr{Ομιλία Γιαννάκη Ομήρου στην Κεντρική Συγκέντρωση   }     & 1154 & 698 \\
    CY & 2004 & KOP & \gr{Κίνημα Οικολόγων Περιβαλλοντιστών} & \gr{Εκλογικο Μανιφεστο Των Ευρωεκλογων Του 2004 ΟμοσπονδΙασ Πρασινων Ευρωπαικων Κομματων}   & 1466 & 827 \\
    CZ & 2004 & CSSD & Cesk\'{a} strana soci\'{a}lne demokratick\'{a} & Za Evropu bezpec\'{i}, m\'{i}ru, prosperity a  soci\'{a}ln\'{i}ch  jistot  & 1138 & 664 \\
    CZ & 2009 & CSSD & Cesk\'{a} strana soci\'{a}lne demokratick\'{a} & Jistota 2009  & 876 & 627 \\
    CZ & 2004 & KDU-CSL & Krestansk\'{a} a demokratick\'{a} unie – Ceskoslovensk\'{a} strana lidov\'{a} & Evropsk\'{y} volebn\'{i} program KDU - CSL   & 2602 & 1443 \\
    CZ & 2009 & KDU-CSL & Krestansk\'{a} a demokratick\'{a} unie – Ceskoslovensk\'{a} strana lidov\'{a} & Volebn\'{i} Program Pro Volby Do EP 2009-2014  & 1754 & 1173 \\
    CZ & 2004 & KSCM & Komunistick\'{a} strana Cech a Moravy & S v\'{a}mi a pro v\'{a}s doma i v EU   & 1771 & 1155 \\
    CZ & 2009 & KSCM & Komunistick\'{a} strana Cech a Moravy & Otevren\'{y} volebn\'{i} program KSCM pro volby do - Evropsk\'{e}ho parlamentu 2009   & 698 & 519 \\
    CZ & 2009 & NEZ & Politck\'{e} Hnut\'{i} Nezt\'{a}vislt\'{i} & Volby do Evropsk\'{e}ho parlamentu 2009   & 785 & 596 \\
    CZ & 2004 & ODS & Obcansk\'{a} demokratickt\'{a} strana & Stejnt\'{e} \u{s}ance pro v\v{s}echny - Program pro volby do Evropsk\'{e}ho Parlamentu  & 1439 & 976 \\
    CZ & 2009 & ODS & Obcansk\'{a} demokratickt\'{a} strana & Volebnt\'{i} Program ODS  & 5608 & 2865 \\
    CZ & 2009 & SNK-ED & SNK Evrop\v{s}t\'{i} demokrat\'{e} & Spolecne uka\v{z}me Evrope sebevedomou tv\'{a}r Cesk\'{e} republiky, kter\'{a} um\'{i} vyu\v{z}\'{i}t sv\'{y}ch \v{s}anc\'{i}!   & 2285 & 1365 \\
    DE & 2004 & B90/GR\"{U}NEN & B\"{u}ndnis 90/Die Gr\"{u}nen & Europa Besser Machen - Du Entscheidest!  & 24984 & 7243 \\
    DE & 2009 & B90/GR\"{U}NEN & B\"{u}ndnis 90/Die Gr\"{u}nen & F\"{u}r ein besseres Europa!  & 1263 & 756 \\
    DE & 2004 & CDU & Christlich Demokratische Union Deutschlands & Europa-Manifest der CDU  & 1773 & 999 \\
    DE & 2009 & CDU & Christlich Demokratische Union Deutschlands & Starkes Europa – Sichere Zukunft  & 3771 & 1759 \\
    DE & 2004 & CSU & Christlich-Soziale Union in Bayern e. V. & F\"{u}r ein starkes Bayern in Europa  & 1904 & 1062 \\
    DE & 2009 & CSU & Christlich-Soziale Union in Bayern e. V. & CSU-Europawahlprogramm 2009  & 3217 & 1462 \\
    DE & 2004 & FDP & Freie Demokratische Partei & Wir k\"{o}nnen Europa besser! - F\"{u}r ein freies und faires Europa& 6600 & 2664 \\
    DE & 2009 & FDP & Freie Demokratische Partei & Ein Europa der Freiheit - f\"{u}r die Welt des 21. Jahrhunderts  & 6523 & 2829 \\
    DE & 2004 & DIELINKE & Partei des Demokratischen Sozialismus - DIE LINKE & Alternativen sind machbar:  F\"{u}r ein soziales, demokratisches und friedliches Europa!   & 12869 & 4777 \\
    DE & 2009 & DIELINKE & Partei des Demokratischen Sozialismus - DIE LINKE & Solidarit\"{a}t, Demokratie, Frieden - Gemeinsam f\"{u}r den Wechsel in Europa! & 9718 & 3835 \\
    DE & 2009 & REP & Die Republikaner & F\"{u}r die deutsche Republik – Raus aus dieser EU!  & 444 & 320 \\
    DE & 2004 & SPD & Sozialdemokratische Partei Deutschlands & Europamanifest der SPD  & 1965 & 1019 \\
    DE & 2009 & SPD & Sozialdemokratische Partei Deutschlands & Europamanifest & 5853 & 2404 \\
    DK & 2004 & A & Socialdemokraterne & Socialdemokraternes Visioner for Fremtidens Europa & 2362 & 1199 \\
    DK & 2009 & A & Socialdemokraterne & F\ae llesskab & 2758 & 1283 \\
    DK & 2004 & B & Det Radikale Venstre - Danmarks social-liberale parti & Program til Europa-Parlamentsvalg 2004   & 1422 & 830 \\
    DK & 2009 & B & Det Radikale Venstre - Danmarks social-liberale parti & Europa  & 2338 & 1178 \\
    DK & 2004 & C & Det Konservative Folkeparti & Sund konservativ fornuft i Europa & 948 & 530 \\
    DK & 2009 & C & Det Konservative Folkeparti & Konservatives EP-valgprogram & 2847 & 1283 \\
    DK & 2004 & F & Socialistisk Folkeparti & Fremtidens Europa - SFs valgprogram til Europaparlamentsvalg 2004  & 4151 & 1804 \\
    DK & 2009 & F & Socialistisk Folkeparti & Et ansvarligt Europa   & 473 & 338 \\
    DK & 2009 & J & Juni Bev\ae gelsen & F\r{a} Tilsendt Hanne Dahls Nye Bog Helt Gratis & 417 & 263 \\
    DK & 2009 & N & Folkebev\ae gelsen mod EU & Valggrundlag - opstillingsgrundlag og rammer  & 1019 & 572 \\
    DK & 2004 & O & Dansk Folkeparti & Den Europ\ae iske Union    & 791 & 509 \\
    DK & 2009 & O & Dansk Folkeparti & Den Europ\ae iske Union  & 1452 & 816 \\
    DK & 2004 & V & Venstre, Danmarks liberale parti & En st\ae rk stemme i det ny Europa – Venstres Valgprogram til EP valg 2004  & 2687 & 1360 \\
    DK & 2009 & V & Venstre, Danmarks liberale parti & Venstres handlingsprogram til Europa-Parlamentsvalget 2009  & 4287 & 1709 \\
    EE & 2004 & EKRP-EKD & Erakond Eesti Kristlikud Demokraadid-Eesti Kristlik Rahvapartei & Kaitse Eesti Krooni, Vali Rahvaliit & 1062 & 794 \\
    EE & 2004 & IL & Erakond Isamaaliit & Eesti Eest Euroopas!  & 987 & 782 \\
    EE & 2004 & K & Eesti Keskerakond & Eesti Keskerakonna Valimisprogramm Euroopa Parlamendi Valimisteks   & 848 & 696 \\
    EE & 2009 & K & Eesti Keskerakond & Eesti Vajab Vahetust!   & 1060 & 859 \\
    EE & 2004 & RE & Eesti Reformierakond & Reformierakonna Platvorm Euroopa Parlamendi  Valimisteks   & 726 & 559 \\
    EE & 2009 & RE & Eesti Reformierakond & Plaan Eesti Majanduskasvu Taastamiseks  & 1421 & 1022 \\
    EE & 2004 & RESP & Erakond Res Publica & Res Publica Teekaart Euroopas  & 4258 & 2675 \\
    EE & 2009 & RESP & Isamaa ja Res Publica Liit & Isamaa Ja Res Publica Liidu Programm Europarlamendi Valimistel    & 833 & 639 \\
    EE & 2004 & RM-SDE & Rahvaerakond M\~{o}\~{o}dukad-Sotsiaaldemokraatlik Erakond & Sotsiaaldemokraatliku Erakonna P\~{o}him\~{o}tted Ja Lubadused T\"{o}\"{o}ks Euroopa Parlamendis   & 877 & 704 \\
    EE & 2009 & RM-SDE & Rahvaerakond M\~{o}\~{o}dukad-Sotsiaaldemokraatlik Erakond & Inimesed Eelk\~{o}ige: Uus Suund Euroopale    & 1397 & 1102 \\
    ES & 2009 & BNG & Bloque Nacionalista Galego & Imos A Europa. V\'{e}s?& 5552 & 1840 \\
    ES & 2009 & CDS & Centro Democrático y Social/Coalici\'{o}n Foro & Programa Electoral Para Las Elecciones Europeas -2009  & 595 & 366 \\
    ES & 2009 & CIUCDCUDC & Convergència i Uni\'{o} & Programa Electoral Ciu Eleccions Europees 2009  & 22238 & 4931 \\
    ES & 2009 & ERC & Esquerra Republicana de Catalunya & Programma Electoral - Eleccions Al Parlament  Europeu 2009  & 4461 & 1741 \\
    ES & 2004 & IU & Izquierda Unida & Programa De Izquierda Unida   & 12489 & 3908 \\
    ES & 2009 & IU & Izquierda Unida & Programa Electoral Elecciones Europeas 2009. Izquierda Unida  & 16479 & 4534 \\
    ES & 2009 & Los Verdes & Confederaci\'{o}n de los Verdes & Programa Electoral Los Verdes & 18814 & 5034 \\
    ES & 2004 & PNV-EAJ & Partido Nacionalista Vasco-Euzko Alderdi Jeltzalea & Una Nueva Europa Ampliada  Abierta A Las Personas Y Al Mundo  & 22489 & 4968 \\
    ES & 2009 & PNV-EAJ & Partido Nacionalista Vasco-Euzko Alderdi Jeltzalea & Programa Electoral Europeas-09   & 7285 & 2699 \\
    ES & 2004 & PP & Partido Popular & Programa Electoral Elleciones Europeas  & 6244 & 2140 \\
    ES & 2009 & PP & Partido Popular & Programa Electoral Extenso Elecciones Al Parlamento Europeo  & 17745 & 4591 \\
    ES & 2004 & PSOE & Partido Socialista Obrero Espa\~{n}ol & Manifiesto Europeas 2004 & 4120 & 1744 \\
    ES & 2009 & PSOE & Partido Socialista Obrero Espa\~{n}ol & Manifiesto-Programa  Electoral Psoe ‘Europeas 2009’  & 5566 & 2201 \\
    ES & 2009 & UPD & Uni\'{o}n Progreso y Democracia & Programa Electoral    & 5971 & 2302 \\
    FI & 2004 & KD & Suomen Kristillisdemokraatit & Kristillisdemokraattien  & 313 & 285 \\
    FI & 2009 & KD & Suomen Kristillisdemokraatit & Tehtävä EU:ssa  & 4867 & 3050 \\
    FI & 2004 & KESK & Suomen Keskusta & Keskustan Eurooppa-kannanotto   & 2510 & 1732 \\
    FI & 2009 & KESK & Suomen Keskusta & Urhoutta Eurooppaan  & 3444 & 2347 \\
    FI & 2004 & KOK & Kansallinen Kokoomus & ”Jotta Suomella menisi paremmin” - Kokoomuksen eurovaalijulistus   & 1847 & 1325 \\
    FI & 2009 & KOK & Kansallinen Kokoomus & Kokoomuksen eurovaaliohjelma 2009   & 985 & 805 \\
    FI & 2009 & PERUS & Perussuomalaiset & Perussuomalaisten Eu-Vaaliohjelma 2009  & 1626 & 1114 \\
    FI & 2004 & RKP/SFP & Suomen ruotsalainen kansanpuolue/Svenska folkpartiet i Finland & Eurooppa Koskee Sinua   & 1343 & 1026 \\
    FI & 2009 & RKP/SFP & Suomen ruotsalainen kansanpuolue/Svenska folkpartiet i Finland & Moninaisuus tuo lis\"{a}arvoa. RKP – yhteinen tekij\"{a} & 749 & 602 \\
    FI & 2004 & SDP & Suomen Sosialidemokraattinen Puolue & Ihmisten Eurooppaan  & 1491 & 1129 \\
    FI & 2009 & SDP & Suomen Sosialidemokraattinen Puolue & Euroopan Parlamentin Vaalien - Vaaliohjelma 2009 & 2331 & 1667 \\
    FI & 2004 & VAS & Vasemmistoliitto & Meid\"{a}n Eurooppa   & 574 & 470 \\
    FI & 2009 & VAS & Vasemmistoliitto & Parempi Eurooppa on mahdollinen  & 1474 & 1097 \\
    FI & 2004 & VIHR & Vihre\"{a} liitto & Vihre\"{a}n liiton EU-ohjelma   & 198 & 179 \\
    FI & 2009 & VIHR & Vihre\"{a} liitto & Green new deal - uusi vihre\"{a} sopimus Euroopalle  & 2115 & 1543 \\
    FR & 2009 & EE & Europe Écologie & Le Contrat Ecologiste Pour L'Europe  & 8427 & 3058 \\
    FR & 2009 & FG & Front de Gauche & D\'{e}claration de principes du Front de Gauche pour Changer d'Europe & 1508 & 855 \\
    FR & 2004 & FN & Front National & Les Abberations de l'Europe  & 6120 & 2424 \\
    FR & 2009 & FN & Front National & «Leur» Europe N’est Pas La Notre ! Voila L’europe Que Nous Voulons  & 1344 & 803 \\
    FR & 2009 & Libertas & Libertas & Le Projet & 490 & 321 \\
    FR & 2009 & LO & Lutte ouvrière & Lutte Ouvrière dans les élections européennes  & 837 & 482 \\
    FR & 2009 & MODEM & Mouvement Démocrate & Nous l'Europe & 1683 & 870 \\
    FR & 2004 & PCF & Parti communiste fran\c{c}ais & L'Europe: oui. Mais pas celle-l\`{a}!  & 2310 & 1037 \\
    FR & 2004 & PRG & Parti Radical de Gauche & De nouveaux caps pour l’Europe  & 1313 & 735 \\
    FR & 2004 & PS & Parti socialiste & Une Ambition Socialiste pour L'Europe& 4676 & 1853 \\
    FR & 2009 & PS & Parti socialiste & L’Europe face \`{a} la crise: la relance des socialistes  & 1119 & 640 \\
    FR & 2004 & UDF & Union pour la D\'{e}mocratie Française & Nous avons besoin d'Europe  & 8721 & 2941 \\
    FR & 2004 & UMP & Union pour un mouvement populaire & Avec l'Europe, Voyons la France en Grand!  & 1873 & 945 \\
    FR & 2009 & UMP & Union pour un mouvement populaire & 30 Propositions pour une Europe Qui Prot\`{e}ge et Qui Agit & 4748 & 1841 \\
    GR & 2004 & KKE & \gr{Κομμουνιστικό Κόμμα Ελλάδας} & \gr{∆ιακηρυξη Της Κεντρικης Επιτροπης Του ΚΚΕ} & 2810 & 1599 \\
    GR & 2009 & KKE & \gr{Κομμουνιστικό Κόμμα Ελλάδας} & \gr{Διακηρυξη Της Κεντρικης Επιτροπης Του Κκε Για Τις}   & 4179 & 2218 \\
    GR & 2009 & LAOS & \gr{Λαικός Ορθόδοξος Συνδεσμός} & \gr{Ευρωεκλογες 2009}  & 1163 & 705 \\
    GR & 2004 & ND & \gr{Νέα Δημοκρατία} & \gr{Πολιτικα Κειμενα} & 9031 & 3225 \\
    GR & 2009 & ND & \gr{Νέα Δημοκρατία} & \gr{Νεα Δημοκρατία Η Αυθεντική Ευρωπαϊκή Επιλογή}   & 1686 & 1158 \\
    GR & 2009 & OP & \gr{Οικολόγοι Πράσινοι} & \gr{Διακήρυξη για τις Ευρωεκλογές 2009}  & 1392 & 906 \\
    GR & 2004 & PASOK & \gr{Πανελλήνιο Σοσιαλιστικό Κίνημα} & \gr{Ευρωεκλογες 2004 - Το Όραµα, Οι Θέσεις, Οι Δεσµεύσεις µας}   & 2037 & 1049 \\
    GR & 2009 & PASOK &  \gr{Πανελλήνιο Σοσιαλιστικό Κίνημα} & \gr{Ψηφιζουμε Για Την Ευρώπη - Αποφασιζουμε Για Την Ελλάδα}  & 2236 & 1130 \\
    GR & 2004 & SYRIZA & \gr{Συνασπισμός Ριζοσπαστικής Αριστεράς - Ενωτικό Κοινωνικό Μέτωπο} & \gr{Συνασπισμός Της Αριστεράς Των Κινημάτων Και Της Οικολογίας} & 2531 & 1328 \\
    GR & 2009 & SYRIZA & \gr{Συνασπισμός Ριζοσπαστικής Αριστεράς - Ενωτικό Κοινωνικό Μέτωπο} & \gr{Διακηρυξη Για Τισ Ευρωεκλογεσ}& 1050 & 694 \\
    HU & 2004 & FIDESZ-MPP & Fidesz – Magyar Polg\'{a}ri Sz\"{o}vets\'{e}g & Csak egyr\"{u}tt siker\"{u}lhet! & 23616 & 9144 \\
    HU & 2009 & FIDESZ-MPP & Fidesz – Magyar Polg\'{a}ri Sz\"{o}vets\'{e}g & El\^{o}sz\'{o}   & 64743 & 18442 \\
    HU & 2009 & JOBBIK & Jobbik Magyarorsz\'{a}g\'{e}rt Mozgalom & Magyarorsz\'{a}g a magyarok\'{e}  & 14393 & 7280 \\
    HU & 2004 & MDF & Magyar Demokrata F\'{o}rum & „A norm\'{a}lis Magyarorsz\'{a}g\'{e}rt!”  & 1514 & 1096 \\
    HU & 2009 & MDF & Magyar Demokrata F\'{o}rum & Mi\'{e}rt IGEN az MDF list\'{a}j\'{a}ra j\'{u}nius 7-\'{e}n?  & 2681 & 1614 \\
    HU & 2004 & MSZP & Magyar Szocialista P\'{a}rt & A Sikeres Eur\'{o}pai Magyarorsz\'{a}g\'{e}rtM & 772 & 543 \\
    HU & 2009 & MSZP & Magyar Szocialista P\'{a}rt & \'{U}jult erovel & 1570 & 1066 \\
    HU & 2004 & SZDSZ & Szabad Demokrat\'{a}k Sz\"{o}vets\'{e}ge & Egy \'{U}j, Kibov\'{i}tett Eur\'{o}pa, Mely Nyitott \'{A}llampolg\'{a}rai & 7478 & 3776 \\
    HU & 2009 & SZDSZ & Szabad Demokrat\'{a}k Sz\"{o}vets\'{e}ge & 200 001 Szabad, Demokrata Szavaz\'{o} & 456 & 375 \\
    IE & 2004 & FF & Fianna F\'{a}il & Fianna F\'{a}il 2004  & 4707 & 1510 \\
    IE & 2009 & FF & Fianna F\'{a}il & Europe, we are better working together  & 8014 & 2265 \\
    IE & 2004 & FG & Fine Gael & Fine Gael European Parliament Elections 2004 & 5861 & 1872 \\
    IE & 2009 & FG & Fine Gael & Securing Ireland's Future in Europe  & 6404 & 1882 \\
    IE & 2004 & GREENS & Green Party & Manifesto 2004 - European and Local Elections & 3948 & 1595 \\
    IE & 2009 & GREENS & Green Party & A Green New Deal for Europe  & 2445 & 1033 \\
    IE & 2004 & LAB & Labour Party & Making the Difference in Europe  & 3080 & 1116 \\
    IE & 2009 & LAB & Labour Party & Putting people, jobs and fairness at the heart of Europe  & 4441 & 1533 \\
    IE & 2004 & SF & Sinn F\'{e}in & An Ireland of Equals in a Europe of Equals & 12062 & 2610 \\
    IE & 2009 & SF & Sinn F\'{e}in & Europe '09 & 5008 & 1369 \\
    IE & 2009 & SP & Socialist Party & We Want a Europe Fit for Workers  & 3332 & 1412 \\
    IT & 2009 & Altra & Altra Italia & Programma Unitario Per Le Elezioni Europee  & 2181 & 1171 \\
    IT & 2004 & AN & Alleanza Nationale & Programma - Alleanza Nazionale  & 2015 & 1047 \\
    IT & 2009 & Auton. & L'Autonomia & Nasce il Polo dell'Autonomia  & 382 & 271 \\
    IT & 2004 & DSULIVO & Uniti nell'Ulivo & L'Europa contro le nostre paure   & 14761 & 3910 \\
    IT & 2004 & FI & Forza Italia & Elezioni  Per  Il  Parlamento  Europeo  & 3591 & 1371 \\
    IT & 2009 & IDV & Italia dei Valori & Torniamo In Europa  & 244 & 181 \\
    IT & 2004 & LN & Lega Nord & Programma Per Le Elezioni Europee 2004  & 6306 & 2401 \\
    IT & 2009 & LN & Lega Nord & Proposte e Obiettivi  & 21632 & 5568 \\
    IT & 2009 & PDL & Il Popolo della Libert\`{a} & Elezioni 2009: Manifesto del Partito Popolare Europeo & 777 & 501 \\
    IT & 2004 & PRC & Partito della Rifondazione Comunista & La Sinistra, L'altra Europa  & 29371 & 6610 \\
    IT & 2009 & SEL & Sinistra e Liberta & Sinistra A Liberta - Programma Elettorale & 3963 & 1729 \\
    IT & 2009 & UDC & Unione dei Democratici Cristiani e di Centro & UDC 2009 & 356 & 259 \\
    LT & 2009 & DP & Darbo partija & Geroves Lietuvai Europoje – Svarbiausias Yra Tavo Balsas !   & 681 & 547 \\
    LT & 2004 & LiCS & Liberalu ir centro sajunga & “Padarykime Europa Naudinga Lietuvai”   & 3325 & 1856 \\
    LT & 2009 & LiCS & Liberalu ir centro sajunga & Liberalu Ir Centro Sajungos Rinkimu I Europos Parlamenta   & 4282 & 2333 \\
    LT & 2004 & LKD & Lietuvos krik\u{s}cionys demokratai & 2004 Metu Rinkimu I Europos Parlamenta Programa  & 2699 & 1662 \\
    LT & 2009 & LLRA & Lietuvos lenku rinkimu akcija & Lietuvos Lenku Rinkimu Akcijos Kandidatu I Europos Parlamenta Rinkimu Deklaracija   & 1291 & 914 \\
    LT & 2009 & LRLS & Lietuvos Respublikos Liberalu sajudis & Programa  2009 – 2013 M. Europos Parlamento Kadencijai  & 2599 & 1441 \\
    LT & 2004 & LSDP & Lietuvos socialdemokratu partija & Su Europa - Už Lietuva Veikime Kartu!   & 2490 & 1534 \\
    LT & 2009 & LSDP & Lietuvos socialdemokratu partija & Lietuvos Socialdemokratu Partijos Rinkimu I Europos Parlamenta 2009 Metais Programa   & 4766 & 2433 \\
    LT & 2009 & LVLS & Lietuvos valstieciu liaudininku sajunga & Lietuvos Valstieciu Liaudininku Sajungos (Lvls) Rinkimu I Europos Parlamenta  Programa  & 1877 & 1239 \\
    LT & 2004 & NS & Naujoji Sajunga (Socialliberalai) & Naujosios Sajungos Programa 2004 Metu Europos Parlamento Rinkimams  & 7545 & 3399 \\
    LT & 2004 & TS & Tevynes Sajunga & Tevynes Sajungos Rinkimu I Europos Parlamenta Programa & 5537 & 2954 \\
    LT & 2009 & TS-LKD & Tevynes sajunga - Lietuvos krik\u{s}\u{s}cionys demokratai & Tevynes Sajungos-Lietuvos Krik\u{s}cioniu Demokratu Rinkimu I Europos Parlamenta Programines Nuostatos  & 873 & 634 \\
    LT & 2009 & TT & Tvarka ir teisingumas - Liberalu Demokratu Partija & 2009 Metu Europos Parlamento Rinkimu Programa   & 855 & 643 \\
    LV & 2004 & JL & Jaunais Laiks & Jaunais laiks priek\u{s}vele\u{s}anu programma 2004.gada Eiropas Parlamenta vele\u{s}anam  & 349 & 306 \\
    LV & 2009 & JL & Jaunais Laiks & Jaunais laiks priek\u{s}vele\u{s}anu programma 2009.gada Eiropas Parlamenta vele\u{s}anam  & 375 & 295 \\
    LV & 2004 & LC & Latvijas Cel\u{s} & Savieniba "Latvijas cel\u{s}" priek\u{s}vele\u{s}anu programma 2004.gada Eiropas Parlamenta vele\u{s}anam   & 367 & 309 \\
    LV & 2009 & LPP/LC &  Latvijas Pirma partija/Latvijas Cel\u{s} & Partija "LPP/LC" priek\u{s}vele\u{s}anu programma 2009.gada Eiropas Parlamenta vele\u{s}anam   & 352 & 297 \\
    LV & 2004 & PCTVL & Par cilveka tiesibam vienota Latvija & Politisko organizaciju apvieniba "Par cilveka tiesibam vienota Latvija" priek\u{s}vele\u{s}anu programma 2004.gada Eiropas Parlamenta vele\u{s}anam   & 357 & 302 \\
    LV & 2009 & PCTVL & Par cilveka tiesibam vienota Latvija & PCTVL - Par cilveka tiesibam vienota Latvija priek\u{s}vele\u{s}anu programma 2009.gada Eiropas Parlamenta vele\u{s}anam  & 371 & 298 \\
    LV & 2009 & PS & Pilsoniska Savieniba & "Pilsoniska savieniba" priek\u{s}vele\u{s}anu programma 2009.gada Eiropas Parlamenta vele\u{s}anam   & 390 & 329 \\
    LV & 2009 & SC & Saskanas Centrs & Politisko partiju apvieniba "Saskanas Centrs" priek\u{s}vele\u{s}anu programma 2009.gada Eiropas Parlamenta vele\u{s}anam   & 377 & 307 \\
    LV & 2004 & TB/LNNK & Tevzemei un Brivibai/LNNK & Apvieniba "Tevzemei un Brivibai"/LNNK priek\u{s}vele\u{s}anu programma 2004.gada Eiropas Parlamenta velešanam   & 434 & 353 \\
    LV & 2009 & TB/LNNK & Tevzemei un Brivibai/LNNK & Apvieniba "Tevzemei un Brivibai"/LNNK priek\u{s}vele\u{s}anu programma 2009.gada Eiropas Parlamenta velešanam   & 463 & 394 \\
    LV & 2004 & TP & Tautas Partija & Tautas partija priek\u{s}vele\u{s}anu programma 2004.gada Eiropas Parlamenta vele\u{s}anam   & 349 & 294 \\
    LV & 2009 & TP & Tautas Partija & Tautas partija priek\u{s}vele\u{s}anu programma 2009.gada Eiropas Parlamenta vele\u{s}anam  & 406 & 346 \\
    LV & 2004 & ZZS & Zalo un Zemnieku Savieniba & Zalo un Zemnieku savieniba priek\u{s}vele\u{s}anu programma 2004.gada Eiropas Parlamenta vele\u{s}anam & 230 & 209 \\
    NL & 2004 & CDA & Christen-Democratisch App\`{e}l & Verkiezingsmanifest CDA 2004 & 1042 & 560 \\
    NL & 2009 & CDA & Christen-Democratisch App\`{e}l & Kracht en Ambitie  & 6278 & 2271 \\
    NL & 2004 & CUSGP & ChristenUnie-Staatskundig Gereformeerde Partij & Geloofwaardige keuzes - Manifest voor Christelijke politiek in Europa  & 6431 & 2540 \\
    NL & 2009 & CUSGP & ChristenUnie-Staatskundig Gereformeerde Partij & Samenwerking Ja, Superstaat Nee  & 9119 & 2894 \\
    NL & 2004 & D66 & Democraten '66 & Een succesvol Europa  & 3651 & 1505 \\
    NL & 2009 & D66 & Democraten '66 & Europa gaat om mensen!  & 10035 & 3120 \\
    NL & 2004 & GL & GroenLinks & Eigenwijs Europees  & 16119 & 5296 \\
    NL & 2009 & GL & GroenLinks & Nieuwe Energie voor Europa  & 11997 & 4197 \\
    NL & 2004 & LPF & Lijst Pim Fortuyn & ..... Is U iets Gevraagd ?   & 1427 & 782 \\
    NL & 2004 & PVDA & Partij van de Arbeid & Een Sterk en Sociaal Europa   & 5669 & 2080 \\
    NL & 2009 & PVDA & Partij van de Arbeid & Verkiezingsprogramma Europees Parlement 2009-2014  & 8552 & 2818 \\
    NL & 2009 & PVV & Partij voor de Vrijheid & Partij voor de Vrijheid - Verkiezingsprogramma Europees Parlement 2009  & 234 & 157 \\
    NL & 2004 & SP & Socialistische Partij & Wie zwijgt stemt toe! & 8343 & 2888 \\
    NL & 2009 & SP & Socialistische Partij & Een Beter Europa Begint in Nederland  & 6659 & 2304 \\
    NL & 2004 & VVD & Volkspartij voor Vrijheid en Democratie & Een nieuw, Uitgereid Europa, open voor zijn burgers en open voor de wereld  & 7552 & 2313 \\
    NL & 2009 & VVD & Volkspartij voor Vrijheid en Democratie & Voor een werkend Europa  & 1892 & 965 \\
    PL & 2009 & PDP-CL & Porozumienie dla Przyszlosci -CentroLewica & Europa To Ludzie   & 1665 & 1056 \\
    PL & 2004 & PiS & Prawo i Sprawiedliwosc & Deklaracja Krakowska  & 504 & 418 \\
    PL & 2009 & PiS & Prawo i Sprawiedliwosc & Nowoczesna Solidarna Bezpieczna Polska  & 3909 & 2078 \\
    PL & 2004 & PO & Platforma Obywatelska & Program Europejski Platformy Obywatelskiej  & 996 & 725 \\
    PL & 2009 & PO & Platforma Obywatelska & Projekt dokumentu wyborczego EPL 2009r.   & 13178 & 4912 \\
    PL & 2004 & PSL & Polskie Stronnictwo Ludowe & Zadbamy O Polske !  & 765 & 515 \\
    PL & 2009 & PSL & Polskie Stronnictwo Ludowe & Narodowe  Priorytety  Europejskiej  Polityki  PSL  & 3014 & 1354 \\
    PL & 2004 & SLD & Sojusz Lewicy Demokratycznej & Manifest Europejski SLD    & 555 & 444 \\
    PL & 2009 & SLD & Sojusz Lewicy Demokratycznej & Po pierwsze, czlowiek  & 5714 & 2599 \\
    PL & 2009 & SRP & Samoobrona Rzeczpospolitej Polskiej & Przedstawiciele Samoobrony w Parlamencie Europejskim  & 498 & 379 \\
    PL & 2004 & UW & Unia Wolnosci & Ruszyla kampania wyborcza Unii Wolnosci    & 231 & 180 \\
    PT & 2004 & BE & Bloco de Esquerda & Refundar a Europa Mudar Portugal   & 1913 & 1003 \\
    PT & 2009 & BE & Bloco de Esquerda & Compromisso Eleitoral Da Candidatura Do Bloco \`{A}s Europeias    & 3461 & 1629 \\
    PT & 2009 & CDS-PP & Centro Democr\'{a}tico e Social – Partido Popular & Manifesto Eleitoral Europeias 2009  & 1439 & 825 \\
    PT & 2004 & CDU-PCP/PEV & Partido Comunista Portugu\^{e}s/Partido Ecologista "Os Verdes" & Declara\c{c}\~{a}o Program\'{a}tica2004  & 5023 & 1767 \\
    PT & 2009 & CDU-PCP/PEV & Partido Comunista Portugu\^{e}s/Partido Ecologista "Os Verdes" & Declara\c{c}\~{a}o Program\'{a}tica do PCP para as Elei\c{c}\~{o}es Europeias de 2009 & 4701 & 1642 \\
    PT & 2004 & PPD/PSD & Partido Social Democrata & For\c{c}a Portugal   & 2370 & 1214 \\
    PT & 2009 & PPD/PSD & Partido Social Democrata & Pelo Interesse Nacional   & 690 & 449 \\
    PT & 2004 & PS & Partido Socialista & Pela Europa, pelos portugueses  & 5553 & 2118 \\
    PT & 2009 & PS & Partido Socialista & As Pessoas Primeiro - Um Novo Rumo Para A Europa & 3903 & 1625 \\
    SE & 2004 & C & Centerpartiet & Smalare men vassare!  & 2953 & 1336 \\
    SE & 2009 & C & Centerpartiet & Europas f\"{o}renta krafter  & 1043 & 630 \\
    SE & 2009 & FP & Folkpartiet Liberalerna & Ja till Europa  & 1985 & 1089 \\
    SE & 2009 & JL & Junilistan & Junilistans valplattform 2009  & 548 & 380 \\
    SE & 2004 & KD & Kristdemokraterna & Inf\"{o}r valet till Europaparlamentet 13 juni 2004 & 7580 & 2933 \\
    SE & 2009 & KD & Kristdemokraterna & Ett tryggt Europa – v\r{a}r v\"{a}g dit. & 699 & 498 \\
    SE & 2004 & M & Moderata samlingspartiet & Europasamarbetet kan g\"{o}ra Sverige bättre   & 1420 & 751 \\
    SE & 2009 & M & Moderata samlingspartiet & Tid f\"{o}r ansvar  & 1478 & 807 \\
    SE & 2004 & MP & Milj\"{o}partiet de Gröna & Ja till samarbete, nej till EU-stat - f\"{o}r ett gr\"{o}nt och solidariskt Europa  & 3406 & 1565 \\
    SE & 2009 & MP & Milj\"{o}partiet de Gröna & Valmanifest - Gr\"{o}nt Klimatval 2009  & 284 & 240 \\
    SE & 2009 & PP & Piratpartiet & Principprogram version 3.3 & 1349 & 815 \\
    SE & 2004 & S & Sveriges Socialdemokratiska arbetarpart & Valmanifest 2004  & 638 & 414 \\
    SE & 2009 & S & Sveriges Socialdemokratiska arbetarpart & Valmanifest - Jobben först & 735 & 432 \\
    SE & 2004 & V & V\"{a}nsterpartiet & V\"{a}nsterpartiets EU-Valplattform  & 1529 & 927 \\
    SE & 2009 & V & V\"{a}nsterpartiet & Valplattform inför EU-parlamentsvalet  & 2141 & 1182 \\
    SI & 2009 & LDS & Liberalna demokracija Slovenije & Poslanica LDS za evropske volitve   & 788 & 600 \\
    SI & 2004 & NSI & Nova Slovenija – kr\v{s}canska ljudska stranka & Volitve V Evropski Parlament  & 492 & 391 \\
    SI & 2009 & NSI & Nova Slovenija – kr\v{s}canska ljudska stranka &  Nova Slovenija Kr\v{s}\`{e}anski Ljudska Stranka & 587 & 441 \\
    SI & 2009 & SD & Socialni demokrati & Manifest Stranke evropskih socialdemokratov  & 5870 & 2491 \\
    SI & 2004 & SDS & Slovenska demokratska stranka & Spletna Stran - Program  & 1653 & 1066 \\
    SI & 2009 & SDS & Slovenska demokratska stranka & Nova pot - 20 let slovenske pomladi  & 328 & 245 \\
    SI & 2004 & SLS & Slovenska ljudska stranka & »Vec Slovenije V Evropi, Vec Evrope V Sloveniji«  & 2285 & 1291 \\
    SI & 2009 & SLS & Slovenska ljudska stranka & SLO: SLS + SKD Slovenska Ljudska Stranka  & 161 & 136 \\
    SI & 2009 & Zares & Zares – socialno-liberalni & Vzemimo Evropo Zares  & 14802 & 5045 \\
    SI & 2004 & ZLSD & Zdru\v{z}ena lista socialnih demokratov & V Evropi za dobro Slovenije!  & 2073 & 1185 \\
    SK & 2004 & KDH & Krestanskodemokratick\'{e} hnutie & Volebn\'{y} program KDH do volieb do Eur\'{o}pskeho parlamentu   & 1464 & 1002 \\
    SK & 2009 & KDH & Krestanskodemokratick\'{e} hnutie & Volebn\'{y} program KDH  do Eur\'{o}pskeho parlamentu   & 1735 & 1135 \\
    SK & 2004 & LS-HZDS & Ludov\'{a} strana - Hnutie za demokratick\'{e} Slovensko & Odpovede na ot\'{a}zky: Irena Belohorsk\'{a}, kandid\'{a}tka na poslanca EP za HZDS  & 788 & 563 \\
    SK & 2009 & LS-HZDS & Ludov\'{a} strana - Hnutie za demokratické Slovensko & Slovensko – Stabilné Srdce Eur\'{o}py & 5084 & 2641 \\
    SK & 2004 & SDKU-DS & Slovensk\'{a} demokratick\'{a} a krestansk\'{a} \'{u}nia - Demokratick\'{a} strana & Manifest SDK\'{U} pre nov\'{u} Eur\'{o}pu & 1805 & 1020 \\
    SK & 2009 & SDKU-DS & Slovensk\'{a} demokratick\'{a} a krestansk\'{a} \'{u}nia - Demokratick\'{a} strana & Za Prosperuj\'{u}ce Slovensko V Silnej Európe & 5312 & 2317 \\
    SK & 2004 & SMER-SD & Smer – soci\'{a}lna demokracia & silnej\v{s}ie Slovensko v soci\'{a}lnej Eur\'{o}pe& 2150 & 1121 \\
    SK & 2009 & SMER-SD & Smer – soci\'{a}lna demokracia & Soci\'{a}lna Eur\'{o}pa – Odpoved Na Kr\'{i}zu & 461 & 303 \\
    SK & 2004 & SMK-MKP & Strana madarskej komunity - Magyar K\"{o}z\"{o}ss\'{e}g P\'{a}rtja & Hely\"{u}nk Eur\'{o}p\'{a}ban  & 2506 & 1556 \\
    SK & 2009 & SMK-MKP & Strana madarskej komunity - Magyar K\"{o}z\"{o}ss\'{e}g P\'{a}rtja & Na\v{s}a bud\'{u}cnost v Eur\'{o}pe  & 3944 & 2117 \\
    SK & 2009 & SNS & Slovensk\'{a} n\'{a}rodn\'{a} strana & Jaroslav Pa\v{s}ka: Priority na najbli\v{z}\v{s}\'{i}ch 5 rokov v Eur\'{o}pskom parlamente    & 180 & 153 \\
    UK & 2009 & BNP & British National Party & 2009 Manifesto for the European Elections  & 964 & 489 \\
    UK & 2004 & CON & Conservative Party & Putting Britain First  & 7128 & 2070 \\
    UK & 2009 & CON & Conservative Party & Vote for Change& 4742 & 1611 \\
    UK & 2009 & DUP & Democratic Unionist Party & Strong Leadership in Challenging Times & 385 & 278 \\
    UK & 2009 & GREEN & Green Party of England and Wales & "it's the economy, stupid" & 7831 & 2389 \\
    UK & 2004 & LAB & Labour Party & Britain is working  & 4289 & 1273 \\
    UK & 2009 & LAB & Labour Party & Winning the fight for Britain's future & 4910 & 1357 \\
    UK & 2004 & LD & Liberal Democrats & Making Europe Work For You & 7986 & 2162 \\
    UK & 2009 & LD & Liberal Democrats & Stronger Together, poorer apart  & 5355 & 1590 \\
    UK & 2004 & PC & Plaid Cymru – the Party of Wales & Fighting Hard For Wales  & 2184 & 932 \\
    UK & 2009 & PC & Plaid Cymru – the Party of Wales & European Manifesto & 2914 & 1232 \\
    UK & 2009 & SDLP & Social Democratic and Labour Party & A Vision For Europe - Ambition For You & 7055 & 2223 \\
    UK & 2009 & SF & Sinn F\'{e}in & Sinn F\'{e}in European Election Manifesto 2009 & 4920 & 1372 \\
    UK & 2004 & SNP & Scottish National Party & Vote for Scotland  & 3447 & 1248 \\
    UK & 2009 & SNP & Scottish National Party & We've got what it takes & 3764 & 1211 \\
    UK & 2009 & UKIP & UK Independence Party & UKIP Manifesto 2009  & 295 & 197 \\
    UK & 2009 & UUP & Ulster Unionist Party & Vote For Change  & 4742 & 1611 \\
    \bottomrule
\multicolumn{7}{l}{* Refers to the number of words after the documents were cleaned}	
\end{longtable}
\end{landscape}

\newpage

\section*{Appendix C: Document preparation}

\subsection*{Document Selection}

We obtained the manifestos from the \textit{Euromanifestos Project} website.\footnote{\url{http://www.ees-homepage.net/}} For all countries, text files were available for the 2009 manifestos, while for the 2004 manifestos, only some parties in Germany and the United Kingdom were available in this format. We thus used the stored portable document file, which we converted into UTF-8 text files, to assure compatibility and preservation of non-English characters. When conversion from .pdf was not possible due to the file being saved as an image, we used optical character recognition (OCR) software. While OCR will never convert a text 100\% faithfully, sufficient results can be gained, especially as the software we used allowed us to manually correct mistakes and instances were the software was not sure. For some countries, not all the released manifestos were stored in the database, or the stored document was something other than a true Euromanifesto, in which case we looked for the document in other online sources. Both the resulting .txt and .pdf version of these source documents can be found among our replication files.

\subsection*{Pre-processing}

From all text files, we removed headers and footers, page numbering, section headings, graphs, numbers, currency symbols and tables. We then imported these texts into Wordfreq (cite) to make the frequency tables for each country. From these frequency tables, we then deleted stop-words as they carry minimal information value \cite[332]{Slapin2008}. While not all studies using Wordscores apply stop-words, a significant number do \cite{Ruedin2013,Ruedin2013a,Slapin2008}. Moreover, the practise seems to be common in automatic content analysis \cite{Grimmer2013}, and seems especially suited for Wordscores, as it falsely assumes all scored words to carry the same informative value. However, a word such as ‘immigration’ adds information to a text in a way words like ‘the’ or ‘and’ do not. Nevertheless, as these words occur often in all texts, their score will be close to the mean of the reference texts, and will thus cause the scores for the virgin texts to cluster around the mean. As such, they are indistinguishable from truly centrist words, causing parties to appear more centric than they really are \cite[360--361]{Lowe2008}. Removing these words thus increases the discriminative power of Wordscores. Here, we follow \citeasnoun{Ruedin2013a}, and remove the 20 most frequently occurring words for each country in both 2004 and 2009. We do not use stemming, as this decreases the effectiveness of the method \cite{Ruedin2013a} and because it is not beneficial for all languages. This is especially the case for languages in which compound words are common, such as in German or Finnish, where stemming may lead to a reduction of information. Table \ref{tab:title3} shows  the 20 most frequently occurring words that were dropped for Great Britain. Most of these words can easily be considered non-informative, as they are either adjectives, adverbs or propositions. Even a word as \textit{european} or \textit{europe} can be argued to function mostly as an adjective as would be expected in a manifesto for European elections. The .dta files with these words removed may be found in the replication files.

\begin{table}[ht]
\centering
\caption{Words dropped for Great Britain} \label{tab:title3}
\begin{tabular}{llll}
\toprule
\multicolumn{2}{c}{2004} & \multicolumn{2}{c}{2009} \\
\cmidrule(lr){1-2}
\cmidrule(lr){3-4}
Word & Count & Word & Count\\
\midrule
the      & 2626 & the      & 2785 \\
to       & 1337 & and      & 1814 \\
and      & 1335 & to       & 1770 \\
of       & 1110 & of       & 1252 \\
in       & 844  & in       & 1115 \\
a        & 641  & a        & 795  \\
eu       & 555  & for      & 739  \\
for      & 543  & we       & 707  \\
that     & 448  & that     & 527  \\
is       & 419  & is       & 476  \\
be       & 344  & eu       & 459  \\
we       & 329  & will     & 453  \\
european & 327  & our      & 399  \\
on       & 316  & on       & 394  \\
our      & 256  & european & 340  \\
europe   & 255  & are      & 305  \\
are      & 250  & be       & 300  \\
will     & 240  & as       & 299  \\
has      & 232  & europe   & 294  \\
it       & 230  & with     & 292  \\
\bottomrule
\end{tabular}
\end{table}

\subsection*{Wordcount}

The table below shows the word count for the documents. Using the wordscores package for Stata, we calculated the mean and standard deviation for the total words in the documents and the unique words (referring to words only occurring in a single document). In addition, \textit{New} indicates whether the 2004 European election was the first election the country participated in. Documents from the new countries were significantly shorter in 2004, but showed an increase in 2009, while the number of unique words changed little. The number of documents analysed was higher in 2004 than in 2009, which is mostly to due the availability of an existing digital copy. The number of words per manifesto differs significantly per country and also within countries as shown by the standard deviation. This implies that the size and scope of documents differ and that when performing an analysis, scholars need to be aware of what the document under investigation covers and whether all documents are the same. 

\begin{table}[htbp]
\centering
\caption {Total and Unique word count for the used documents} \label{tab:title0} 

\begin{tabular}{lcrrrrrrrrrr}
\toprule
&& \multicolumn{5}{c}{2004} &  \multicolumn{5}{c}{2009} \\
\cmidrule(lr){3-7}
\cmidrule(lr){8-12}
&&& \multicolumn{2}{c}{Total}&\multicolumn{2}{c}{Unique}&&\multicolumn{2}{c}{Total}&\multicolumn{2}{c}{Unique}\\
\cmidrule(lr){4-5}
\cmidrule(lr){6-7}
\cmidrule(lr){9-10}
\cmidrule(lr){11-12}
    Country & New & Obs & Mean & SD  & Mean  & SD & Obs & Mean & SD & Mean& SD \\
    \midrule
    AT & No & 4 & 2037 & 1231 & 1100 & 580 &    6 & 2708 & 2046 & 1285 & 864 \\
    BE(FR) & No & 4 & 8709 & 5753 & 2658 & 1111    & 5 & 10363 & 3389 & 3229 & 523 \\
    BE(NL) & No & 6 & 6217 & 5149 & 2137 & 1405    & 7 & 7966 & 4128 & 2711 & 1077 \\
    CY & Yes & 5 & 1511 & 635 & 855 & 344 &    4 & 1258 & 365 & 759 & 200 \\
    CZ & Yes & 4 & 1738 & 632 & 1060 & 326 &    6 & 2001 & 1876 & 1191 & 890 \\
    DE & No & 6 & 8349 & 9228 & 2961 & 2567 &    7 & 4398 & 3219 & 1909 & 1216 \\
    DK & No & 6 & 2060 & 1274 & 1039 & 509 &    8 & 1949 & 1348 & 930 & 515 \\
    EE & Yes & 6 & 1460 & 1376 & 1035 & 808 &   4 & 1178 & 283 & 906 & 204 \\
    ES & No & 4 & 11336 & 8241 & 3190 & 1513 &    10 & 10471 & 7523 & 3024 & 1,627 \\
    FI & No & 7 & 1182 & 858 & 878 & 580 &    8 & 2199 & 1369 & 1528 & 823 \\
    FR & No & 6 & 4169 & 2887 & 1656 & 896 &   8 & 2520 & 2722 & 1109 & 909 \\
    GR & No & 4 & 4102 & 3301 & 1800 & 976 &    6 & 1951 & 1171 & 1135 & 567 \\
    HU & Yes & 4 & 8345 & 10614 & 3640 & 3932    & 5 & 16769 & 27399 & 5755 & 7605 \\
    IE & No & 5 & 5932 & 3576 & 1741 & 556 &    6 & 4941 & 2033 & 1582 & 432 \\
    IT & No & 5 & 11209 & 11281 & 3068 & 2273    & 7 & 4219 & 7797 & 1383 & 1932 \\
    LT & Yes & 5 & 4319 & 2171 & 2281 & 840    & 8 & 2153 & 1598 & 1273 & 752 \\
    LV & Yes & 6 & 348 & 66 & 296 & 47 &    7 & 391 & 36 & 324 & 36 \\
    NL & No & 8 & 6279 & 4795 & 2246 & 1482    & 8 & 6846 & 4025 & 2341 & 1268 \\
    PL & Yes & 5 & 610 & 288 & 456 & 196 &    6 & 4663 & 4544 & 2063 & 1597 \\
    PT & No & 4 & 3715 & 1839 & 1526 & 510    & 5 & 2839 & 1700 & 1234 & 561 \\
    SE & No & 6 & 2921 & 2504 & 1321 & 890    & 9 & 1140 & 645 & 675 & 323 \\
    SI & Yes & 4 & 1626 & 800 & 983 & 405    & 6 & 3756 & 5831 & 1493 & 1945 \\
    SK & Yes & 5 & 1743 & 660 & 1052 & 354    & 6 & 2786 & 2294 & 1444 & 1069 \\
    UK & No & 5 & 5007 & 2464 & 1537 & 546    & 12 & 3990 & 2447 & 1297 & 691 \\
        \bottomrule
    \end{tabular}
\end{table}

\newpage

\begin{landscape}
\section*{Appendix D: Data sources and question wording}
 \small

    \begin{longtable}{p{3.4cm}p{4.4cm}p{4.4cm}p{4.4cm}p{4.4cm}}
    \toprule
      & \multicolumn{1}{c}{\textbf{LR - Left-Right}} & \multicolumn{1}{c}{\textbf{EU - EU Integration}} & \multicolumn{1}{c}{\textbf{EC - Economic}} & \multicolumn{1}{c}{\textbf{SO - Social}} \\
    \midrule
        &&&&
    \endhead

    Benoit \& Laver Expert Survey \cite{Benoit2006}& Left-Right - Please locate each party on a general left-right dimension, taking all aspects of party policy into account & \dag EU Authority (AT, BE, UK, DK, FI, DE, GR, IT, NL, NI, PT, ES, SE), EU Larger \& Stronger (FR), \dag EU Strengthening (IE)  & Economic (Spending vs. Taxes)  & Social  \\
    &   &  &   & \\
& Left (1) & Favours (1)& Promotes raising taxes to increase public services (1) & Favours liberal policies on matters such as abortion, homosexuality, and euthanasia (1)  \\
      &   &  &   & \\
&  Right (20) & Opposes (20) & Promotes cutting public services to cut taxes (20) & Opposes liberal policies on matters such as abortion, homosexuality, and euthanasia (20) \\
      
&   & Countries excluded are CZ, EE, HU, LV, LT, PL, SK, SI, CY &   &  \\

    Chapel Hill Expert Survey 2002 \cite**{Hooghe2010}& LRGEN = position of the party in 2002 in terms of its broad ideological stance, where & POSITION = overall orientation of the party leadership towards European integration in 2002, where & LRECON = position of the party in 2002 in terms of its ideological stance on economic issues (role of government in economy), where & GALTAN = position of the party in 2002 in terms of its ideological stance on democratic freedoms and rights (role of government in life choices), where \\
    &&&&\\
    & 0 indicates that a party is at the extreme left of the ideological spectrum &1 = Strongly opposed to European integration&0 indicates that a party is at the extreme left of the ideological spectrum&0 indicates that a party is at the extreme left of the ideological spectrum\\
        &&&&\\
    & 5 means that it is at the center &4 = Neutral, no stance on the issue of European integration&5 means that it is at the center&5 means that it is at the center\\
        &&&&\\
    & 10 indicates that it is at the extreme right &7 = Strongly in favour of European integration&10 indicates that it is at the extreme right&10 indicates that it is at the extreme right\\
&   &  &   &  \\  
&   &  &   &  \\
&   &  &   &  \\
&   &  &   &  \\

    Chapel Hill Expert Survey 2010 \cite**{Bakker2012}& LRGEN = position of the party in 2010 in terms of its overall ideological stance & POSITION = overall orientation of the party leadership towards European integration in 2010 & LRECON = position of the party in 2010 in terms of its ideological stance on economic issues & GALTAN = position of the party in 2010 in terms of its ideological stance on democratic freedoms and rights \\
          &&&&\\
      & 0 = extreme left & 1 = strongly opposed & 0 = extreme left & 0 = extreme left \\
      & (-) & (-) & (-) & (-) \\
      & 5 = center & 4 = neutral & 5 = center & 5 = center \\
      & (-) & (-) & (-) & (-) \\
      & 10 = extreme right & 7 = strongly in favour & 10 extreme right & 10 extreme right \\
      &   &   &   &  \\
         \hline
                        &   &   &   &  \\
	Euromanifestos Project 2004 \cite{Braun2010}& LEFT - placement of Euromanifesto according to the coder on a left-right scale & \dag EU - placement of Euromanifesto according to coder on a pro-anti-EU-integration scale & STATE - placement of Euromanifesto according to coder on a state interventionism vs. free enterprise scale & LIB - placement of Euromanifesto according to coder on a libertarian-authoritarian scale. \\
				                  &   &   &   &  \\
      & 1=left  & 1 = pro & 1=state interventionism  & 1=libertarian   \\
	  & 10=right  & 10 = anti   &  10=free enterprise & 10=authoritarian \\
			                  &   &   &   &  \\
   	Euromanifestos Project 2009 \cite{Braun2010}& LEFT - Left - Right & \dag INTEGRATION -  Pro EU-Integration - Anti-EU-Integration & STATE - State Interventionism - Free Enterprise & LIBERTA - Libertarian - Authoritarian  \\
   	&&&&\\
& Coder rating on a 10-point-scale  & Coder rating on a 10-point-scale  & Coder rating on a 10-point-scale  & Coder rating on a 10-point-scale \\
   	&&&&\\
		            \hline
		                  &   &   &   &  \\
   EU Profiler 2009 \cite{Trechsel2010} \ddag & Modified Left-Right - using items 1, 2, 3, 5, 6, 7, 8, 9, 10, 11, 14, 16, 18, 19 and 20, with missing values recoded to 4 (Neutral) & Original EU Integration (Y axis), using items 12, 21, 22, 23, 24, 26 and 27 & Scale composed of items 1, 2, 11, 14, 16, and 18 & Scale composed of items 5, 6, 7, 8, 9, 10, 19, 20 and 25 \\
               &   &   &   &  \\
    \bottomrule
                   &   &   &   &  \\
        \multicolumn{5}{l}{\dag Denotes variables that have been reversed for subsequent analysis}  \\
        \multicolumn{5}{l}{\ddag EU Profiler data were scaled according to \citeasnoun{Gemenis2013c}.}\\
\end{longtable}
\end{landscape}

\newpage

\begin{landscape}
\section*{Appendix E: Documents selected for the Martin-Vanberg transformation}

\small

\begin{longtable}{m{1.4cm}m{0.7cm}m{1.4cm}m{1.4cm}m{1.4cm}m{1.4cm}m{1.4cm}m{1.4cm}m{1.4cm}m{1.4cm}m{1.4cm}m{1.4cm}m{1.4cm}m{1.4cm}}
\toprule
\multicolumn{1}{c}{Country*} &
&
\multicolumn{4}{c}{BL} &
\multicolumn{4}{c}{CHES} &
\multicolumn{4}{c}{EMP}\\ 
\cmidrule(lr){3-6}
\cmidrule(lr){7-10}
\cmidrule(lr){11-14}
& &
\multicolumn{1}{c}{LR} & \multicolumn{1}{c}{EU} & \multicolumn{1}{c}{EC} & \multicolumn{1}{c}{SO} & \multicolumn{1}{c}{LR} & \multicolumn{1}{c}{EU} & \multicolumn{1}{c}{EC} & \multicolumn{1}{c}{SO} & \multicolumn{1}{c}{LR} & \multicolumn{1}{c}{EU} & \multicolumn{1}{c}{EC} & \multicolumn{1}{c}{SO}\\
\midrule
\endhead
    \multirow{2}{*}{AT} & low & GR\"{U}N & FP\"{O}   & GR\"{U}N & GR\"{U}N & GR\"{U}N & FP\"{O}   & GR\"{U}N & GR\"{U}N & GR\"{U}N & FP\"{O}   & GR\"{U}N & GR\"{U}N \\
    & high & FP\"{O}   & GR\"{U}N & \"{O}VP   & FP\"{O}   & FP\"{O}   & \"{O}VP   & \"{O}VP   & FP\"{O}   & FP\"{O}   & \"{O}VP   & FP\"{O}   & FP\"{O} \\
     \multirow{2}{*}{BE (FR)} & low & ECOLO & MR & ECOLO & ECOLO & ECOLO & ECOLO & ECOLO & ECOLO & PS    & MR & PS    & PS \\
     & high & MR & CDH   & MR & CDH   & MR & CDH   & MR & CDH   & CDH   & ECOLO & CDH   & ECOLO \\
     \multirow{2}{*}{BE (NL)} & low & GROEN & VB    & GROEN & GROEN & GROEN & VB    & GROEN & GROEN & SPA   & VB    & SPA   & VLD \\
     & high & VB    & CDV   & VLD   & VB    & VB    & CDV   & VB    & VB    & VLD   & GROEN & VLD   & SPA \\
     \multirow{2}{*}{CY} & low & AKEL  & -     & AKEL  & DISY  & -     & -     & -     & -     & KOP   & KOP   & AKEL  & KOP \\
     & high & DISY  & -     & DISY  & AKEL  & -     & -     & -     & -     & DIKO  & DISY  & DIKO  & EDEK \\
    \multirow{2}{*}{CZ} & low & KSCM  & -     & KSCM  & CSSD  & KSCM  & KSCM  & KSCM  & ODS   & KSCM  & ODS   & KSCM  & CSSD \\
     & high & ODS   & -     & ODS   & KDUCSL & ODS   & CSSD  & KDUCSL & KSCM  & ODS   & CSSD  & ODS   & KSCM \\
     \multirow{2}{*}{DK} & low & F     & O     & F     & B     & F     & O     & F     & F     & F     & O     & F     & V \\
     & high & O     & V     & C     & C     & O     & V     & V     & O     & O     & V     & O     & O \\
     \multirow{2}{*}{EE} & low & K     & -     & SDE   & SDE   & -     & -     & -     & -     & SDE   & RE    & SDE   & RESP \\
     & high & RE    & -     & RESP  & EKRP & -     & -     & -     & -     & RE    & SDE   & RESP  & IL \\
     \multirow{2}{*}{FI} & low & VAS   & KESK  & VAS   & VIHR  & VAS   & KD    & VAS   & VIHR  & VAS   & KD    & VAS   & KESK \\
     & high & KOK   & SDP   & KOK   & KD    & KOK   & KOK   & KOK   & KD    & KOK   & KESK  & KOK   & KD \\
     \multirow{2}{*}{FR} & low &   -    & FN    & PCF   & PS    & PCF   & FN    & PCF   & PS    & PCF   & FN    & PCF   & PRG \\
     & high &  -     & UDF   & FN    & FN    & FN    & PS    & FN    & FN    & FN    & PCF   & UDF   & FN \\
     \multirow{2}{*}{DE} & low & LINKE & CSU   & LINKE & B90GR\"{U} & LINKE & LINKE & LINKE & B90GR\"{U} & LINKE & CSU   & LINKE & SPD \\
     & high & CSU   & B90GR\"{U} & FDP   & CDU   & CSU   & CDU   & FDP   & CSU   & CSU   & LINKE & CDU   & CDU \\
     \multirow{2}{*}{GR} & low & KKE   & KKE   & KKE   & SYRIZA & SYRIZA & KKE   & KKE   & SYRIZA & KKE   & KKE   & SYRIZA & SYRIZA \\
     & high & ND    & ND    & ND    & ND    & ND    & PASOK & ND    & ND    & ND    & ND    & ND    & KKE \\
     \multirow{2}{*}{HU} & low & MSZP  & -     & FIDESZ & SZDSZ & MSZP  & FIDESZ & FIDESZ & SZDSZ & FIDESZ & SZDSZ & MDF   & SZDSZ \\
     & high & FIDESZ & -     & SZDSZ & FIDESZ & FIDESZ & SZDSZ & SZDSZ & FIDESZ & SZDSZ & MSZP  & SZDSZ & MSZP \\
     \multirow{2}{*}{IE} & low & GREENS & GREENS & SF    & GREENS & GREENS & SF    & SF    & GREENS & SF    & SF    & SF    & SF \\
     & high & FF    & FG    & FF    & FF    & FG    & FG    & FG    & FF    & FG    & FF    & FG    & FF \\
     \multirow{2}{*}{IT} & low & PRC   & LN    & PRC   & PRC   & PRC   & LN    & PRC   & DSULIVO & PRC   & PRC   & PRC   & FI \\
     & high & AN    & DSULIVO & FI    & AN    & AN    & DSULIVO & FI    & AN    & FI    & DSULIVO & FI    & LN \\
    \multirow{2}{*}{LV} & low & PCTVL & -     & PCTVL & JL    & PCTVL & PCTVL & PCTVL & TP    & PCTVL & ZZS   & PCTVL & LC \\
     & high & TP    & -     & TP    & LC    & TBLNNK & JL    & TP    & LC    & TBLNNK & JL    & LC    & JL \\
     \multirow{2}{*}{LT} & low & LSDP  & -     & LSDP  & LICS  & LSDP  & LKD   & NS    & LICS  & LSDP  & LSDP  & LKD   & LICS \\
     & high & LICS  & -     & LICS  & LKD   & TS    & TS    & TS    & LKD   & TS    & NS    & TS    & LKD \\
    \multirow{2}{*}{NL} & low & SP    & LPF   & SP    & GL    & SP    & LPF   & SP    & D66   & SP    & LPF   & SP    & VVD \\
     & high & LPF   & D66   & VVD   & CUSGP & LPF   & D66   & LPF   & CUSGP & LPF   & VVD   & CDA   & SP \\
    \multirow{2}{*}{PL} & low & SLDUP & -     & SLDUP & SLDUP & SLDUP & PSL   & PSL   & SLDUP & SLDUP & PIS   & PSL   & UW \\
     & high & PIS   & -     & PO    & PIS   & PIS   & UW    & PO    & PIS   & PIS   & UW    & PO    & PIS \\
    \multirow{2}{*}{PT} & low & BE    & BE    & CDU   & BE    & CDU   & CDU   & CDU   & CDU   & CDU   & CDU   & CDU   & BE \\
     & high & PSD   & PS    & PSD   & PSD   & PSD   & PS    & PSD   & PSD   & PSD   & PS    & PSD   & CDU \\
     \multirow{2}{*}{SK} & low & SMER  & -     & SMER  & SMER  & SMER  & SMER  & SMER  & SDKUDS & SMER  & KDH   & SMER  & SMER \\
     & high & SDKUDS & -     & KDH   & KDH   & KDH   & SDKUDS & SDKUDS & KDH   & KDH   & SMER  & KDH   & LSHZDS \\
     \multirow{2}{*}{SI} & low & ZLSD  & -     & ZLSD  & ZLSD  & ZLSD  & SLS   & ZLSD  & ZLSD  & ZLSD  & SLS   & ZLSD  & SDS \\
     & high & NSI   & -     & NSI   & NSI   & NSI   & SDS   & SLS   & NSI   & NSI   & NSI   & NSI   & SLS \\
     \multirow{2}{*}{ES} & low & PSOE  & PP    & IU    & IU    & IU    & IU    & IU    & IU    & IU    & IU    & IU    & IU \\
     & high & PP    & PSOE  & PP    & PP    & PP    & PSOE  & PP    & PP    & PNVEAJ & PSOE  & PNVEAJ & PP \\
    \multirow{2}{*}{ SE} & low & V     & V     & V     & V     & V     & MP    & V     & MP    & V     & V     & V     & MP \\
     & high & M     & M     & M     & KD    & M     & M     & M     & KD    & KD    & M     & M     & V \\
     \multirow{2}{*}{GB} & low & PC    & CON   & PC    & LD    & SNP   & CON   & SNP   & LD    & PC    & CON   & PC    & SNP \\
     & high & CON   & LD    & CON   & CON   & CON   & LD    & CON   & CON   & CON   & PC    & CON   & CON \\
     \multirow{2}{*}{NI} & low & SF    & SDLP  & SF    & SF    & -     & -     & -     & -     & SF    & SUP   & DUP   & SF \\
     & high & UUP   & DUP   & UUP   & DUP   & -     & -     & -     & -     & UUP   & SDLP  & UUP   & DUP \\
    \bottomrule
    \multicolumn{13}{l}{*Low and high refer to the party with either the lowest score or the highest score on a dimension}
\end{longtable}
\end{landscape}

\newpage

\section*{Appendix F: Investigating the Martin \& Vanberg transformation}

In their original article \citeasnoun{Martin2008} (hereafter MV) advise in a footnote to calculate the difference between the exogenous assigned scores and the score as used the their transformation to calculate the size of the trade-off scholars have to make between increased accuracy of the dictionary and internal consistency and the ability make valid comparisons. While this step is not necessary to validate the applicability of the MV transformation in our study as we do not compare our scores against the reference scores, we decided to calculate these differences in order to test the transformation and give a preliminary assessment of the trade-off for scholars who want to use the transformation in the future. To calculate the trade-off, we input the reference documents a second time as the virgin documents. The difference between the transformed score and the exogenous assigned score then indicates the degree of trade-off. In addition, it provides the user with an extra tool to assess whether the actual word usage of the texts is reflected in the exogenous assigned score. A large difference then means that the exogenous score is not equal to what is reflected in the words. This difference can be either negative or positive, depending on the direction (either lower or higher on the dimension of interest). To give an idea of how this works, we calculate the difference on the EU integration dimension in the Netherlands using the reference scores from the Benoit \& Laver dataset.

\begin{table}[htbp] 
  \centering
 \caption {Differences for the Netherlands on the EU integration dimension} \label{tab:title1} 
    \begin{tabular}{lrrrr} 
    \toprule Party & Exogenous score  & MV altered score & Difference & \% Difference \\  
    \midrule 
    LPF & 5.1667 & 5.1667  & 0 & 0 \\  
    SP & 5.4706 & 7.407 & 1.9364 & 35.41 \\    
    CU-SGP & 7.3572 & 8.7889 &  1.4317 & 19.41 \\
    VVD & 8.4 & 9.7341 & 1.3341 & 15.88 \\
    CDA & 11.3 & 12.1469  & 0.8469 & 7.49 \\
    GL & 11.4737 & 11.882 & 0.4083 & 3.59 \\
    PVDA & 13.5263 & 13.2584 & $-0.2679$ & $-2.01$ \\
    D66 & 13.9 & 13.9 & 0 & 0 \\
    \bottomrule
    \end{tabular}
\end{table}

As Table \ref{tab:title1} shows, the scores of the anchor texts (LPF and D66) are fully recovered, while the scores of the texts in between have changed. These changes range from $-2.01\%$ for the PvdA to $35.41\%$ for the SP, indicating that the words in the documents indicate a respectively lower score for the PvdA and a higher score for the SP then what is suggested by the exogenous reference scores. Nevertheless, the SP document, which shows the most significant difference, retains its position relative to the other parties as the CU-SGP score also increases. However, a reversal does take place between the CDA and GL. Based on the exogenous scores, the GL document is more positive about European integration than the CDA, while the MV transformation switches these positions. Besides the PvdA, all parties receive a higher score than exogenous assigned, ranging from a small 3.59\% voor GL to a 35.41\% for the SP. While \citeasnoun{Martin2008} do not give a criterion as to what the maximum amount of difference should be, we consider the differences between the exogenous scores and the scores given by the MV transformation to be sufficiently large to warrant closer inspection. We therefore extend our calculation and include all countries and dimensions, to rule out any possibility of these differences arising out of peculiarities of this specific example.\\

As the table below shows, the results of this analysis show a similar pattern. However, in some cases the positions of the parties are switched and large differences such as the  $35.41\%$ for the SP above are not uncommon. Therefore, if scholars choose to use the MV transformation in the future, we would strongly advise them to calculate these differences. Not only will this help them to assess the size of the trade-off, the MV calculated score for the reference documents will also be a more valid score to compare the transformed scores for the virgin texts against. Additionally, they can be used as a (partial) check on how well the exogenous assumed relative distances between the reference texts are shown in the actual word use \cite{Martin2008}. Especially with large differences this can warrant a closer inspection of the exogenous assigned score for the party and why it differences from the actual word use.

\begin{landscape}

\begin{longtable}{llrrrrrrrrrrrr}
\caption{Difference between exogenous and calculated scores, in percentages} \label{tab:title2} \\
    \multirow{2}{*}{Country} & \multirow{2}{*}{Party}  & \multicolumn{4}{c}{Benoit \& Laver} &    \multicolumn{4}{c}{CHES} &  \multicolumn{4}{c}{EMP} \\
\cmidrule(r){3-6}
\cmidrule(r){7-10}
\cmidrule(r){11-14}
     &  & LR & EU & EC & SO & LR & EU & EC & SO & LR & EU & EC & SO \\
    \midrule
		&&&&&&&&&&&&&\\
		\endfirsthead
      \multirow{2}{*}{Country} & \multirow{2}{*}{Party}  & \multicolumn{4}{c}{Benoit \& Laver} &    \multicolumn{4}{c}{CHES} &  \multicolumn{4}{c}{EMP} \\
\cmidrule(r){3-6}
\cmidrule(r){7-10}
\cmidrule(r){11-14}
     &   & LR & EU & EC & SO & LR & EU & EC & SO & LR & EU & EC & SO \\
    \midrule
		&&&&&&&&&&&&&\\
		\endhead
    \multirow{1}{*}{AT} & FP\"{O} & $-0.03$ & $-0.05$ & $8.66$ & 0.00 &    0.00 & 0.00 & 9.91 & 0.00 &    0.00 & 0.00 & 0.00 & 0.00 \\
     & GR\"{U}NEN & $-0.05$ & 0.02 & $-0.09$ & $-0.13$ &    0.00 & 1.92 & 0.00 & 0.00 &    0.00 & 2.35 & 0.00 & 0.00 \\
     & \"{O}VP & $-8.86$ & 0.81 & $-0.02$ & $-10.44$ &    $-9.29$ & 0.00 & 0.00 & $-11.41$ &    $-11.71$ & 0.00 & $-11.76$ & $-9.86$ \\
     & SP\"{O} & $-3.02$ & 0.95 & 2.22 & $-2.47$ &    $-2.27$ & 2.06 & 1.90 & $-3.78$ &    $-2.48$ & 17.74 & $-4.50$ & $-2.62$ \\
		&&&&&&&&&&&&&\\
    \multirow{1}{*}{BE (FR)} & CDH & $-18.39$ & $-0.03$ & $-12.28$ & 0.03 &    $-14.05$ & 0.00 & $-16.97$ & 0.00 &  0.00 & $-2.27$ & 0.00 & 0.02 \\
     & ECOLO & 0.00 & 2.05 & $-0.02$ & $-0.10$ &    0.00 & 0.00 & 0.00 & 0.00 &    $-20.15$ & 0.00 & 8.25 & 0.25 \\
     & MR & 0.00 & 0.04 & 0.00 & 20.38 &    0.00 & 1.50 & 0.00 & 17.21 &    $-33.68$ & 0.00 & $-38.26$ & $-0.04$ \\
     & PS & 25.81 & 3.77 & 22.60 & 43.32 &    10.02 & 1.04 & 29.10 & 20.10 &    0.00 & $-2.57$ & 0.00 & 0.00 \\
		&&&&&&&&&&&&&\\
    \multirow{1}{*}{BE (NL)} & CDV & $-5.25$ & $-0.03$ & $-4.89$ & $-8.46$ & $-3.52$ & 0.00 & $-4.64$ & $-7.03$ &  $-10.98$ & $-2.91$ & $-11.22$ & 7.21 \\
     & GROEN & 0.00 & 3.24 & $-0.04$ & 0.04 &    0.00 & 3.71 & 0.00 & 0.00 &    $-20.21$ & 0.00 & $-11.97$ & 8.14 \\
     & NVA & 6.93 & $-4.36$ & 2.61 & 10.42 &    7.84 & $-3.93$ & 5.46 & 9.66 &    $-9.95$ & $-7.13$ & $-16.83$ & 5.55 \\
     & SPA & 10.84 & 2.85 & 7.07 & 16.63 &    10.49 & 4.84 & 15.88 & 5.34 &    0.00 & $-0.17$ & 0.00 & 0.00 \\
     & VB & 0.02 & $-0.06$ & $-0.83$ & 0.00 &    0.00 & 0.00 & 0.00 & 0.00 &    $-3.59$ & 0.00 & $-10.79$ & 4.47 \\
     & VLD & $-0.91$ & 2.96 & 0.01 & $-7.30$ &    $-0.41$ & 4.26 & 0.71 & $-6.62$ &    0.00 & $-0.12$ & 0.00 & 0.00 \\
		&&&&&&&&&&&&&\\
    \multirow{1}{*}{CY} & AKEL & 0.00 & $-$ & 0.00 & 0.00 &    $-$ & $-$ & $-$ & $-$ &    9.42 & 0.70 & 0.00 & 4.44 \\
     & DIKO & $-4.04$ & $-$ & $-3.17$ & 0.76    & $-$ & $-$ & $-$ & $-$ &    0.00 & 1.31 & 0.00 & $-6.88$ \\
     & DISY & 0.00 & $-$ & 0.00 & 0.00 &    $-$ & $-$ & $-$ & $-$    & $-3.54$ & 0.00 & $-4.05$ & $-11.62$ \\
     & EDEK & $-23.13$ & $-$ & $-3.82$ & 1.77    & $-$ & $-$ & $-$ & $-$    & $-5.97$ & 2.39 & 0.40 & 0.00 \\
     & KOP & 2.19 & $-$ & 1.85 & $-0.40$    & $-$ & $-$ & $-$ & $-$ &    0.00 & 0.00 & $-7.24$ & 0.00 \\
		&&&&&&&&&&&&&\\
    \multirow{1}{*}{CZ} & CSSD & 1.83 & $-$ & 2.40 & $-0.02$    & 1.86 & 0.00 & 3.36 & $-0.36$    & 2.18 & 0.00 & 1.10 & 0.00 \\
     & KDU-CSL & $-0.21$ & $-$ & 0.48 & 0.01    & $-0.13$ & $-0.68$ & 0.00 & $-1.07$    & $-0.34$ & $-0.56$ & $-0.29$ & $-0.10$ \\
     & KSCM & $-0.04$ & $-$ & 0.10 & 0.21    & 0.00 & 0.00 & 0.00 & 0.00    & 0.00 & $-0.32$ & 0.00 & 0.00 \\
     & ODS & 0.00 & $-$ & $-0.03$ & 0.55    & 0.00 & 1.26 & 1.22 & 0.00    & 0.00 & 0.00 & 0.00 & $-$ \\
		&&&&&&&&&&&&&\\
    \multirow{1}{*}{DE} & B90/GR\"{U}NEN & $-15.73$ & $-0.01$ & $-1.63$ & $-0.07$ & $-15.61$ & 1.59 & $-2.06$ & 0.00 & $-11.31$ & 4.96 & $-9.09$ & $-5.99$ \\
     & CDU & $-1.50$ & 1.23 & 6.95 & 0.00 &    0.51 & 0.00 & 7.03 & 0.87 &    0.17 & $-3.66$ & 0.00 & 0.00 \\
     & CSU & 0.00 & $-0.03$ & 7.90 & 2.76 &    0.00 & 0.89 & 7.84 & 0.00 &    0.00 & 0.00 & 0.57 & 2.96 \\
     & FDP & $-6.09$ & $-1.31$ & 0.01 & 32.09 &    $-5.05$ & $-2.40$ & 0.00 & 18.33 &    $-6.37$ & $-0.78$ & $-3.74$ & $-3.19$ \\
     & PDS/DIELINKE & 0.09 & 1.06 & $-0.03$ & 18.18 &    0.00 & 0.00 & 0.00 & 5.02 &    0.00 & 0.00 & 0.00 & $-11.29$ \\
     & SPD & 0.10 & $-3.78$ & 6.65 & 15.74 &    0.28 & 0.28 & 6.81 & 6.89 &    2.71 & $-1.75$ & $-1.15$ & 0.00 \\
		&&&&&&&&&&&&&\\
    \multirow{1}{*}{DK} & A & $-9.23$ & 5.48 & $-2.51$ & $-4.47$ &    $-11.73$ & 4.15 & $-0.74$ & $-15.46$ &   $-11.13$ & 0.44 & $-5.83$ & $-0.78$ \\
     & B & $-10.90$ & 2.67 & $-3.71$ & 0.00 &    $-12.94$ & 4.67 & 0.87 & $-12.58$ &    $-14.37$ & 3.70 & $-12.51$ & $-0.59$ \\
     & C & $-6.78$ & 2.53 & 0.00 & 0.00 &    $-8.76$ & 3.75 & 6.00 & $-9.88$ &    $-9.45$ & $-1.53$ & $-10.43$ & 4.87 \\
     & F & $-0.07$ & 32.98 & 0.00 & 9.81 &    0.00 & 30.51 & 0.00 & 0.00 &    0.00 & 1.69 & 0.00 & 4.34 \\
     & O & $-0.02$ & $-0.07$ & $-4.85$ & 9.97 &    0.00 & 0.00 & 7.91 & 0.00 &    0.00 & 0.00 & 0.00 & 0.00 \\
     & V & $-13.90$ & $-0.03$ & $-6.39$ & $-7.63$ &    $-15.61$ & 0.00 & 0.00 & $-17.21$ &    $-17.72$ & 0.00 & $-17.37$ & 0.00 \\
		&&&&&&&&&&&&&\\
    \multirow{1}{*}{EE} & EKRP-EKD & 0.09 & $-$ & 1.47 & 0.00 &    $-$ & $-$ & $-$ & $-$ &    $-$ & $-$ & $-$ & $-$ \\
     & IL & 0.41 & $-$ & 3.65 & $-0.48$ &    $-$ & $-$ & $-$ & $-$ &    0.41 & 0.21 & 3.12 & 0.00 \\
     & K & $-0.03$ & $-$ & 0.73 & -1.23 &    $-$ & $-$ & $-$ & $-$ &    0.05 & 0.04 & 2.35 & $-6.11$ \\
     & RE & $-0.02$ & $-$ & 2.16 & -0.93 &    $-$ & $-$ & $-$ & $-$ &   0.00 & 0.00 & 2.43 & $-0.29$ \\
     & RESP & $-2.46$ & $-$ & $-0.02$ & $-0.06$ &    $-$ & $-$ & $-$ & $-$     $-$ & $-1.17$ & 0.00 & 0.00 \\
     & SDE & 1.35 & $-$ & 0.05 & 0.00 &    $-$ & $-$ & $-$ & $-$ &    0.00 & 0.00 & 0.00 & $-0.55$ \\
		&&&&&&&&&&&&&\\
    \multirow{1}{*}{ES} & IU & $-$ & 0.03 & $-0.01$ & 0.06 &    0.00 & 0.00 & 0.00 & 0.00 &    0.00 & 0.00 & 0.00 & 0.00 \\
     & PNV-EAJ & 3.13 & $-4.06$ & $-1.27$ & $-4.12$ &    $-3.12$ & $-2.88$ & $-2.58$ & $-3.32$ &    0.00 & $-0.22$ & 0.00 & $-1.15$ \\
     & PP & $-0.02$ & 0.01 & $-0.01$ & $-0.01$ &    0.00 & 0.01 & 0.00 & 0.00 &    6.25 & 0.64 & 7.86 & 0.00 \\
     & PSOE & $-0.03$ & 0.03 & $-0.33$ & $-6.68$ &    $-1.67$ & 0.00 & $-0.74$ & $-3.51$ &    $-0.37$ & 0.00 & 1.79 & $-1.33$ \\
		&&&&&&&&&&&&&\\
    \multirow{1}{*}{FI} & KD & 7.91 & $-25.89$ & 2.81 & $-0.02$ &    7.34 & 0.00 & 6.12 & 0.00 &    7.68 & 0.00 & 1.30 & 0.00 \\
     & KESK & 3.60 & $-0.02$ & 8.04 & $-8.25$ &    1.56 & 8.38 & 5.98 & $-7.82$ &    3.54 & 0.00 & 5.68 & 0.00 \\
     & KOK & $-0.03$ & $-2.13$ & $-0.01$ & $-2.94$ &    0.00 & 0.00 & 0.00 & $-1.52$ &    0.00 & 3.00 & 0.00 & 2.07 \\
     & RKP-SFP & 3.43 & $-0.20$ & 4.49 & 10.28 &    2.83 & 1.64 & 3.81 & 4.14 &    3.81 & 0.54 & $-$ & $-$ \\
     & SDP & 8.98 & 0.01 & 9.30 & 9.30 &    9.32 & 1.66 & 9.08 & 8.69 &    7.91 & 0.75 & 5.06 & $-3.56$ \\
     & VAS & 0.10 & $-5.38$ & 0.08 & 8.18 &    0.00 & 4.99 & 0.00 & 3.39 &    0.00 & 2.31 & 0.00 & $-7.37$ \\
     & VIHR & $-6.83$ & 1.12 & $-5.02$ & 0.01 &    $-6.34$ & 2.43 & $-6.40$ & 0.00 &    2.80 & 1.95 & $-0.54$ & $-7.99$ \\
		&&&&&&&&&&&&&\\
    \multirow{1}{*}{FR} & FN & $-$ & 0.00 & 0.02 & $-0.02$ &    0.00 & 0.00 & 0.00 & 0.00 &    0.00 & 0.00 & 9.08 & 0.00 \\
     & PCF & $-$ & $-3.48$ & 0.00 & $-28.59$ &    0.00 & 16.30 & 0.00 & $-22.97$ &    0.00 & 0.21 & 0.00 & $-$ \\
     & PRG & $-$ & $-$ & $-$ & $-$ &    $-12.55$ & 8.73 & $-11.64$ & $-27.39$ &    $-12.48$ & 0.00 & 1.96 & 0.00 \\
     & PS & $-$ & $-10.03$ & 18.34 & 0.08 &    12.84 & 3.62 & 2.91 & 0.00 &    11.45 & $-6.74$ & 13.64 & $-$ \\
     & UDF & $-$ & $-0.02$ & $-6.53$ & $-11.54$ &    $-0.98$ & 0.00 & $-7.73$ & $-9.52$ &    $-1.87$ & $-8.68$ & 0.00 & $-$ \\
     & UMP & $-$ & $-19.23$ & $-3.83$ & $-9.68$ &    $-4.75$ & 11.63 & $-4.66$ & $-10.73$ &    $-2.81$ & $-3.27$ & $-$ & $-$ \\
		&&&&&&&&&&&&&\\
    \multirow{1}{*}{GR} & KKE & $-0.08$ & 0.14 & $-0.03$ & $-4.58$ &    3.20 & 0.00 & 0.00 & $-2.87$ &    0.00 & 0.00 & $-$ & 0.00 \\
     & ND & 0.02 & 2.10 & 0.00 & 0.00 &    0.00 & 2.57 & 0.00 & 0.00 &    0.00 & 0.00 & 0.00 & $-10.60$ \\
     & PASOK & $-0.93$ & $-0.03$ & $-2.42$ & -5.12 &    $-0.45$ & 0.00 & $-2.81$ & $-6.94$ &    $-2.65$ & $-4.06$ & $-5.83$ & $-$ \\
     & SYRIZA & 7.18 & $-3.40$ & 4.96 & 0.05 &    0.00 & $-1.73$ & 0.45 & 0.00 &    2.25 & $-2.38$ & 0.00 & 0.00 \\
		&&&&&&&&&&&&&\\
    \multirow{1}{*}{HU} & FIDESZ-MPP & $-0.03$ & $-$ & 0.03 & 0.00 &    0.00 & 0.00 & 0.00 & 0.00 &    0.00 & $-2.10$ & $-0.54$ & $-13.11$ \\
     & MDF & 5.53 & $-$ & $-6.12$ & 9.96 &    6.81 & $-0.10$ & $-3.76$ & 8.73 &    $-1.93$ & $-1.14$ & 0.00 & $-10.30$ \\
     & MSZP & 0.05 & $-$ & $-9.25$ & $-15.63$ &    0.00 & 9.55 & $-6.12$ & $-29.95$ &    5.03 & 0.00 & $-3.29$ & 0.00 \\
     & SZDSZ & 23.11 & $-$ & 0.00 & 0.00 &    22.82 & 0.00 & 0.00 & 0.00 &    0.00 & 0.00 & 0.00 & 0.00 \\
		&&&&&&&&&&&&&\\
    \multirow{1}{*}{IE} & FF & 0.02 & 9.52 & 0.03 & 0.01 &    0.27 & 1.72 & 0.38 & 0.00 &    0.32 & 0.00 & 0.83 & 0.00 \\
     & FG & 0.92 & 0.02 & 0.45 & 5.34 &    0.00 & 0.00 & 0.00 & 4.37 &    0.00 & $-1.00$ & 0.00 & 0.54 \\
     & GREENS & $-0.07$ & 0.03 & $-15.43$ & 0.00 & 0.00 & $-23.05$ & $-21.31$ & 0.00 & $-21.09$ & $-13.63$ & $-16.75$ & $-11.49$ \\
     & LAB & 8.95 & 1.36 & $-0.30$ & 16.23 &    9.07 & $-1.63$ & $-1.58$ & 13.08 &    $-1.51$ & $-1.13$ & $-2.36$ & 0.39 \\
     & SF & 11.94 & 32.25 & 0.00 & 2.88 &    12.77 & 0.00 & 0.00 & 0.17 &    0.00 & 0.00 & 0.00 & 0.00 \\
		&&&&&&&&&&&&&\\
    \multirow{1}{*}{IT} & AN & 0.03 & 0.83 & $-6.32$ & 0.00 &    0.00 & 6.73 & $-2.63$ & 0.00 &    7.12 & $-24.13$ & $-7.46$ & 6.21 \\
     & DS/ULIVO & $-12.74$ & 0.03 & $-6.11$ & $-5.73$ &   $-7.78$ & 0.00 & $-9.92$ & 0.00 &  $-6.07$ & 0.00 & $-5.41$ & 9.22 \\
     & FI & $-8.87$ & 5.67 & 0.00 & $-10.91$ &    $-8.16$ & 7.15 & 0.00 & $-9.65$ &    0.00 & 12.03 & 0.00 & 0.00 \\
     & LN & $-11.62$ & $-0.08$ & $-3.62$ & $-12.46$    & $-10.36$ & 0.00 & $-4.08$ & $-11.18$  & $-4.40$ & $-12.75$ & $-4.13$ & 0.00 \\
     & PRC & $-0.09$ & 7.39 & $-0.11$ & $-0.07$    & 0.00 & 9.49 & 0.00 & $-2.05$    & 0.00 & 0.00 & 0.00 & 0.00 \\
		&&&&&&&&&&&&&\\
    \multirow{1}{*}{LT} & LICS & $-0.02$ & $-$ & 0.01 & 0.02    & 3.38 & 0.68 & 0.03 & 0.00    & 1.41 & 1.23 & 0.87 & 0.00 \\
     & LKD & 0.93 & $-$ & 1.24 & 0.01    & 2.54 & 0.00 & $-6.94$ & 0.00    & 1.52 & 1.13 & 0.00 & 0.00 \\
     & LSDP & 0.06 & $-$ & $-0.06$ & $-2.99$    & 0.00 & 0.52 & $-8.86$ & $-0.60$    & 0.00 & 0.00 & 0.08 & 0.30 \\
     & NS & 7.78 & $-$ & 6.37 & 3.65    & 15.84 & 1.44 & 0.00 & $-0.95$    & 6.08 & 0.00 & 0.13 & $-0.07$ \\
     & TS & 0.11 & $-$ & 1.47 & $-3.46$    & 0.00 & 0.00 & 0.00 & $-1.38$    & 0.00 & 1.91 & 0.00 & $-1.41$ \\
		&&&&&&&&&&&&&\\
    \multirow{1}{*}{LV} & JL & $-2.77$ & $-$ & $-2.22$ & $-0.04$    & $-1.89$ & 0.00 & $-3.94$ & 1.62    & $-1.49$ & 0.00 & $-0.59$ & 0.00 \\
    & LC & $-0.66$ & $-$ & $-0.42$ & $-0.03$    & $-0.25$ & 1.00 & $-2.56$ & 0.00    & $-1.17$ & 0.38 & 0.00 & $-$ \\
     & PCTVL & $-0.15$ & $-$ & 0.00 & $-2.09$    & 0.00 & 0.00 & 0.00 & 0.58    & 0.00 & 0.97 & 0.00 & $-$ \\
     & TB/LNNK & $-0.22$ & $-$ & 0.14 & $-0.61$    & 0.00 & 1.20 & $-0.95$ & 0.76    & 0.00 & 0.25 & $-$ & 1.27 \\
     & TP & $0.03$ & $-$ & $-0.03$ & $-0.99$ &    0.48 & 2.16 & 0.00 & 0.00 &    1.03 & 1.79 & $-0.23$ & $-$ \\
     & ZZS & 2.13 & $-$ & 0.96 & $-0.27$ &    4.76 & 1.24 & 1.75 & 1.64 &    4.97 & 0.00 & $-$ & $-$ \\
		&&&&&&&&&&&&&\\
    \multirow{1}{*}{NL} & CDA & $-7.82$ & 7.49 & 5.59 & 15.46 &    $-8.25$ & 5.30 & $-8.24$ & 10.55    & $-6.77$ & $-4.41$ & 0.00 & 3.17 \\
     & CU-SGP & $-17.19$ & 19.41 & $-4.36$ & $-0.02$ & $-17.48$ & 12.69 & $-18.14$ & 0.00  & $-15.03$ & 18.46 & $-10.61$ & 0.21 \\
     & D66 & $-11.50$ & 0.00 & 0.71 & 22.23    & $-11.01$ & 0.00 & $-13.59$ & 0.00    & $-13.05$ & $-4.48$ & $-9.22$ & 12.95 \\
     & GL & $-18.64$ & 3.59 & $-12.79$ & 0.05  & $-16.94$ & 3.10 & $-26.68$ & $-32.66$   & $-14.52$ & $-6.20$ & $-15.43$ & 10.33 \\
     & LPF & $-0.01$ & 0 & 14.93 & 17.57    & $-0.06$ & 0.00 & 0.00 & 11.79    & 0.00 & 0.00 & 3.94 & $-16.60$ \\
     & PVDA & $-10.31$ & $-2.01$ & 1.54 & 8.10    & $-9.99$ & $-0.60$ & $-11.90$ & $-2.76$    & $-6.74$ & $-1.04$ & 2.35 & $-5.93$ \\
     & SP & -0.15 & 35.41 & $-0.03$ & 2.43 &    0.00 & 27.02 & 0.00 & $-7.21$    & 0.00 & 31.51 & 0.00 & 0.00 \\
     & VVD & $-14.00$ & 15.88 & 0.02 & $-15.41$ &  $-14.29$ & 11.70 & $-13.98$ & $-1.57$    & $-13.09$ & 0.00 & $-7.36$ & 0.00 \\
		&&&&&&&&&&&&&\\
    \multirow{1}{*}{PL} & PIS & 0.00 & 0.00 & 0.19 & $-0.03$ &    0.00 & $-0.14$ & 3.23 & 0.00 &    0.00 & 0.00 & 2.48 & 0.00 \\
     & PO & $-1.93$ & $-1.93$ & 0.00 & $-1.98$ &    $-1.42$ & $-3.50$ & 0.00 & $-2.60$ &    $-2.43$ & $-3.53$ & 0.00 & 0.70 \\
     & PSL & $-4.09$ & $-4.09$ & $-10.47$ & 0.80 &    $-3.58$ & 0.00 & 0.00 & 1.20 &    $-8.49$ & $-2.05$ & 0.00 & $-$ \\
     & SLD-UP & 0.04 & 0.04 & 0.00 & 0.03 &    0.00 & $-3.87$ & 5.96 & 0.00 &    0.00 & $-3.43$ & 4.63 & 10.66 \\
     & UW & $-0.73$ & $-0.73$ & 7.20 & $-17.42$ &    $-1.23$ & 0.00 & 8.33 & $-31.82$    & $-2.30$ & 0.00 & 6.95 & 0.00 \\
		&&&&&&&&&&&&&\\
    \multirow{1}{*}{PT} & BE & 0.08 & 0.00 & $-13.57$ & 0.11 &    $-$ & $-$ & $-$ & $-$ &    $-18.64$ & 0.86 & $-10.04$ & 0.00 \\
     & CDU & 17.82 & 8.90 & $-0.11$ & 18.61 &    0.00 & 0.00 & 0.00 & 0.00 &    0.00 & 0.00 & 0.00 & 0.00 \\
     & PS & 4.52 & 0.00 & 0.96 & 7.64 &    0.09 & 0.00 & 1.40 & 5.35 &    1.20 & 0.00 & 2.84 & 18.43 \\
     & PSD & $-0.02$ & 6.13 & 0.00 & 0.00 &    0.00 & 4.84 & 0.00 & 0.00 &    0.00 & 6.75 & 0.00 & 2.88 \\
		&&&&&&&&&&&&&\\
    \multirow{1}{*}{SE} & C & 6.32 & 13.24 & 7.23 & 11.58 &    6.09 & $-0.98$ & 7.51 & 2.27 &    8.68 & 14.05 & 4.72 & $-4.78$ \\
     & KD & $-2.82$ & $-4.21$ & $-0.16$ & $-0.02$ &    $-2.74$ & $-6.24$ & $-0.66$ & 0.00    & 0.00 & $-4.42$ & $-1.29$ & $-3.57$ \\
     & M & 0.02 & 0.00 & $-0.02$ & 8.41 &    0.00 & 0.00 & 0.00 & 1.42 &    3.71 & 0.00 & 0.00 & 0.00 \\
     & MP & 12.67 & 38.07 & 12.31 & 30.48 &    13.11 & 0.00 & 10.34 & 0.00    & 23.30 & 35.47 & 8.52 & 1.22 \\
     & S & 3.31 & 9.03 & 4.39 & 5.13 &    2.64 & 4.90 & 4.25 & $-1.74$ &    3.84 & 8.69 & 2.76 & $-6.15$ \\
     & V & $-0.09$ & $-0.02$ & $-0.04$ & 0.05 &    0.00 & $-33.98$ & 0.00 & $-20.93$    & 0.00 & 0.00 & 0.00 & 0.00 \\
		&&&&&&&&&&&&&\\
    \multirow{1}{*}{SI} & NSI & $-0.02$ & $-$ & 0.00 & $-0.01$ &   0.00 & 0.00 & 1.20 & 0.00 &    0.00 & 0.00 & 0.00 & $-0.61$ \\
     & SDS & $-3.71$ & $-$ & $-1.75$ & $-3.53$ &    $-2.32$ & 0.00 & 0.64 & $-2.35$    & $-3.42$ & $-0.30$ & $-2.79$ & 0.00 \\
     & SLS & $-5.09$ & $-$ & $-1.10$ & $-5.61$ &    $-4.10$ & 0.00 & 0.00 & $-4.68$    & $-6.45$ & 0.00 & $-4.23$ & 0.00 \\
     & ZLSD & 0.07 & $-$ & 0.00 & $-0.08$ &    0.00 & -1.48 & 0.00 & 0.00    & 0.00 & $-0.72$ & 0.00 & 0.85 \\
		&&&&&&&&&&&&&\\
    \multirow{1}{*}{SK} & KDH & 0.01 & $-$ & $-0.03$ & $-0.02$ &    0.00 & $-0.67$ & 3.32 & 0.00    & 0.00 & 0.00 & 0.00 & $-0.99$ \\
     & LSHZDS & $-5.66$ & $-$ & $-3.90$ & $-9.14$ &    $-6.99$ & $-1.51$ & $-6.83$ & 0.32    & $-4.07$ & 2.93 & $-6.30$ & 0.00 \\
     & SDKUDS & $-3.25$ & $-$ & $-2.14$ & $-4.29$ &    $-3.31$ & 0.00 & 0.00 & 0.00    & $-2.88$ & 0.35 & $-3.18$ & $-1.12$ \\
     & SMER & $-0.03$ & $-$ & 0.02 & 0.02 &    0.00 & 0.00 & 0.00 & $-1.57$ &    0.00 & 0.00 & 0.00 & 0.00 \\
     & SMK-MKP & 1.84 & $-$ & 4.51 & 10.11 &    9.08 & 11.60 & 1.29 & $-7.02$    & 3.25 & 12.40 & $-5.16$ & 0.83 \\
		&&&&&&&&&&&&&\\
   \multirow{1}{*}{UK} & CON & $-0.02$ & 0.10 & 0.01 & 0.02 &    0.00 & 0.00 & 0.00 & 0.00    & 0.00 & 0.00 & 0.00 & 0.00 \\
     & LAB & 14.18 & 0.78 & 20.02 & $-5.84$ &    8.94 & 2.07 & 9.87 & $-1.60$    & 21.02 & $-12.60$ & 27.25 & 7.66 \\
     & LD & 30.74 & 0.03 & 40.65 & 0.07 &    18.73 & 0.00 & 17.56 & 0.00    & 22.54 & $-14.73$ & 18.21 & 20.66 \\
     & PC & $-0.02$ & 23.04 & 0.04 & $-11.52$ &    $-17.42$ & 14.65 & $-20.92$ & $-17.86$    & 0.00 & 0.00 & 0.00 & 9.79 \\
     & SNP & 17.86 & 12.72 & 18.38 & $-3.85$ &    0.00 & 6.87 & 0.00 & $-6.10$    & 31.37 & $-8.42$ & 37.88 & 0.00 \\
		&&&&&&&&&&&&&\\
    \bottomrule
\end{longtable}
\end{landscape}

\newpage

\begin{landscape}

\section*{Appendix G: Concordance Correlations}

The four tables below show the concordance correlations between the wordscores estimates for the virgin texts and the expert scores, for all the different combinations of exogenous reference score, type of benchmark, transformation, and rescaling.\\

\centering
    \begin{longtable}[c]{lllcrcrrcr}
    \caption{Left-Right Dimension}\\
    \toprule
    Reference & Benchmark & Transformation & Rescale & rho\_c & \# of Observations   & CI\_low & CI\_high & Pearson's \textit{r}     & C\_b \\
    \midrule
    BL    & CHES  & LBG   & pc    & 0.624 & 133   & 0.527 & 0.704 & 0.687 & 0.907 \\
    BL    & EMP   & LBG   & pc    & 0.653 & 151   & 0.561 & 0.73  & 0.69  & 0.946 \\
    BL    & EUP   & LBG   & pc    & 0.497 & 147   & 0.395 & 0.587 & 0.595 & 0.836 \\
    BL    & CHES  & LBG   & wd    & 0.638 & 138   & 0.529 & 0.726 & 0.644 & 0.99 \\
    BL    & EMP   & LBG   & wd    & 0.587 & 158   & 0.48  & 0.676 & 0.617 & 0.951 \\
    BL    & EUP   & LBG   & wd    & 0.438 & 154   & 0.316 & 0.545 & 0.49  & 0.893 \\
    BL    & CHES  & MV    & pc    & 0.624 & 133   & 0.527 & 0.704 & 0.687 & 0.907 \\
    BL    & EMP   & MV    & pc    & 0.653 & 151   & 0.561 & 0.73  & 0.69  & 0.946 \\
    BL    & EUP   & MV    & pc    & 0.507 & 147   & 0.407 & 0.596 & 0.607 & 0.836 \\
    BL    & CHES  & MV    & wd    & 0.267 & 138   & 0.123 & 0.401 & 0.296 & 0.904 \\
    BL    & EMP   & MV    & wd    & 0.213 & 158   & 0.089 & 0.329 & 0.263 & 0.809 \\
    BL    & EUP   & MV    & wd    & 0.079 & 154   & -0.052 & 0.207 & 0.095 & 0.825 \\
    CHES  & CHES  & LBG   & pc    & 0.597 & 134   & 0.494 & 0.683 & 0.653 & 0.915 \\
    CHES  & EMP   & LBG   & pc    & 0.673 & 147   & 0.583 & 0.747 & 0.71  & 0.948 \\
    CHES  & EUP   & LBG   & pc    & 0.464 & 142   & 0.351 & 0.564 & 0.546 & 0.851 \\
    CHES  & CHES  & LBG   & wd    & 0.642 & 138   & 0.533 & 0.731 & 0.643 & 1 \\
    CHES  & EMP   & LBG   & wd    & 0.565 & 158   & 0.454 & 0.658 & 0.595 & 0.949 \\
    CHES  & EUP   & LBG   & wd    & 0.445 & 154   & 0.323 & 0.552 & 0.497 & 0.896 \\
    CHES  & CHES  & MV    & pc    & 0.597 & 134   & 0.494 & 0.683 & 0.653 & 0.915 \\
    CHES  & EMP   & MV    & pc    & 0.673 & 147   & 0.583 & 0.747 & 0.71  & 0.948 \\
    CHES  & EUP   & MV    & pc    & 0.464 & 142   & 0.351 & 0.564 & 0.546 & 0.851 \\
    CHES  & CHES  & MV    & wd    & 0.314 & 138   & 0.186 & 0.431 & 0.377 & 0.832 \\
    CHES  & EMP   & MV    & wd    & 0.215 & 158   & 0.09  & 0.333 & 0.262 & 0.821 \\
    CHES  & EUP   & MV    & wd    & 0.176 & 154   & 0.044 & 0.303 & 0.209 & 0.844 \\
    EMP   & CHES  & LBG   & pc    & 0.485 & 138   & 0.365 & 0.59  & 0.535 & 0.906 \\
    EMP   & EMP   & LBG   & pc    & 0.590  & 158   & 0.487 & 0.677 & 0.62  & 0.951 \\
    EMP   & EUP   & LBG   & pc    & 0.428 & 154   & 0.317 & 0.527 & 0.508 & 0.841 \\
    EMP   & CHES  & LBG   & wd    & 0.235 & 138   & 0.169 & 0.299 & 0.607 & 0.387 \\
    EMP   & EMP   & LBG   & wd    & 0.320  & 158   & 0.25  & 0.387 & 0.667 & 0.48 \\
    EMP   & EUP   & LBG   & wd    & 0.298 & 154   & 0.221 & 0.372 & 0.591 & 0.505 \\
    EMP   & CHES  & MV    & pc    & 0.485 & 138   & 0.365 & 0.59  & 0.535 & 0.906 \\
    EMP   & EMP   & MV    & pc    & 0.590  & 158   & 0.487 & 0.677 & 0.62  & 0.951 \\
    EMP   & EUP   & MV    & pc    & 0.428 & 154   & 0.317 & 0.527 & 0.508 & 0.841 \\
    EMP   & CHES  & MV    & wd    & 0.070  & 138   & 0.04  & 0.099 & 0.409 & 0.17 \\
    EMP   & EMP   & MV    & wd    & 0.093 & 158   & 0.06  & 0.126 & 0.446 & 0.208 \\
    EMP   & EUP   & MV    & wd    & 0.083 & 154   & 0.046 & 0.12  & 0.361 & 0.229 \\
    \bottomrule
         \multicolumn{9}{l}{ * wd = whole dimension, pc = per country}
\end{longtable}
\end{landscape}

\begin{landscape}
\centering
    \begin{longtable}[c]{lllcrcrrcr}
  \caption{EU Integration Dimension}\\
    \toprule
    Reference & Benchmark & Transformation & Rescale & rho\_c & \# of Observations   & CI\_low & CI\_high & Pearson's \textit{r}     & C\_b \\
    \midrule
    BL    & CHES  & LBG   & pc    & 0.518 & 98    & 0.365 & 0.644 & 0.539 & 0.961 \\
    BL    & EMP   & LBG   & pc    & 0.452 & 107   & 0.289 & 0.588 & 0.458 & 0.987 \\
    BL    & EUP   & LBG   & pc    & 0.489 & 104   & 0.332 & 0.619 & 0.499 & 0.979 \\
    BL    & CHES  & LBG   & wd    & 0.452 & 138   & 0.309 & 0.575 & 0.453 & 0.998 \\
    BL    & EMP   & LBG   & wd    & 0.466 & 159   & 0.335 & 0.579 & 0.467 & 0.999 \\
    BL    & EUP   & LBG   & wd    & 0.403 & 154   & 0.263 & 0.526 & 0.406 & 0.993 \\
    BL    & CHES  & MV    & pc    & 0.518 & 98    & 0.365 & 0.644 & 0.539 & 0.961 \\
    BL    & EMP   & MV    & pc    & 0.452 & 107   & 0.289 & 0.588 & 0.458 & 0.987 \\
    BL    & EUP   & MV    & pc    & 0.489 & 104   & 0.332 & 0.619 & 0.499 & 0.979 \\
    BL    & CHES  & MV    & wd    & 0.248 & 138   & 0.087 & 0.396 & 0.251 & 0.987 \\
    BL    & EMP   & MV    & wd    & 0.312 & 159   & 0.170  & 0.44  & 0.323 & 0.966 \\
    BL    & EUP   & MV    & wd    & 0.216 & 154   & 0.063 & 0.359 & 0.22  & 0.983 \\
    CHES  & CHES  & LBG   & pc    & 0.430  & 134   & 0.294 & 0.55  & 0.462 & 0.931 \\
    CHES  & EMP   & LBG   & pc    & 0.345 & 148   & 0.202 & 0.474 & 0.361 & 0.957 \\
    CHES  & EUP   & LBG   & pc    & 0.406 & 142   & 0.269 & 0.527 & 0.431 & 0.944 \\
    CHES  & CHES  & LBG   & wd    & 0.508 & 138   & 0.389 & 0.611 & 0.566 & 0.899 \\
    CHES  & EMP   & LBG   & wd    & 0.405 & 159   & 0.279 & 0.516 & 0.447 & 0.906 \\
    CHES  & EUP   & LBG   & wd    & 0.334 & 154   & 0.211 & 0.446 & 0.4   & 0.834 \\
    CHES  & CHES  & MV    & pc    & 0.430  & 134   & 0.294 & 0.55  & 0.462 & 0.931 \\
    CHES  & EMP   & MV    & pc    & 0.345 & 148   & 0.202 & 0.474 & 0.361 & 0.957 \\
    CHES  & EUP   & MV    & pc    & 0.406 & 142   & 0.269 & 0.527 & 0.431 & 0.944 \\
    CHES  & CHES  & MV    & wd    & 0.361 & 138   & 0.231 & 0.478 & 0.421 & 0.858 \\
    CHES  & EMP   & MV    & wd    & 0.256 & 159   & 0.123 & 0.381 & 0.289 & 0.886 \\
    CHES  & EUP   & MV    & wd    & 0.140  & 154   & 0.002 & 0.273 & 0.16  & 0.873 \\
    EMP   & CHES  & LBG   & pc    & 0.370  & 138   & 0.237 & 0.489 & 0.423 & 0.875 \\
    EMP   & EMP   & LBG   & pc    & 0.296 & 159   & 0.165 & 0.416 & 0.337 & 0.878 \\
    EMP   & EUP   & LBG   & pc    & 0.401 & 154   & 0.278 & 0.511 & 0.455 & 0.88 \\
    EMP   & CHES  & LBG   & wd    & 0.202 & 138   & 0.141 & 0.261 & 0.577 & 0.35 \\
    EMP   & EMP   & LBG   & wd    & 0.180  & 159   & 0.125 & 0.235 & 0.517 & 0.348 \\
    EMP   & EUP   & LBG   & wd    & 0.223 & 154   & 0.165 & 0.279 & 0.624 & 0.357 \\
    EMP   & CHES  & MV    & pc    & 0.370  & 138   & 0.237 & 0.489 & 0.423 & 0.875 \\
    EMP   & EMP   & MV    & pc    & 0.296 & 159   & 0.165 & 0.416 & 0.337 & 0.878 \\
    EMP   & EUP   & MV    & pc    & 0.401 & 154   & 0.278 & 0.511 & 0.455 & 0.88 \\
    EMP   & CHES  & MV    & wd    & 0.082 & 138   & 0.047 & 0.117 & 0.41  & 0.201 \\
    EMP   & EMP   & MV    & wd    & 0.075 & 159   & 0.043 & 0.107 & 0.374 & 0.199 \\
    EMP   & EUP   & MV    & wd    & 0.093 & 154   & 0.058 & 0.126 & 0.45  & 0.206 \\
    \bottomrule
      \multicolumn{9}{l}{ * wd = whole dimension, pc = per country}
\end{longtable}
\end{landscape}

\begin{landscape}
\centering
    \begin{longtable}[c]{lllcrcrrcr}
  \caption{Economic Dimension}\\
    \toprule
    Reference & Benchmark & Transformation & Rescale & rho\_c & \# of Observations   & CI\_low & CI\_high & Pearson's \textit{r}     & C\_b \\
    \midrule
    BL    & CHES  & LBG   & pc    & 0.449 & 138   & 0.330  & 0.553 & 0.52  & 0.863 \\
    BL    & EMP   & LBG   & pc    & 0.424 & 158   & 0.303 & 0.531 & 0.472 & 0.898 \\
    BL    & EUP   & LBG   & pc    & 0.433 & 154   & 0.322 & 0.532 & 0.526 & 0.823 \\
    BL    & CHES  & LBG   & wd    & 0.576 & 138   & 0.453 & 0.677 & 0.579 & 0.995 \\
    BL    & EMP   & LBG   & wd    & 0.481 & 158   & 0.356 & 0.589 & 0.498 & 0.966 \\
    BL    & EUP   & LBG   & wd    & 0.527 & 154   & 0.415 & 0.623 & 0.585 & 0.901 \\
    BL    & CHES  & MV    & pc    & 0.449 & 138   & 0.330  & 0.553 & 0.52  & 0.863 \\
    BL    & EMP   & MV    & pc    & 0.424 & 158   & 0.303 & 0.531 & 0.472 & 0.898 \\
    BL    & EUP   & MV    & pc    & 0.433 & 154   & 0.322 & 0.532 & 0.526 & 0.823 \\
    BL    & CHES  & MV    & wd    & 0.242 & 138   & 0.140  & 0.339 & 0.367 & 0.659 \\
    BL    & EMP   & MV    & wd    & 0.209 & 158   & 0.123 & 0.292 & 0.359 & 0.583 \\
    BL    & EUP   & MV    & wd    & 0.192 & 154   & 0.109 & 0.272 & 0.355 & 0.541 \\
    CHES  & CHES  & LBG   & pc    & 0.463 & 134   & 0.345 & 0.566 & 0.542 & 0.854 \\
    CHES  & EMP   & LBG   & pc    & 0.431 & 147   & 0.308 & 0.539 & 0.49  & 0.878 \\
    CHES  & EUP   & LBG   & pc    & 0.411 & 142   & 0.295 & 0.516 & 0.51  & 0.807 \\
    CHES  & CHES  & LBG   & wd    & 0.553 & 138   & 0.428 & 0.658 & 0.563 & 0.983 \\
    CHES  & EMP   & LBG   & wd    & 0.401 & 158   & 0.273 & 0.515 & 0.436 & 0.919 \\
    CHES  & EUP   & LBG   & wd    & 0.397 & 154   & 0.279 & 0.503 & 0.479 & 0.828 \\
    CHES  & CHES  & MV    & pc    & 0.467 & 128   & 0.347 & 0.572 & 0.545 & 0.857 \\
    CHES  & EMP   & MV    & pc    & 0.438 & 142   & 0.313 & 0.547 & 0.494 & 0.886 \\
    CHES  & EUP   & MV    & pc    & 0.451 & 136   & 0.333 & 0.554 & 0.545 & 0.827 \\
    CHES  & CHES  & MV    & wd    & 0.114 & 138   & -0.04 & 0.263 & 0.124 & 0.919 \\
    CHES  & EMP   & MV    & wd    & 0.092 & 158   & -0.038 & 0.218 & 0.111 & 0.828 \\
    CHES  & EUP   & MV    & wd    & 0.049 & 154   & -0.068 & 0.164 & 0.066 & 0.737 \\
    EMP   & CHES  & LBG   & pc    & 0.348 & 138   & 0.216 & 0.467 & 0.4   & 0.87 \\
    EMP   & EMP   & LBG   & pc    & 0.427 & 158   & 0.307 & 0.534 & 0.477 & 0.896 \\
    EMP   & EUP   & LBG   & pc    & 0.383 & 154   & 0.268 & 0.487 & 0.469 & 0.815 \\
    EMP   & CHES  & LBG   & wd    & 0.202 & 138   & 0.123 & 0.279 & 0.429 & 0.472 \\
    EMP   & EMP   & LBG   & wd    & 0.271 & 158   & 0.190  & 0.348 & 0.501 & 0.541 \\
    EMP   & EUP   & LBG   & wd    & 0.320  & 154   & 0.226 & 0.408 & 0.495 & 0.647 \\
    EMP   & CHES  & MV    & pc    & 0.348 & 138   & 0.216 & 0.467 & 0.4   & 0.87 \\
    EMP   & EMP   & MV    & pc    & 0.427 & 158   & 0.307 & 0.534 & 0.477 & 0.896 \\
    EMP   & EUP   & MV    & pc    & 0.383 & 154   & 0.268 & 0.487 & 0.469 & 0.815 \\
    EMP   & CHES  & MV    & wd    & 0.038 & 138   & -0.007 & 0.083 & 0.144 & 0.266 \\
    EMP   & EMP   & MV    & wd    & 0.106 & 158   & 0.056 & 0.156 & 0.333 & 0.32 \\
    EMP   & EUP   & MV    & wd    & 0.093 & 154   & 0.033 & 0.154 & 0.242 & 0.386 \\
    \bottomrule
    \multicolumn{9}{l}{ * wd = whole dimension, pc = per country}
\end{longtable}
\end{landscape}

\begin{landscape}
\centering
    \begin{longtable}[c]{lllcrcrrcr}
  \caption{Social Dimension}\\
    \toprule
    Reference & Benchmark & Transformation & Rescale & rho\_c & \# of Observations   & CI\_low & CI\_high & Pearson's \textit{r}     & C\_b \\
    \midrule
    BL    & CHES  & LBG   & pc    & 0.569 & 138   & 0.459 & 0.662 & 0.617 & 0.923 \\
    BL    & EMP   & LBG   & pc    & 0.217 & 151   & 0.077 & 0.348 & 0.243 & 0.892 \\
    BL    & EUP   & LBG   & pc    & 0.475 & 154   & 0.359 & 0.576 & 0.522 & 0.909 \\
    BL    & CHES  & LBG   & wd    & 0.609 & 138   & 0.495 & 0.703 & 0.62  & 0.982 \\
    BL    & EMP   & LBG   & wd    & 0.243 & 151   & 0.096 & 0.381 & 0.257 & 0.948 \\
    BL    & EUP   & LBG   & wd    & 0.54  & 154   & 0.419 & 0.642 & 0.544 & 0.993 \\
    BL    & CHES  & MV    & pc    & 0.569 & 138   & 0.459 & 0.662 & 0.617 & 0.923 \\
    BL    & EMP   & MV    & pc    & 0.217 & 151   & 0.077 & 0.348 & 0.243 & 0.892 \\
    BL    & EUP   & MV    & pc    & 0.475 & 154   & 0.359 & 0.576 & 0.522 & 0.909 \\
    BL    & CHES  & MV    & wd    & 0.279 & 138   & 0.173 & 0.379 & 0.403 & 0.694 \\
    BL    & EMP   & MV    & wd    & 0.052 & 151   & -0.057 & 0.16  & 0.076 & 0.681 \\
    BL    & EUP   & MV    & wd    & 0.226 & 154   & 0.113 & 0.334 & 0.302 & 0.751 \\
    CHES  & CHES  & LBG   & pc    & 0.552 & 134   & 0.435 & 0.651 & 0.588 & 0.939 \\
    CHES  & EMP   & LBG   & pc    & 0.161 & 141   & 0.014 & 0.301 & 0.181 & 0.891 \\
    CHES  & EUP   & LBG   & pc    & 0.445 & 142   & 0.318 & 0.556 & 0.483 & 0.921 \\
    CHES  & CHES  & LBG   & wd    & 0.585 & 138   & 0.464 & 0.684 & 0.587 & 0.996 \\
    CHES  & EMP   & LBG   & wd    & 0.154 & 151   & 0.019 & 0.283 & 0.183 & 0.842 \\
    CHES  & EUP   & LBG   & wd    & 0.455 & 154   & 0.323 & 0.57  & 0.465 & 0.978 \\
    CHES  & CHES  & MV    & pc    & 0.552 & 134   & 0.435 & 0.651 & 0.588 & 0.939 \\
    CHES  & EMP   & MV    & pc    & 0.161 & 141   & 0.014 & 0.301 & 0.181 & 0.891 \\
    CHES  & EUP   & MV    & pc    & 0.445 & 142   & 0.318 & 0.556 & 0.483 & 0.921 \\
    CHES  & CHES  & MV    & wd    & 0.208 & 138   & 0.085 & 0.324 & 0.274 & 0.759 \\
    CHES  & EMP   & MV    & wd    & -0.005 & 151   & -0.118 & 0.108 & -0.007 & 0.705 \\
    CHES  & EUP   & MV    & wd    & 0.166 & 154   & 0.027 & 0.299 & 0.188 & 0.887 \\
    EMP   & CHES  & LBG   & pc    & 0.228 & 138   & 0.075 & 0.37  & 0.244 & 0.936 \\
    EMP   & EMP   & LBG   & pc    & 0.094 & 151   & -0.048 & 0.232 & 0.106 & 0.883 \\
    EMP   & EUP   & LBG   & pc    & 0.177 & 154   & 0.033 & 0.313 & 0.193 & 0.916 \\
    EMP   & CHES  & LBG   & wd    & 0.070  & 138   & 0.019 & 0.122 & 0.233 & 0.303 \\
    EMP   & EMP   & LBG   & wd    & 0.080  & 151   & 0.009 & 0.151 & 0.181 & 0.443 \\
    EMP   & EUP   & LBG   & wd    & 0.056 & 154   & 0.005 & 0.106 & 0.177 & 0.316 \\
    EMP   & CHES  & MV    & pc    & 0.228 & 138   & 0.075 & 0.37  & 0.244 & 0.936 \\
    EMP   & EMP   & MV    & pc    & 0.094 & 151   & -0.048 & 0.232 & 0.106 & 0.883 \\
    EMP   & EUP   & MV    & pc    & 0.177 & 154   & 0.033 & 0.313 & 0.193 & 0.916 \\
    EMP   & CHES  & MV    & wd    & 0.022 & 138   & -0.011 & 0.055 & 0.114 & 0.194 \\
    EMP   & EMP   & MV    & wd    & 0.042 & 151   & -0.007 & 0.091 & 0.138 & 0.305 \\
    EMP   & EUP   & MV    & wd    & 0.019 & 154   & -0.015 & 0.053 & 0.089 & 0.215 \\
    \bottomrule
   \multicolumn{9}{l}{ * wd = whole dimension, pc = per country}
\end{longtable}
\end{landscape}

\newpage

The four graphs below shows scattermatrices between the wordscores using the LBG transformation (as can be found in the tables above) and the 2009 expert scores. The matrices were constructed in R using the \texttt{car} package and show the relations between the six data sets including a density plot over the diagonal axis.\\

\begin{figure}[!htbp]
	\centering
		\includegraphics[scale=.9]{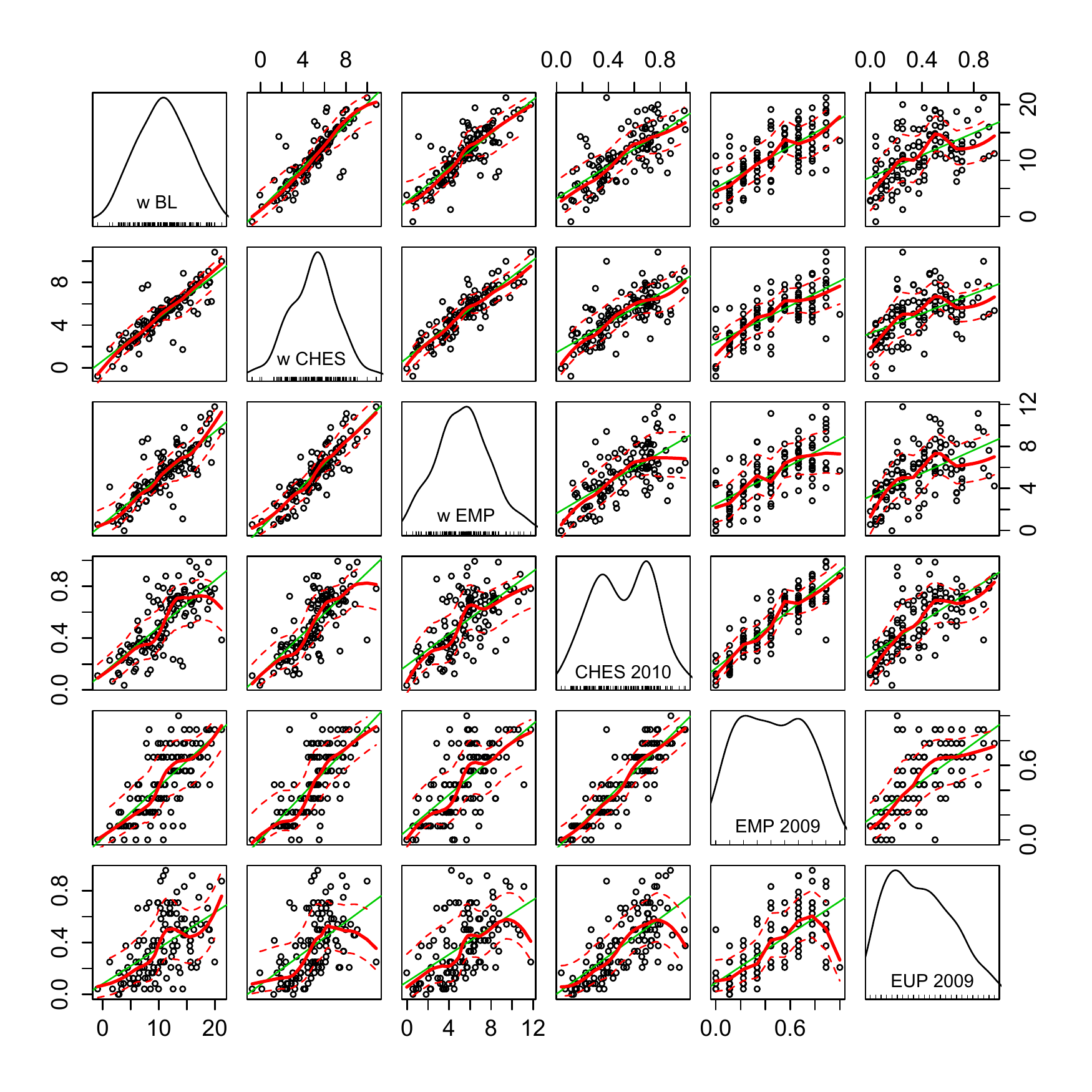}
		  \caption{Left-Right Dimension}
  \label{fig:leftrightmatrix}
  \end{figure}

\begin{figure}[!htbp]
	\centering
		\includegraphics[scale=.9]{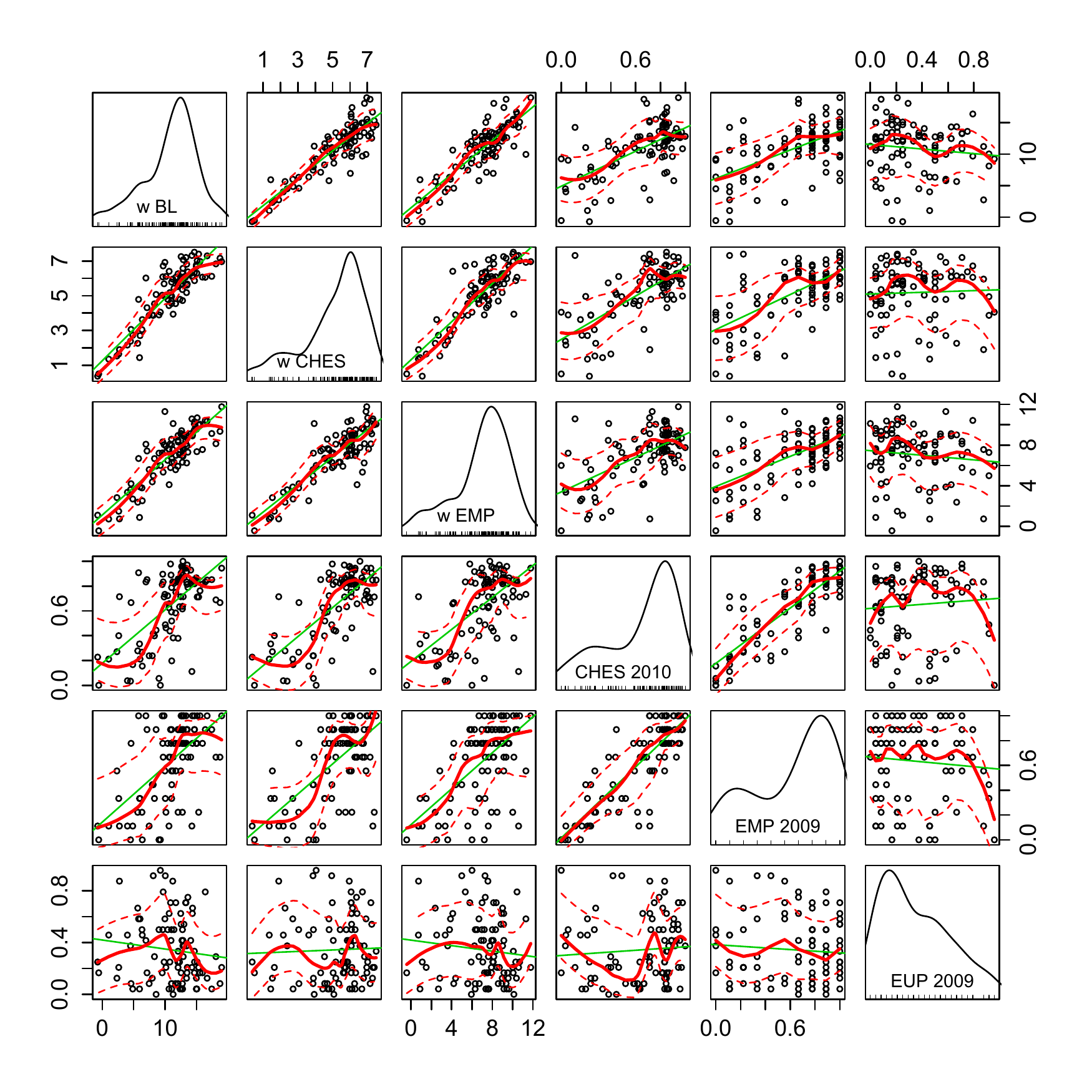}
		\caption{European Integration Dimension}
  \label{fig:euintmatrix}
\end{figure}

\begin{figure}[!htbp]
	\centering
		\includegraphics[scale=.9]{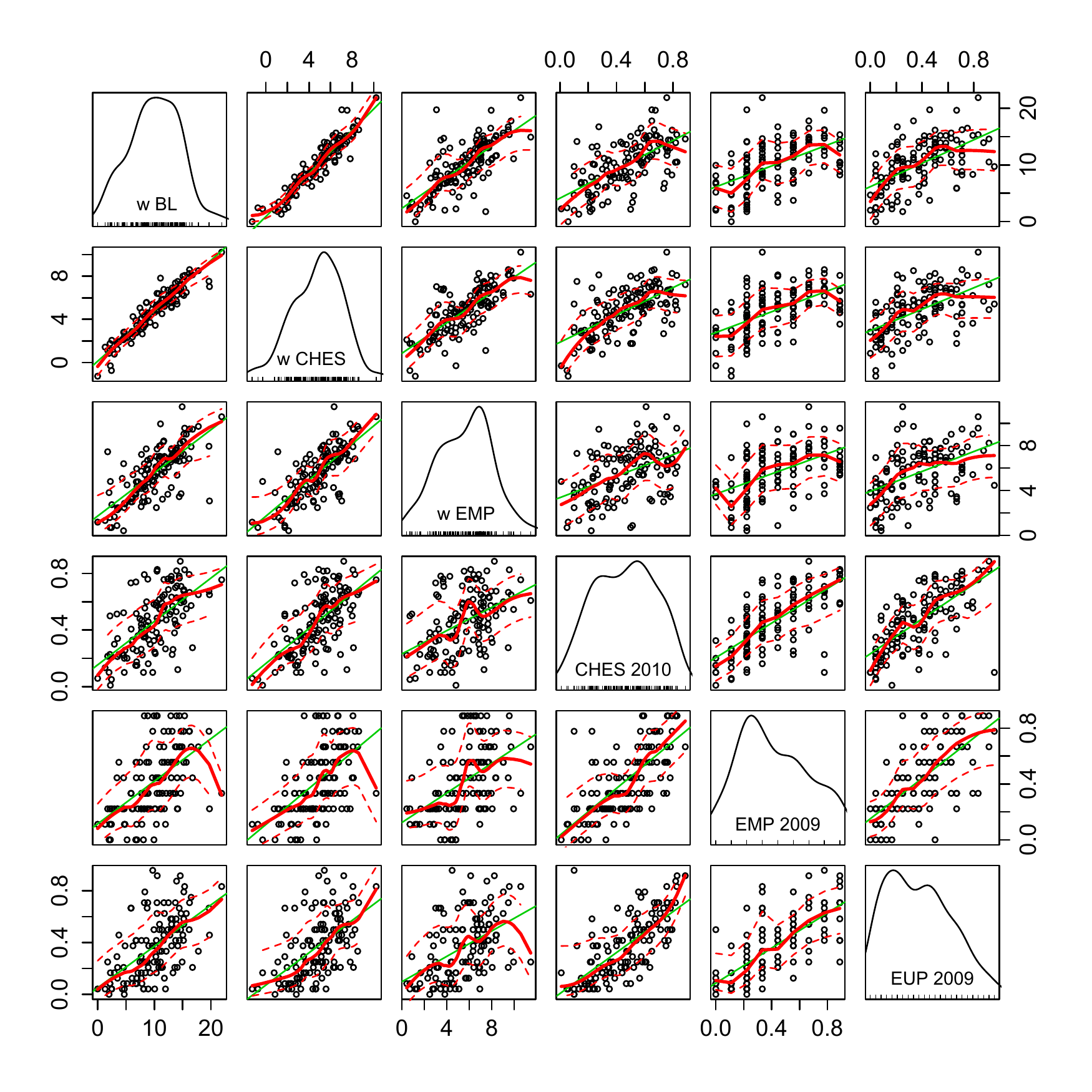}
		\caption{Economic Dimension}
  \label{fig:economicmatrix}
\end{figure}

\begin{figure}[!htbp]
	\centering
		\includegraphics[scale=.9]{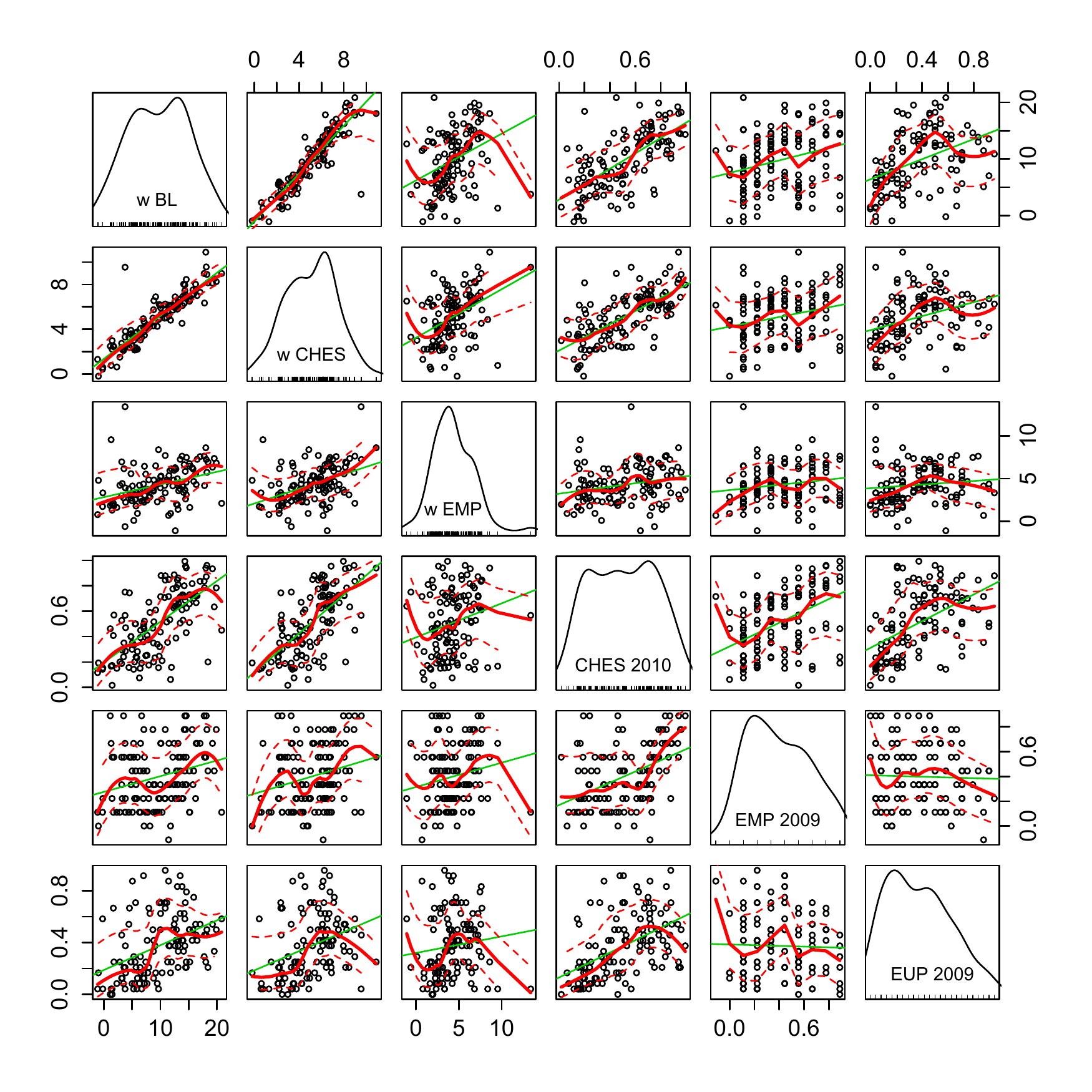}
		\caption{Social Dimension}
  \label{fig:socialmatrix}
\end{figure}

\newpage

\section*{Appendix H - Pearson's R without rescaling}

 \begin{table}[!ht]
 \centering
       \caption{LBG Transformation}
 \begin{tabular}{cccc}
    \toprule
    Dimension & Reference & Benchmark & Pearson's r \\
    \midrule
    EC & BL & CHES & 0.5796 \\
    EC &  BL & EUP & 0.5859 \\
    EC &  BL & EMP & 0.4934 \\
   EC &  CHES & CHES & 0.6045 \\
   EC &  CHES & EUP & 0.5891 \\
    EC & CHES & EMP & 0.5489 \\
    EC & EMP & CHES & 0.4291 \\
    EC & EMP & EUP & 0.4951 \\
    EC & EMP & EMP & 0.5005 \\
    EU & BL & CHES & 0.6217 \\
    EU & BL & EUP & 0.5864 \\
    EU & BL & EMP & 0.5148 \\
     EU & CHES & CHES & 0.6521 \\
     EU & CHES & EUP & 0.6151 \\
    EU & CHES & EMP & 0.5568 \\
     EU & EMP & CHES & 0.5772 \\
 EU & EMP & EUP & 0.6275 \\
 EU & EMP & EMP & 0.5162 \\
    LR & BL & CHES & 0.6928 \\
     LR & BL & EUP & 0.6104 \\
    LR & BL & EMP & 0.6882 \\
    LR & CHES & CHES & 0.6988 \\
     LR & CHES & EUP & 0.5909 \\
     LR & CHES & EMP & 0.7125 \\
    LR & EMP & CHES & 0.6119 \\
     LR & EMP & EUP & 0.5907 \\
     LR & EMP & EMP & 0.6689 \\
    SO & BL & CHES & 0.6211 \\
    SO & BL & EUP & 0.5416 \\
    SO & BL & EMP & 0.2592 \\
    SO & CHES & CHES & 0.6367 \\
    SO & CHES & EUP & 0.5382 \\
    SO & CHES & EMP & 0.2446 \\
    SO & EMP & CHES & 0.2106 \\
    SO & EMP & EUP & 0.1399 \\
    SO & EMP & EMP & 0.1764 \\
    \bottomrule
    \end{tabular}
    \end{table}

\newpage

 \begin{table}[!ht]
 \centering
        \caption{MV Transformation}
 \begin{tabular}{cccc}
    \toprule
    Dimension & Reference & Benchmark & Pearson's r \\
    \midrule
    EC & BL & CHES & 0.3675 \\
    EC & BL & EUP & 0.3548 \\
    EC & BL & EMP & 0.3589 \\
    EC & CHES & CHES & 0.3296 \\
    EC & CHES & EUP & 0.3977 \\
    EC & CHES & EMP & 0.3296 \\
    EC & EMP & CHES & 0.1436 \\
    EC & EMP & EUP & 0.2424 \\
    EC & EMP & EMP & 0.3326 \\
    EU & BL & CHES & 0.4582 \\
    EU & BL & EUP & 0.4921 \\
    EU & BL & EMP & 0.3726 \\
    EU & CHES & CHES & 0.4864 \\
    EU & CHES & EUP & 0.489 \\
    EU & CHES & EMP & 0.4651 \\
    EU & EMP & CHES & 0.4097 \\
    EU & EMP & EUP & 0.4496 \\
    EU & EMP & EMP & 0.3745 \\
    LR & BL & CHES & 0.3614 \\
    LR & BL & EUP & 0.241 \\
    LR & BL & EMP & 0.3508 \\
    LR & CHES & CHES & 0.522 \\
    LR & CHES & EUP & 0.3684 \\
    LR & CHES & EMP & 0.4603 \\
    LR & EMP & CHES & 0.4092 \\
    LR & EMP & EUP & 0.3612 \\
    LR & EMP & EMP & 0.4462 \\
    SO & BL & CHES & 0.4025 \\
    SO & BL & EUP & 0.3015 \\
    SO & BL & EMP & 0.0763 \\
    SO & CHES & CHES & 0.4069 \\
    SO & CHES & EUP & 0.2971 \\
    SO & CHES & EMP & 0.0599 \\
    SO & EMP & CHES & 0.1135 \\
    SO & EMP & EUP & 0.0888 \\
    SO & EMP & EMP & 0.1383 \\
    \bottomrule
    \end{tabular}
    \end{table}

\newpage

\bibliographystyle{apsr}
\bibliography{wordscores}

\end{document}